\documentclass[preprint, 11pt, 3p]{elsarticle}

\usepackage{amssymb}

\usepackage[T1]{fontenc}

\usepackage{amsmath}
\usepackage{amsfonts}

\usepackage{comment}

\usepackage{mathrsfs}

\usepackage{bbm}
\usepackage{bm}

\usepackage{graphicx}
\usepackage{color}

\usepackage{algorithm}
\usepackage{algpseudocode}

\newtheorem{theorem}{Theorem}

\newtheorem{lemma}[theorem]{Lemma}
\newtheorem{remark}{Remark}

\newcommand{\R}{\mathbb{R}}
\newcommand{\C}{\mathbb{C}}

\newcommand{\N}{\mathbb{N}}

\newcommand{\DFT}{\mathrm{DFT}}
\newcommand{\FFT}{\mathrm{FFT}}
\newcommand{\F}{\mathcal{F}}

\renewcommand{\O}{\mathcal{O}}

\newcommand{\bx}{{\bm x}}
\newcommand{\bv}{{\bm v}}
\newcommand{\bw}{{\bm w}}

\newcommand{\sign}{\operatorname{sign}}

\journal{XXX}

\begin{document}

\begin{frontmatter}



\title{The discrete Green's function method for wave packet expansion \\
via the free Schrödinger equation}

\author[inst1,inst2]{Jan-Frederik Mennemann\corref{cor1}}
\ead{jan-frederik.mennemann@univie.ac.at}
\cortext[cor1]{Corresponding author}

\author[inst3]{Sebastian Erne}

\author[inst2,inst3]{Igor Mazets}

\author[inst2]{Norbert J. Mauser}

\affiliation[inst1]{
    organization={{Wolfgang Pauli Institut c/o Fak.~f.~Mathematik, Univ.~Wien}}, 
    addressline={Oskar-Morgenstern-Platz~1}, 
    city={Vienna},
    postcode={1090}, 
    country={Austria}
}

\affiliation[inst2]{
    organization={Research Platform MMM ``Mathematics-Magnetism-Materials'' c/o Fak. f. Mathematik, Univ.~Wien},
    addressline={Oskar-Morgenstern-Platz~1}, 
    city={Vienna},
    postcode={1090}, 
    country={Austria}
}

\affiliation[inst3]{
    organization={Vienna Center for Quantum Science and Technology, Atominstitut, TU~Wien}, 
    addressline={Stadionallee~2}, 
    city={Vienna},
    postcode={1020}, 
    country={Austria}}

\begin{abstract}

We consider the expansion of wave packets governed by the free Schrödinger equation.
This seemingly simple task plays an important role in simulations of various quantum experiments
and in particular in the field of matter-wave interferometry.
The initial tight confinement results in a very fast expansion of the
wave function at later times which significantly complicates an efficient and precise numerical evaluation.
In many practical cases the expansion time is too short for the validity of the
stationary phase approximation and too long for an efficient application 
of Fourier collocation-based methods.
We develop an alternative method based on a discretization of the free-particle propagator.
This simple approach yields highly accurate results which readily follows from the
exceptionally fast convergence of the trapezoidal rule approximation of integrals 
involving smooth and rapidly decaying functions.
We discuss and analyze our approach in detail and demonstrate how to estimate 
the numerical error in the one-dimensional setting.
Furthermore, we show that by exploiting the separability of the Green's function, 
the numerical effort of the multi-dimensional approximation is considerably reduced.
Our method is very fast, highly accurate, and easy to implement on modern hardware.

\end{abstract}




\end{frontmatter}


\section{Introduction}

The free expansion of wave functions describing massive, non-relativistic particles
is an important computational problem because of its high relevance for
a number of applications 
in matter-wave interferometry~\cite{baudon_1999, cronin_2009}, where
atoms are cooled down to temperatures in the microkelvin or 
even nanokelvin range, so that their motion becomes essentially quantum 
and thus dominated by wave phenomena, such as interference.

Initially, the atoms are confined in a magnetic or optical trap undergoing coherent manipulations like beam-splitting 
and beam-recombination operations.
To perform a measurement, the atoms are released from the trap
and their interference pattern is detected after a certain time of flight
in the field of gravity~\cite{dalfovo_1999, bloch_2008}, which can be accounted for using an accelerated reference frame.

If the measurement is performed not with individual atoms or ions \cite{stopp_2021}, 
but with Bose--Einstein condensates consisting of many thousands of atoms~\cite{grond_2010, mazets_2012}, atom-atom interactions are not negligible during the short initial stage of the expansion. 
However, this can be easily taken into account by solving the 
nonlinear Gross-Pitaevskii equation~\cite{bao_2003, mennemann_2015} for a couple 
of milliseconds on a moderately extended spatial grid until the largest part 
of the interaction energy has been converted into kinetic energy.
The subsequent expansion is basically ballistic and the problem of the time-of-flight expansion boils down to solving the free Schr\"odinger equation in three dimensions.

This two-stage approach has been successfully applied in our recent three-dimensional
simulations~\cite{mennemann_2021} of a bosonic Josephson-junction, where
we were able to reproduce the experimental results in~\cite{pigneur_2018} very well.
This problem looks simple, but, in fact, it is not.
First of all, the expansion time is limited by the size of the 
laboratory setup and, therefore, not long enough to ensure the applicability
of the stationary phase principle which yields an asymptotic expression
for the wave function in the far-field limit\footnote{Note that the discrete Green's function method is suboptimal for very small times $t$ as the free particle
propagator becomes singular at $t=0$. 
However, even for relatively small expansion times we find excellent convergence (see Sections 2.1 \& 2.2) 
such that this singularity is practically irrelevant.}.
On the other hand, due to the tight external confinement and the Heisenberg uncertainty principle,
a considerable amount of energy is stored in the trapped quantum gas.
This energy converts into kinetic energy when the atoms are released from the trap
leading to a fast expansion of the atomic cloud in the tightly confined directions.
Depending on the strength of the confinement and the time of flight,
the volume of the atomic cloud increases by a very large factor (up to $\sim 10^3$). 

Previously the free expansion phase was handled by a simple Fourier collocation method
consuming excessive amounts of computational resources.
Imposing periodic boundary conditions, the free wave packet expansion problem can be solved 
using the time splitting spectral \cite{bao_2002} (Fourier split step) method.
To this end, the wave function in the free Schrödinger equation is replaced by
a trigonometric polynomial and the equation is required to hold at the collocation points
$x_j$, $j=0,\dots,J$.
Since the potential in the free Schrödinger equation is zero,
the method reduces to a single time step which is computed in 
$\O(J \log J)$ time using the fast Fourier transform ($\FFT$).
The resulting numerical procedure is what we refer to as the Fourier collocation method.
Unfortunately, due to the vast expansion of the wave function, 
the size of the computational domain and therefore the number of collocation points 
$J$ is required to be very large\footnote{
Note that periodic boundary conditions are well suited to the situation where the system 
is strongly trapped, but in time-of-flight simulations the domain has to be chosen very large 
so that the effect of the unphysical boundary conditions is reduced.}.
In two and in particular in three spatial dimensions the number of required grid points 
becomes astronomically high.
Eventually, the number of required grid points is so large that
the numerical approximations of the initial and the expanded wave function 
cannot even be represented in local memory.
Furthermore, despite the favorable complexity of the $\FFT$-algorithm the computational 
effort in three spatial dimensions is considerable, revealing the need for a better numerical procedure.

The memory problem described above has been addressed in \cite{deuar_2016}.
The method uses two different spatial grids to represent 
the initial wave function $\psi_0$ on $\Omega_0$ and 
the final expanded wave function $\psi$ on $\Omega$.
Both grids employ $J$ grid points but the grid spacing of the final grid 
is enlarged by a factor $m \in \N$.
Consequently, the memory requirements are reduced dramatically which, however,
comes at the numerical costs of $m$ applications of the $\FFT$ of size $J$.
In fact, the algorithm in \cite{deuar_2016} can be seen as a clever way 
of computing only every $m$th value of the numerical approximation in the
Fourier collocation method.

The idea of computing an approximation of the expanded wave function 
on a much coarser spatial grid is motivated by a simple observation.
Wave functions are non-observable quantities. 
The detector measures essentially the density, i.e., 
the square of the absolute value of the wave function. 
If we are only interested in the density, there is no need to resolve 
the fine details in the real and imaginary part of the expanded wave function.
Nonetheless, also the method presented in~\cite{deuar_2016} is based on 
periodic boundary conditions and therefore the final domain $\Omega$ is still 
required to capture the entire non-zero part of the expanded wave function.
In other words, the size of $\Omega$ is determined 
by the fastest moving parts in the initial wave packet.
If $\Omega$ is too small, parts of the wave packet will contaminate the 
numerical solution by periodically reentering the domain from the boundaries.

Alternatively, one might consider the application of a
domain truncation technique like 
complex absorbing potentials~\cite{fevens_1999, WSM2023},
perfectly matched layers~\cite{zheng_2007, mennemann_2014, scrinzi_2014, poetz_2020}
or the recently introduced Fourier contour deformation approach \cite{kaye_2022, kaye_2023}.
However, this idea is not very helpful in solving the wave packet expansion problem since we are particularly interested in the interference pattern forming at late times which implies large spatial scales.
In other words, any domain truncation technique would literally eliminate most or all
valuable information accessible only in the expanded interference pattern.

An effective way to get rid of the above mentioned boundary condition issues
is to consider the integral formulation of the solution.
In particular, we propose to employ a simple discretization of the free particle propagator 
of the free Schrödinger equation.
This approach to solve the free wave packet expansion problem seems so obvious
that it is hard to believe that it has not been used before.
One reason for this could be that
a direct discretization of the single particle Green's function
seems to be too simple to yield accurate results.
Another reason might be that it was believed that the numerical effort to evaluate
the discrete free particle propagator is quadratic in the number
of grid points $J$ even in spatial dimensions higher than one.
It turns out that none of these assumptions are true.

In fact, we show that the most simple discretization of the underlying convolution formula yields 
stunningly accurate results.
In all examples presented below only a very modest number of grid points $J$ is needed 
until the numerical error hits the inevitable barrier caused by rounding errors in the 
double precision arithmetic\footnote{The machine precision of the employed system
using double precision arithmetic is $\epsilon \approx 2.22 \cdot 10^{-16}$.}.
However, the spectacular convergence rate observed in the examples is a well-known effect
in the numerical analysis literature and in particular in the field 
of pseudospectral methods.
It is based on the fact that the 
trapezoidal rule\footnote{
In our application the boundary terms in the trapezoidal quadrature rule vanish 
and hence the approximation coincides with the even more simple rectangular quadrature rule.}
approximation of an integral
for a rapidly decaying and sufficiently smooth function converges at 
a high-order algebraic, spectral, exponential or even super-exponential 
rate with respect to the number of employed discretization points.
In the context of this magical phenomenon we would like to mention the pioneering 
work in~\cite{goodwin_1949} as well as the famous review in \cite{trefethen_2014}.

With regard to the second issue,
we suspect that the separability of the multi-dimensional problem has been overlooked.
Quite obviously, the corresponding $d$-dimensional Green's function can be factorized into
$d$ one-dimensional free particle propagators.
By exploiting this simple observation the numerical effort of the 
multi-dimensional discrete Green's function approximation is reduced tremendously.
The numerical effort to solve the three-dimensional problem is, for example, no longer
in $\O(J^2)$ but in fact only in $\O(J^{4/3})$.
Here, for convenience only, we have assumed that the number of 
grid points $J=J_1 J_2 J_3$ needed to discretize 
the initial wave function $\psi_0$ on $\Omega_0$ coincides approximately with 
the number of grid points $K=K_1 K_2 K_3$ used to approximate the expanded wave function $\psi$ on $\Omega$.
In general, this assumption is not needed. 
It is rather possible to employ two separate spatial grids 
wherein $J_1, \dots, J_d$ and $K_1, \dots, K_d$ are adapted to the problem at hand.
Moreover, unlike in the case of the Fourier collocation method, the 
spatial grid corresponding to the final approximation is not required 
to cover the entire non-zero part of the expanded wave function.
In fact, it is possible to consider any finite $d$-dimensional rectangular domain
allowing to investigate the most interesting part of the wave function only.

The article is organized as follows.
In the remaining part of this introduction we will introduce 
the main problem in mathematical terms and give an important one-dimensional example.
By means of this example we also demonstrate the limitations of the widely 
used stationary phase approximation.
In Section~$2$, we introduce the one-dimensional discrete Green's function approximation 
including an error analysis for two important classes of initial wave functions.
Finally, Section~$3$ covers the discrete Green's function approximation 
for the multi-dimensional problem.
In particular, we present an implementation of the approximation
in three spatial dimensions which is then used to  
solve another set of non-trivial examples.

\subsection{The free wave packet expansion problem}

We consider the free Schrödinger equation
\begin{subequations}
\label{eq:free_schroedinger}
\begin{equation}
\label{eq:free_schroedinger_1}
i \hbar \partial_t \psi(\bx,t)
=
-\frac{\hbar^2}{2 m} \Delta \psi(\bx,t), \quad \psi(\cdot,t=0) = \psi_0
\end{equation}
with the boundary condition
\begin{equation}
\label{eq:free_schroedinger_2}
\lim_{|\bm{x}| \to \infty}|\psi(\bm{x},t)| = 0
\end{equation}
\end{subequations}
for the wave function $\psi: \R^d \times \R \rightarrow \C$ in $d$ spatial dimensions.
The initial wave function $\psi_0$ is assumed to be a smooth function that is either
compactly supported on 
\[
\Omega_0 = [-L_1/2, L_1/2] \times \dots \times [-L_d/2, L_d/2] \subset \R^d
\]
or rapidly decaying for $|\bx| \to \infty$.

We are interested in computing a numerical approximation 
of $\psi = \psi(\cdot,t)$ on
\begin{equation*}
\Omega = [a_1, b_1] \times \dots \times [a_d, b_d] \subset \R^d
\end{equation*}
at a fixed final time $t > 0$.
From the physics point of view, $\Omega$ corresponds 
to the area accessible by the imaging system in an experiment.

Typically, the initial wave function is expected to expand along all coordinate axes, and hence,
$\Omega_0 \subseteq \Omega$
is a reasonable requirement.
However, we will see that this requirement is not needed and that the ability to compute $\psi$
on a finite but otherwise arbitrary rectangular domain
is a valuable feature.

\subsection{Scaling}
In the numerical experiments presented below we measure length in units of 
$\ell_0 = 1 \times 10^{-6}$\,m, 
mass in units of the $^{87}$Rb atom mass $m_0 = 87$ amu
$\approx 1.45 \times 10^{-25}$\,kg and time in units of
$t_0 = m_0 \ell_0^2 / \hbar \approx 1.37 \times 10^{-3}$\,s.
As a result of this scaling we have $\hbar=m=1$ which is employed throughout all equations
presented below.
We note that this scaling is also used in ultracold atom experiments and hence
the numerical simulations are close to real-world observations.

\subsection{Example: Superposition of two phase-shifted Gaussians}
\label{sec:example_gaussians_1d}

\begin{figure}[!ht]
\centering
\includegraphics[width=0.24\textwidth]{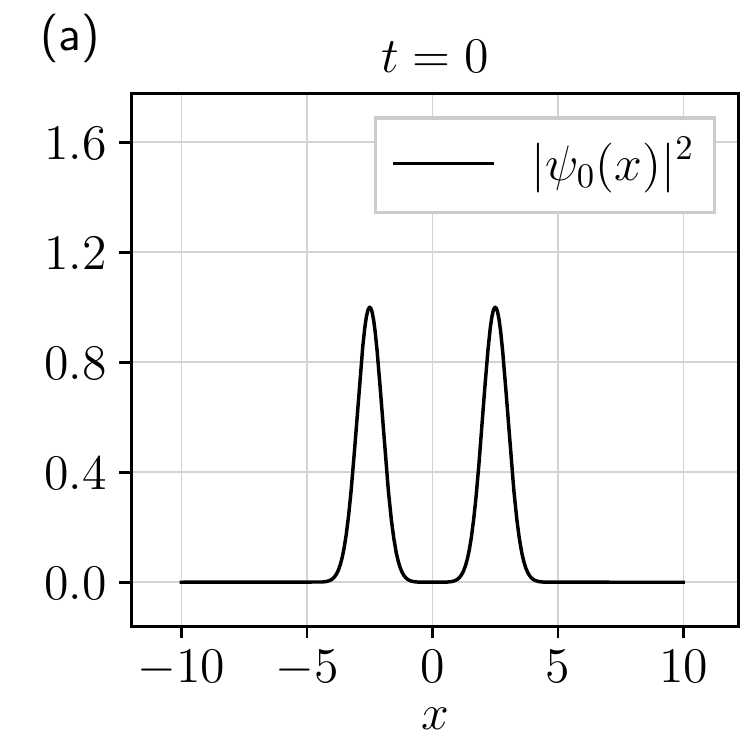}
\includegraphics[width=0.24\textwidth]{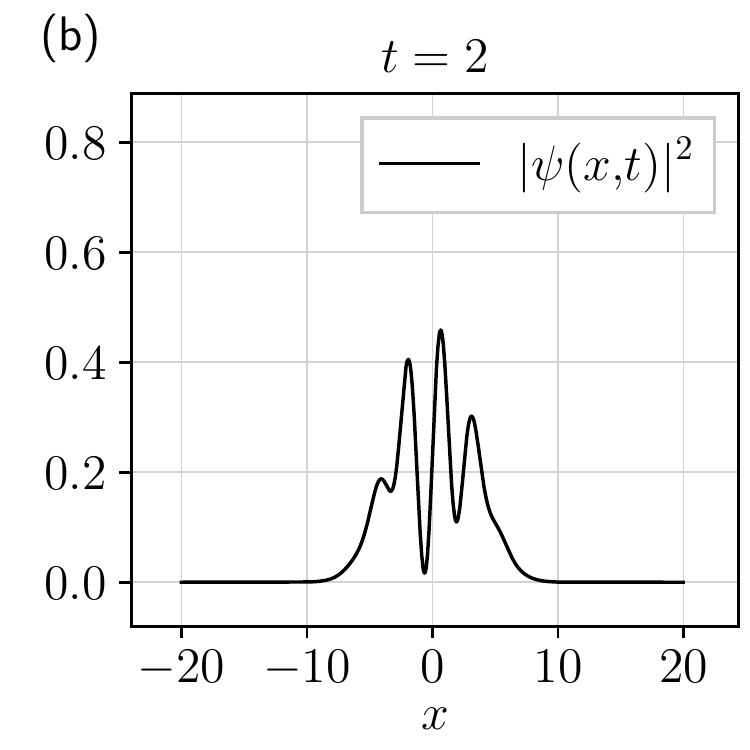}
\includegraphics[width=0.24\textwidth]{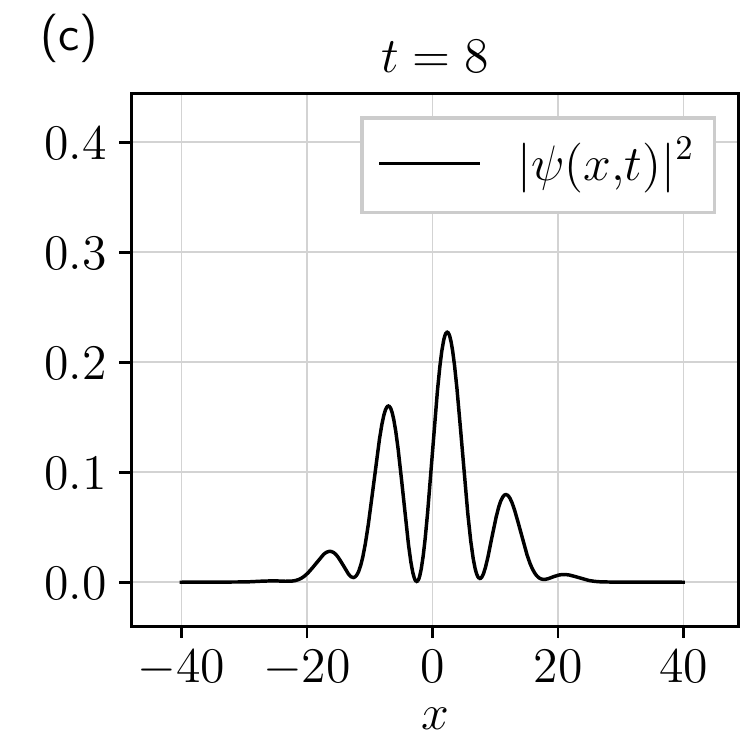}
\includegraphics[width=0.24\textwidth]{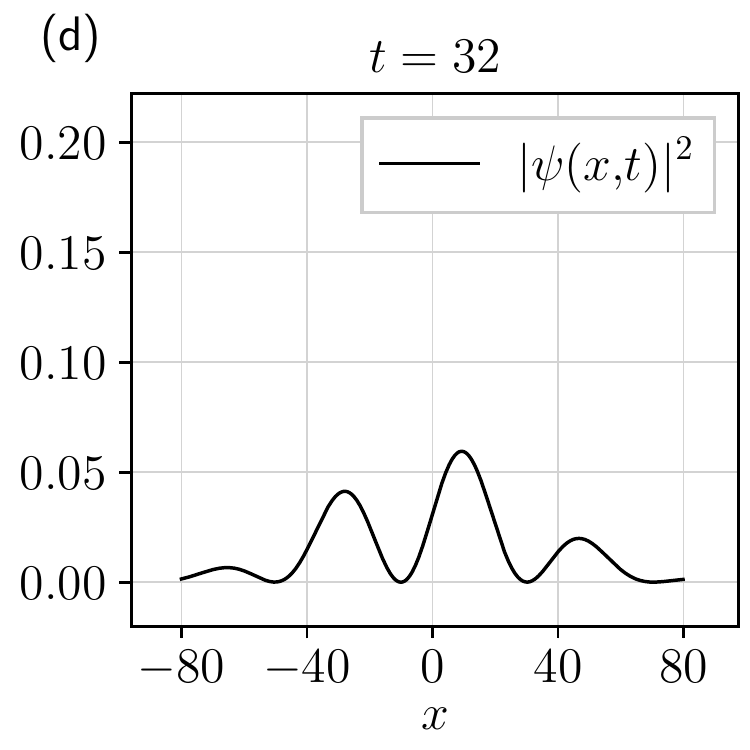}
\\
\includegraphics[width=0.24\textwidth]{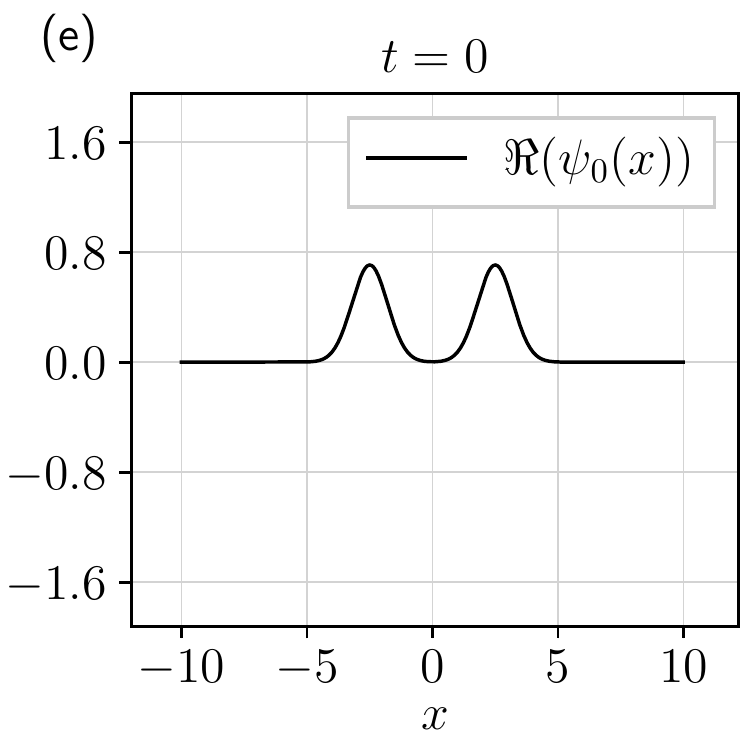}
\includegraphics[width=0.24\textwidth]{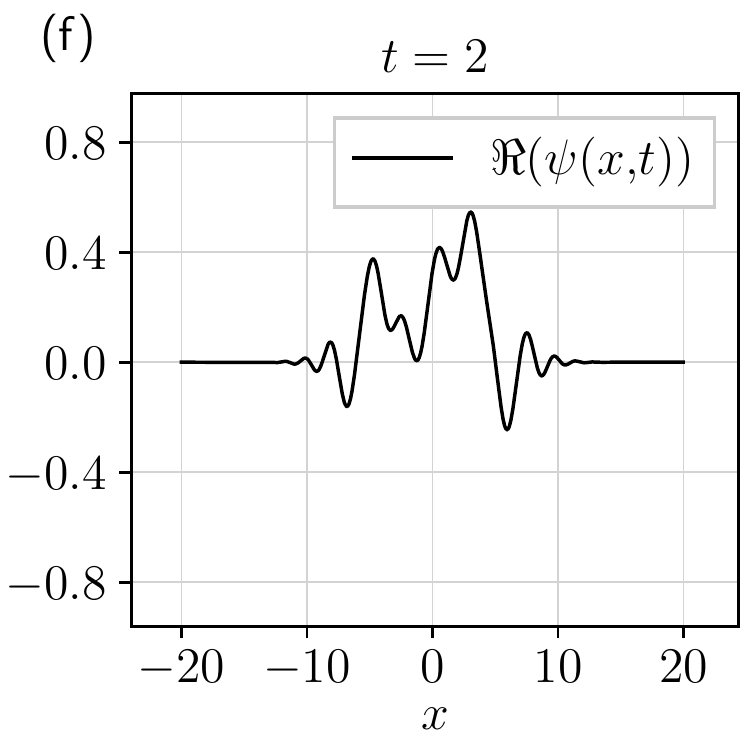}
\includegraphics[width=0.24\textwidth]{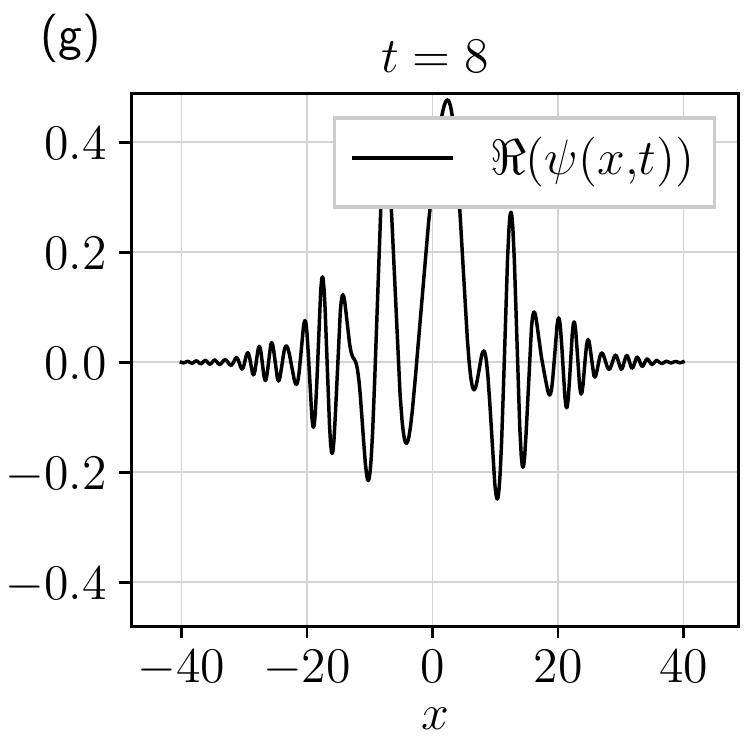}
\includegraphics[width=0.24\textwidth]{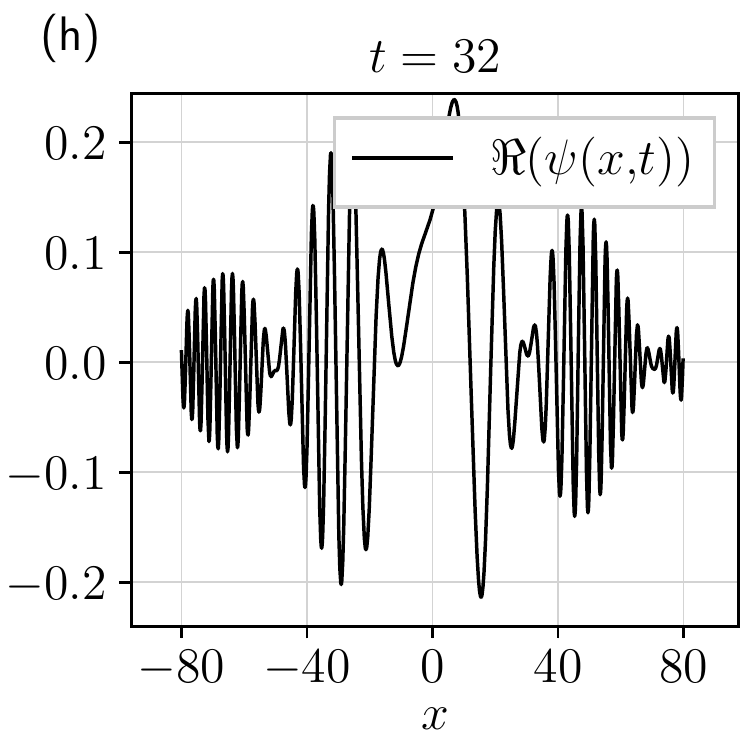}
\\
\includegraphics[width=0.24\textwidth]{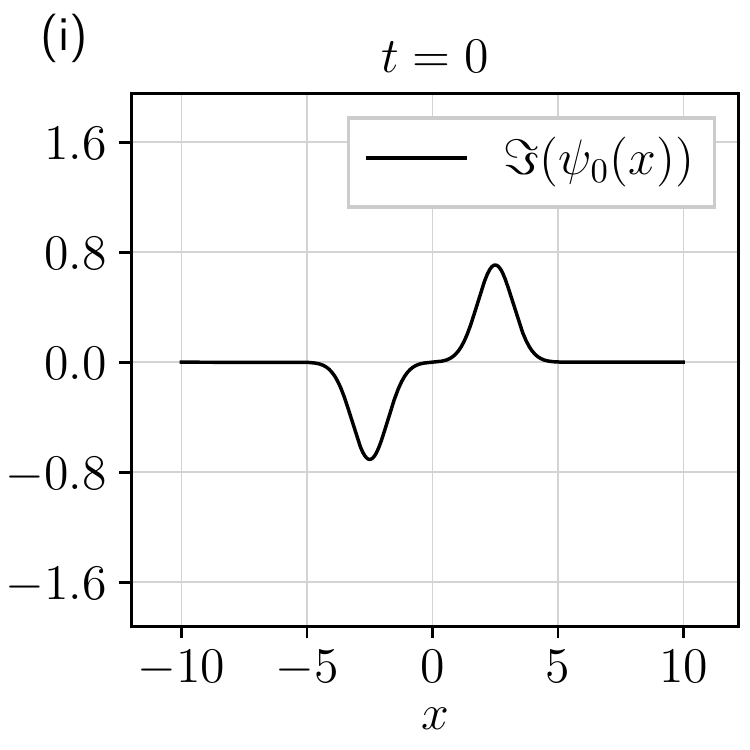}
\includegraphics[width=0.24\textwidth]{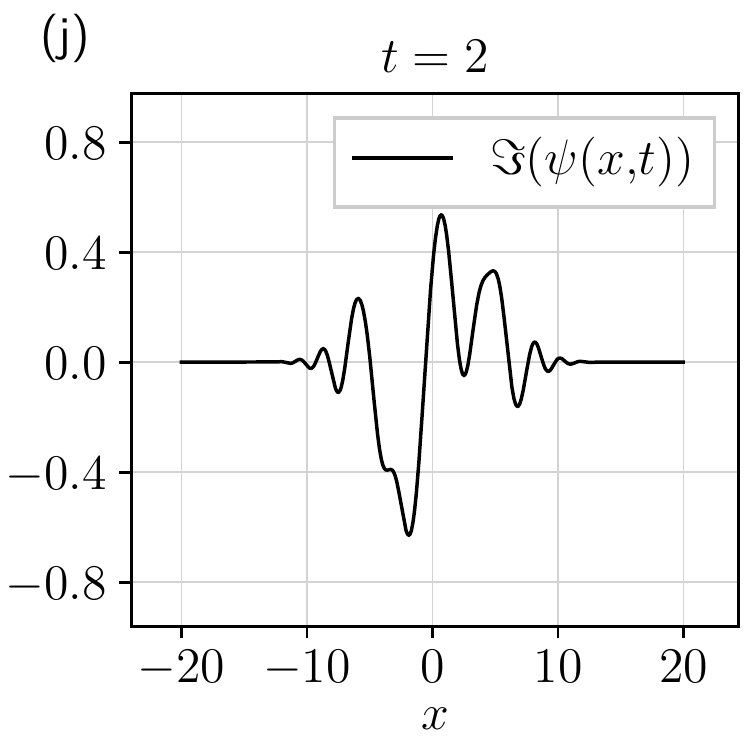}
\includegraphics[width=0.24\textwidth]{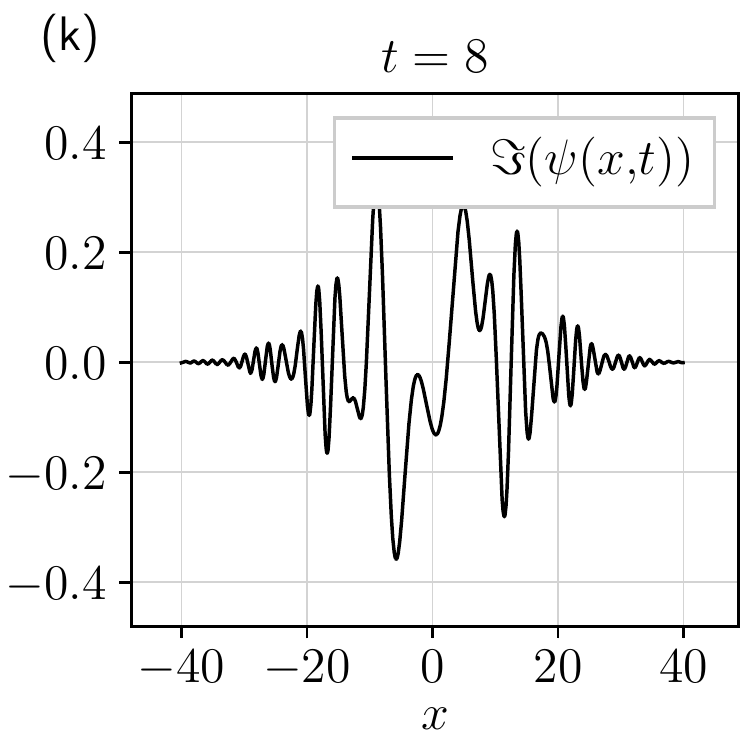}
\includegraphics[width=0.24\textwidth]{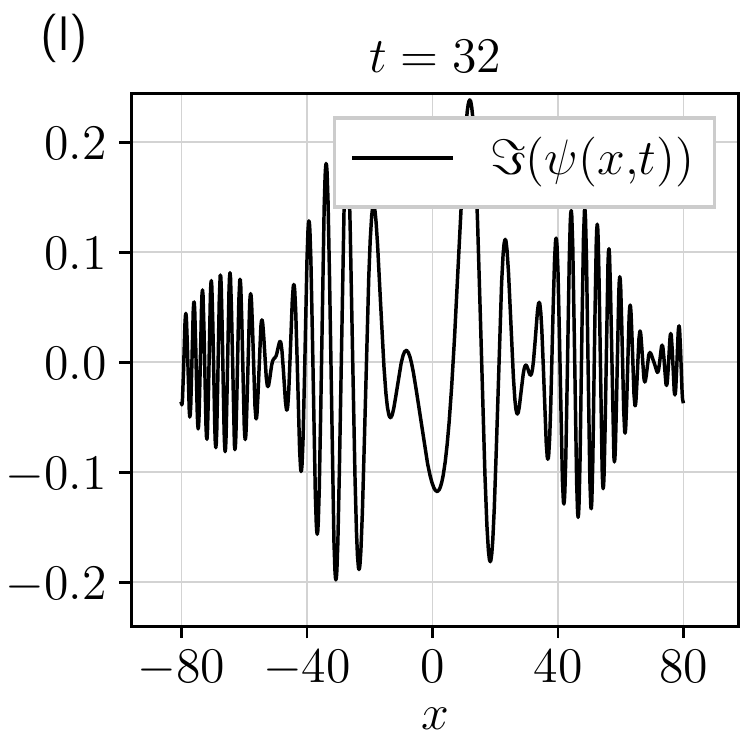}
\caption{
Free expansion of two phase shifted Gaussian wave packets~\eqref{eq:superposition_1d_solution}
at $t=0$, $t=2$, $t=8$ and $t=32$.
}
\label{fig:gaussians_1d_snapshots}
\end{figure}

As an illustration we consider a simple but important example.
The initial wave function is given by a superposition of two 
phase-shifted one-dimensional Gaussian wave packets
\begin{equation}
\label{eq:superposition_1d}
\begin{aligned}
\psi_0(x)
= 
e^{i \pi / 4} \exp \Big(-\frac{(x-\delta)^2}{4 \sigma^2} \Big)
+
e^{-i \pi / 4} \exp \Big(-\frac{(x+\delta)^2}{4 \sigma^2} \Big),
\end{aligned}
\end{equation}
where $\sigma = 1/2$ and $\delta=5/2$.
We note that $\psi_0$ is not normalized to one.
Due to the linearity of the free Schrödinger equation, 
the normalization is actually irrelevant.
The corresponding exact solution to the free Schr\"odinger equation is given by~\cite{schiff_quantum_mechanics}
\begin{equation}
\label{eq:superposition_1d_solution}
\begin{aligned}
\psi(x,t)
= 
\bigg[ \frac{1}{1 + i (t / \tau)} \bigg]^{1/2}
\bigg\{
&e^{i \pi / 4} 
\exp \Big( -\frac{(x-\delta)^2}{4 \sigma^2 [1 + i (t/\tau)]} \Big)
+
e^{-i \pi / 4} 
\exp \Big( -\frac{(x+\delta)^2}{4 \sigma^2 [1 + i (t/\tau)]} \Big)
\bigg\}
\end{aligned}
\end{equation}
for $x \in \R$ and $t \geq 0$ using $\tau = 2 \sigma^2$.

Fig.~\ref{fig:gaussians_1d_snapshots} shows the density, 
the real and the imaginary part of $\psi(x,t)$ in~\eqref{eq:superposition_1d_solution}
using $t=0$, $t=2$, $t=8$ and $t=32$.

\subsection{Asymptotic time evolution}

The solution to the one-dimensional free Schrödinger equation may be written as
\begin{equation}
\label{eq:sol_fse_1d_fourier}
\psi(x, t) 
= 
\frac{1}{2 \pi} \int_{-\infty}^\infty e^{i \xi x} 
e^{-i \frac{1}{2} \xi^2 t} \hat{\psi}_0(\xi) \,d\xi, \;\; x \in \R, \;\; t \geq 0,
\end{equation}
where
\begin{equation*}
\hat{f}(\xi)
=
\F \big\{f\big\}(\xi)
=
\int_{-\infty}^\infty e^{-i \xi x} f(x) \, dx, \;\; \xi \in \R
\end{equation*}
denotes the Fourier transform of an integrable function $f: \R \rightarrow \C$.
Using
\[
f(\xi) = \frac{1}{2 \pi} \hat{\psi}_0(\xi)
\quad \textrm{and} \quad
S_{x,t}(\xi)
=
\frac{1}{t} \xi x - \frac{1}{2} \xi^2
\]
Eq.~\eqref{eq:sol_fse_1d_fourier} reads
\[
\psi(x,t)
=
\int_{-\infty}^\infty 
f(\xi)
e^{i t S_{x,t}(\xi)}  \,d\xi
\]
which by means of the stationary phase principle \cite{bernadini_2009} yields the approximation
\begin{equation*}
\tilde{\psi}(x, t)
\approx
\sqrt{
\frac{2 \pi}{t |S''(\xi_0)|}
}
\exp
\Big\{
i t S_{x,t}(\xi_0) + i \frac{\pi}{4} \sign (S_{x,t}''(\xi_0))
\Big\}
f(\xi_0)
\end{equation*}
or
\begin{equation}
\label{eq:far_field_approximation_1d}
\tilde{\psi}(x, t)
\approx
\frac{1}{\sqrt{2 \pi t}}
e^{- i \frac{\pi}{4}}
e^{
i
\frac{x^2}{2 t}
}
\hat{\psi}_0\big(\tfrac{x}{t}\big).
\end{equation}

\begin{figure}[!t]
\centering
\includegraphics[width=1\textwidth]{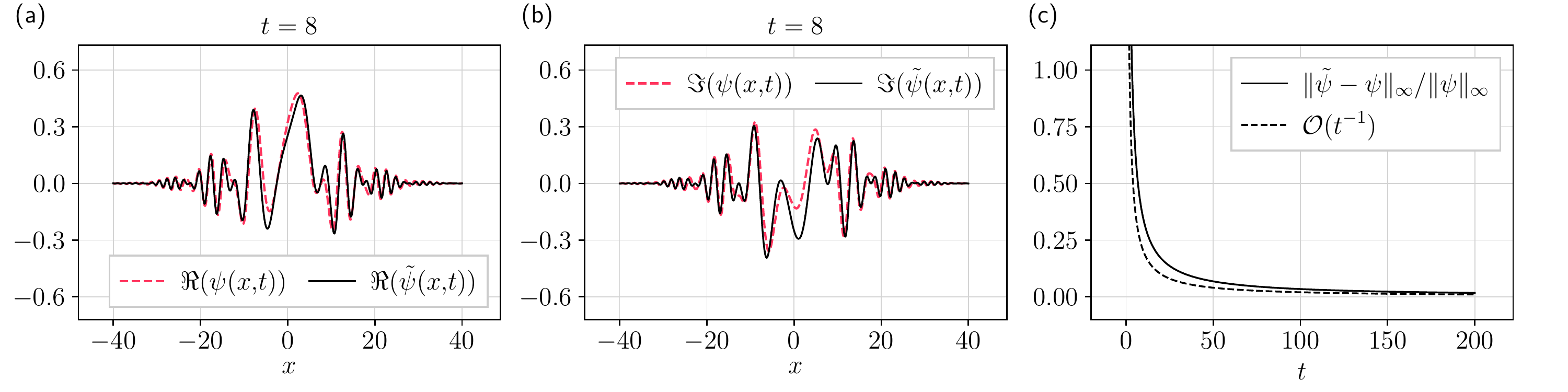}
\caption{
Real part (a) and imaginary part (b) of the asymptotic approximation $\tilde{\psi}$
in \eqref{eq:asymptotic_approx_gaussians_1d} and the exact solution $\psi$ 
in \eqref{eq:superposition_1d_solution} at $t=8$ for the initial wave function $\psi_0$ in 
\eqref{eq:superposition_1d}.
Relative error (c) of the asymptotic approximation $\tilde{\psi}(\cdot, t)$ on the interval 
$\Omega = [-40, 40]$ as a function of time $t$.
}
\label{fig:gaussians_1d_asymptotics_convergence}
\end{figure}

The exact order of convergence is a little tricky to calculate
as the phase factor $S_{x,t}(\xi)$ depends on the parameter $t$ itself.
However, replacing $\psi_0$ in~\eqref{eq:far_field_approximation_1d}
with the initial wave function \eqref{eq:superposition_1d} of the example in the
previous section yields
\begin{align}
\label{eq:asymptotic_approx_gaussians_1d}
\tilde{\psi}(x,t)
&\approx
\sqrt{\tfrac{2}{t}}
e^{- i \frac{\pi}{4}}
e^{
i
\frac{x^2}{2 t}
}
\sigma e^{-\sigma^2 \xi^2} 
\Big[
e^{i \varphi_1} e^{-i \xi \delta}
+
e^{i \varphi_2} e^{i \xi \delta}
\Big],
\end{align}
where we substitute $\xi = x / t$.
The real and imaginary part of the asymptotic approximation \eqref{eq:asymptotic_approx_gaussians_1d} 
and the exact solution \eqref{eq:superposition_1d_solution} 
are shown in
Fig.~\ref{fig:gaussians_1d_asymptotics_convergence}\,(a) and (b) on the interval $[-40, 40]$ for $t=8$.
Their relative difference on the same interval is shown in
Fig.~\ref{fig:gaussians_1d_asymptotics_convergence}\,(c) for $t \in [0, 200]$.
It is clearly visible that the relative error decreases not faster 
than $\O(t^{-1})$ and is still relevant for expansion times $\lesssim 50$\,ms in a real experiment.

\section{Discrete Green's function approximation for the one-dimensional problem}
\label{sec:discrete_green_s_function_approximation_1d}

\begin{figure}[!ht]
\label{fig:G_1d}
\centering
\includegraphics[width=1.0\textwidth]{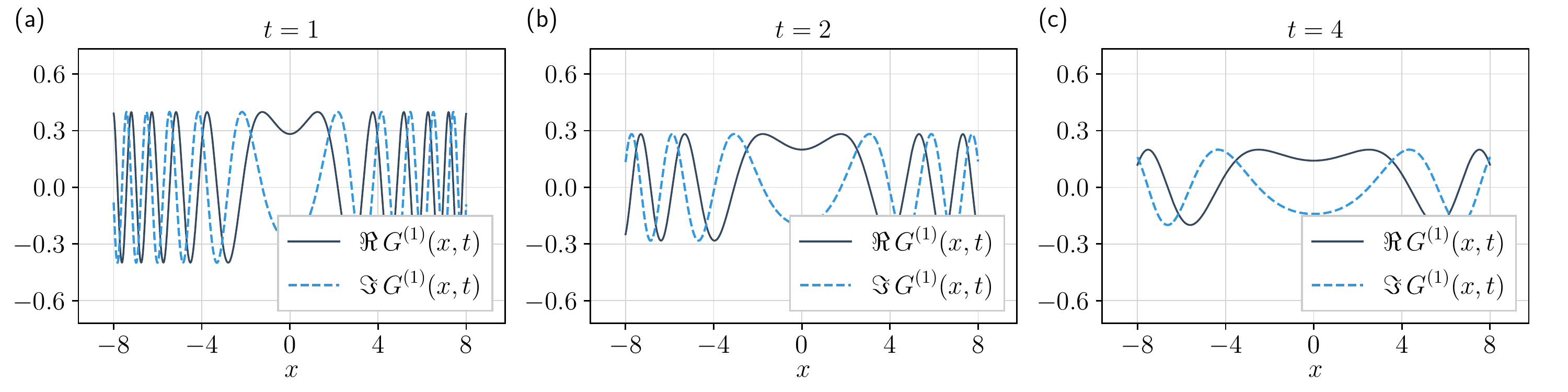}
\caption{
Real and imaginary part of the one-dimensional kernel $G^{(1)}$ in~\eqref{eq:G_1d} 
for $t=1$, $t=2$ and $t=4$.
}
\end{figure}

Since the solution of the multi-dimensional problem can be reduced to the solution 
of several one-dimensional problems, we consider the one-dimensional problem first.

\subsection{Discrete convolution}

The Green's function formalism for the one-dimensional free Schrödinger equation reads
\begin{equation}
\label{eq:psi_x_t_1d_green_s_function}
\psi(x,t)
=
\int_{-\infty}^\infty G^{(1)}(x-x',t) \, \psi_0(x') \,dx', \quad x \in \R, \quad t > 0,
\end{equation}
where
\begin{equation}
\label{eq:G_1d}
G^{(1)}(x, t) 
= 
\frac{e^{-i \pi / 4}}{\sqrt{2 \pi t}} e^{i x^2/(2 t)},
\quad x \in \R
\end{equation}
is the one-dimensional free-particle propagator \cite{andrews_2008}.
Fig.~\ref{fig:G_1d} shows the real and imaginary part of $G^{(1)}$ 
for three different final times $t$.

The initial wave function $\psi_0$ is assumed to decay rapidly. 
Alternatively, $\psi_0$ is assumed to be compactly supported on
$\Omega_0 = [-L/2, L/2]$ for some $L>0$.
In either case we let
\begin{equation*}
\bm{\psi}_0
=
[\psi_0(x_0'), \dots, \psi_0(x_{J-1}')]^\top
\end{equation*}
denote a discrete representation of the initial wave function at the grid points
\begin{equation}
\label{eq:grid_points_x_prime_1d}
x_j'
=
(j-J/2) \triangle x', \quad \triangle x' = L/J, \quad j=0,\dots, J-1
\end{equation}
for an even\footnote{For the sake simplicity, we assume $J$ to be even such that
the grid point $x_{J/2}^\prime=0$ is included 
in the set of grid points~\eqref{eq:grid_points_x_prime_1d}.} 
integer $J \in \N$. 

Our aim is to compute a numerical approximation of $\psi(\cdot,t)$ on the interval $\Omega = [a, b]$.
To this end, we introduce the grid points
\begin{equation}
\label{eq:grid_points_x_1d}
x_k
=
a + k \triangle x, \quad \triangle x = (b-a)/K, \quad k=0,\dots, K-1
\end{equation}
and the approximation
\[
\tilde{\bm{\psi}}
= 
\begin{bmatrix}
\tilde{\psi}_0, \dots, \tilde{\psi}_{K-1}
\end{bmatrix}^\top,
\]
where
\[
\quad
\tilde{\psi}_k \approx \psi(x_k, t), \quad k=0, \dots, K-1
\]
for some 
integer\footnote{While $J$ is required to be even, it is not important 
whether $K$ is an even or an odd integer.
} $K \in \N$.

By means of \eqref{eq:psi_x_t_1d_green_s_function} we obtain
\begin{align*}
\psi(x_k,t)
\approx
\int_{-L/2}^{L/2} G^{(1)}(x_k - x', t) \, \psi_0(x') \,dx',
\end{align*}
which is exact if $\psi_0$ is compactly supported on $\Omega_0 = [-L/2, L/2]$.
The integral is then replaced by the discrete convolution
\begin{equation}
\label{eq:discr_conv_1d_single_value}
\tilde{\psi}_k
=
\triangle x'
\sum_{j=0}^{J - 1} G^{(1)}(x_k - x_j', t) \psi_0(x_j'),
\quad k=0, \dots, K-1.
\end{equation}

In practice, all approximations are computed simultaneously using
\begin{subequations}
\label{eq:discr_conv_1d}
\begin{equation}
\label{eq:discr_conv_1d_1}
\tilde{\bm{\psi}}
=
\triangle x' \,
\bm{G} \bm{\psi}_0
\end{equation}
with the discrete propagator
\begin{equation}
\label{eq:discr_conv_1d_2}
\bm{G} = (G_{kj}) \in \C^{K \times J}, 
\quad G_{kj} = G^{(1)}(x_k - x_j', t),
\quad k=0,\dots,K-1,
\quad j=0,\dots,J-1.
\end{equation}
\end{subequations}

\begin{remark}
In the further course of this paper we will frequently consider the error 
$\|\tilde{\bm{\psi}} - \bm{\psi} \|_\infty$
and the relative error
$\|\tilde{\bm{\psi}} - \bm{\psi} \|_\infty / \| \bm{\psi} \|_\infty$
using the maximum norm $\|\bm{v}\|_\infty = \max_{k} |\bm{v}_k|$ for $\bm{v} \in \C^K$.
Here, 
\[
\bm{\psi} = (\psi_k) \in \C^K, \quad \psi_k = \psi(x_k), \quad k = 0, ..., K-1
\]
denotes the exact solution and 
$\tilde{\bm{\psi}} \in \C^K$ is the approximation in~\eqref{eq:discr_conv_1d}.
We point out that the accuracy of each of the $K$ approximations $\tilde{\psi}_k$, $k=0,...,K-1$ is independent from the parameter $K$ but depends only on the number of
quadrature points $J$.
The parameter $K$, on the other hand, determines the resolution of the expanded
wave function.
In the examples below we use relatively large values of $K$ which allow for a 
nice visualization of the computed approximations.
\end{remark}

\subsection{Application to the example from Section~\ref{sec:example_gaussians_1d}}

\begin{figure}[t]
\centering
\includegraphics[width=\textwidth]{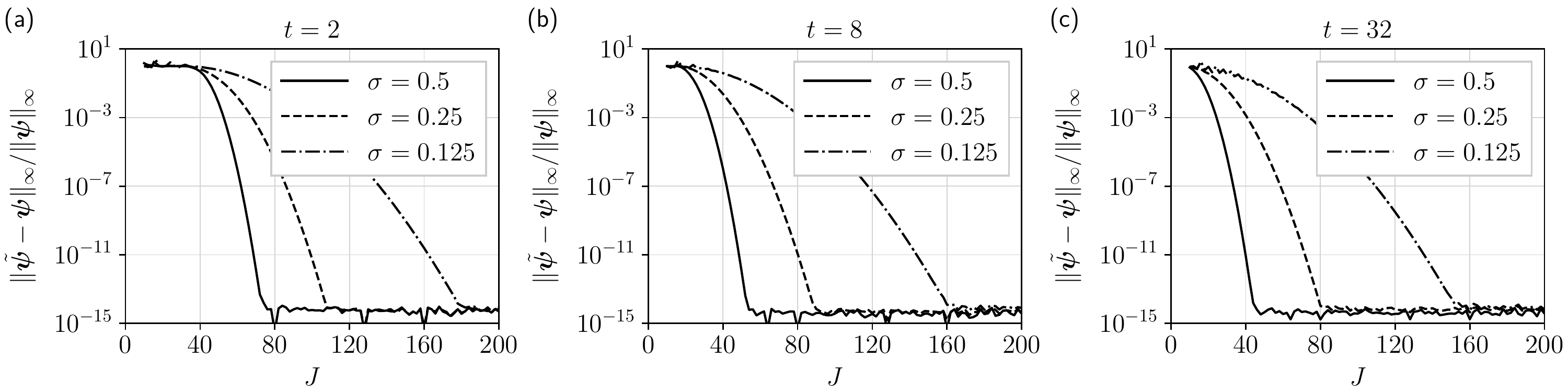}
\caption{
Relative error of the discrete Green's function approximation applied to
the initial wave packet $\psi_0$ in~\eqref{eq:superposition_1d} using two 
phase-shifted Gaussians of width $\sigma$.
}
\label{fig:gaussians_1d_discr_conv_convergence}
\end{figure}

As an example, we apply the discrete Green's function approximation~\eqref{eq:discr_conv_1d}
to the initial wave function in~\eqref{eq:superposition_1d}. 
In particular, we compute numerical approximations
of the expanded wave function $\psi(\cdot, t)$ on the intervals 
$\Omega = [-20, 20]$, $\Omega = [-40, 40]$ and $\Omega = [-80, 80]$ 
corresponding to the final times $t=2$, $t=8$ and $t=32$, respectively.
We always choose $L=20$ for the length of the initial domain $\Omega_0 = [-L/2, L/2]$ and
$K=1024$ for the number of grid points
in the final domain $\Omega$.

Fig.~\ref{fig:gaussians_1d_discr_conv_convergence} shows the relative error as a function of the number of grid points $J$ for the three final 
times $t=2$, $t=8$ and $t=32$.
The results are presented for different values for the widths $\sigma$ of 
the Gaussian wave packets in the initial wave function~\eqref{eq:superposition_1d}.
From the shape of the curves and the fact that we employ a semi-logarithmic scaling 
it is clearly visible that the numerically observed convergence rate is faster than exponential.
Only a very modest number of grid points $J$ is needed until the relative error hits the inevitable barrier caused by rounding errors in the double precision arithmetic.

As expected, the wave packet expansion problem becomes more difficult 
to solve when the smallest feature size (here the width $\sigma$) 
in the initial wave function gets smaller.
The same holds true if the final time $t$ becomes smaller
which is a direct consequence of the fact that the propagator in~\eqref{eq:G_1d} 
becomes singular for $t \to 0$.
Extremely small final times $t$ are, however, not of practical relevance in the free
wave packet expansion problem\footnote{For very small $t$ it is also possible
to employ a Fourier collocation method since the expansion of the wave function is minimal and hence the boundary conditions are of no importance.}.

\subsection{Error analysis for analytic initial wave functions}

In this section we consider rapidly decaying initial wave functions which are defined 
on the real line $\R$ and have an analytic extension to the strip in the complex plane
\begin{equation}
\label{eq:strip_Z}
\mathcal{Z}_c = \{ z \in \C: |\Im(z)| < c \}
\end{equation}
for some $c > 0$.

Our aim is to estimate the error of the discrete Green's 
function approximation~\eqref{eq:discr_conv_1d} at a fixed final time $t>0$.
To this end, we let
\begin{equation}
\label{eq:psi_tilde_of_x}
\tilde{\psi}(x,t)
=
\triangle x'
\sum_{j=0}^{J - 1} G^{(1)}(x - x_j', t) \psi_0(x_j')
\end{equation}
with $\triangle x' = L/J$ and $x_j'= (j-J/2) \triangle x'$
denote the numerical approximation of the solution
\begin{equation*}
\psi(x,t)
=
\int_{-\infty}^\infty G^{(1)}(x-x',t) \, \psi_0(x') \,dx'
\end{equation*}
at $x \in [a,b]$.
Within the Green's function approximation~\eqref{eq:discr_conv_1d}, 
the expression in \eqref{eq:psi_tilde_of_x} is evaluated at $x=x_k$, $k=0,\dots, K-1$
using $x_k = a + k \triangle x$ and $\triangle x = (b-a)/K$.
Since all grid points are located inside the interval $[a,b]$,
the error is bounded by
\begin{equation}
\label{eq:error_bound_1d_general}
\|\tilde{\psi}(\cdot, t) - \psi(\cdot, t) \|_{L^\infty[a,b]}
=
\max_{x \in [a,b]} |\tilde{\psi}(x,t)-\psi(x,t)|.
\end{equation}

In order to evaluate~\eqref{eq:error_bound_1d_general} we first consider the error
\begin{equation}
\label{eq:error_1d_fixed_x}
\mathcal{E} = |\tilde{\psi}(x,t)-\psi(x,t)|
\end{equation}
for a fixed $x \in [a,b]$.
Using
\begin{equation}
\label{eq:f_of_x_prime_real_line}
f:\R \rightarrow \C \;\;  \textrm{with} \;\; x' \mapsto G^{(1)}(x-x', t) \psi_0(x')
\end{equation}
and $h = \triangle x'$
we may write
\[
\mathcal{E}
=
\Big|
h \sum_{j=-J/2}^{J/2-1} f(j h) - \int_{-\infty}^\infty f(x') \, dx'
\Big|
\]
which yields the estimate
\begin{equation}
\label{eq:sum_error_trunc_discr}
\mathcal{E}
\leq
\mathcal{E}_\mathrm{trunc} + \mathcal{E}_\mathrm{discr},
\end{equation}
where we introduced the expressions
\[
\mathcal{E}_\mathrm{trunc}
=
\Big|
h \sum_{j=-J/2}^{J/2-1} f(j h) 
- h \sum_{j=-\infty}^\infty f(j h)
\Big|
\]
and
\begin{equation*}
\mathcal{E}_\mathrm{disrc}
=
\Big|
h \sum_{j=-\infty}^\infty f(j h)
-
\int_{-\infty}^\infty f(x') \, dx'
\Big|.
\end{equation*}

The first expression is the truncation error
\begin{equation*}
\mathcal{E}_\mathrm{trunc}
\leq 
h \sum_{|j| \geq J/2} |f(j h)|
\end{equation*}
which in our application
\begin{equation}
\label{eq:error_trunc}
\mathcal{E}_\mathrm{trunc}
\leq 
\frac{h}{\sqrt{2 \pi t}}
\sum_{|j| \geq J/2} |\psi_0(j h)|
\end{equation}
is independent from $x$.
Since $\psi_0$ is assumed to be a rapidly decaying function,
the truncation error in a typical application is extremely small.

The second term is the discretization error
\begin{equation}
\label{eq:error_I_h_discr}
\mathcal{E}_\mathrm{discr}
=
\Big|
h \sum_{j=-\infty}^\infty f(j h) - \int_{-\infty}^\infty f(x') \,dx'
\Big|.
\end{equation}
The discretization error implicitly depends on $x$ and $t$.
It decreases exponentially fast, provided 
$f$ meets the requirements of the following theorem~\cite{trefethen_2014}:

\begin{theorem}
\label{theo:error_analytic}
Let $f$ be a complex function defined on the whole real line.
Suppose further that $f$ has an analytic extension
to the strip $\mathcal{Z}_c$ in \eqref{eq:strip_Z} for some $c > 0$ 
and $f(z') \to 0$ uniformly as $|z'| \to \infty$ 
in the strip.
Moreover, for some $M$, it satisfies
\begin{equation}
\label{eq:integral_w}
\int_{-\infty}^\infty |f(x'+iy')| \,dx' \leq M
\end{equation}
for all $y' \in (-c, c)$.
Then, for any $h > 0$, 
\[
I_h = h \sum_{j=-\infty}^\infty f(j h)
\]
exists and satisfies
\begin{equation*}
|I_h - I| \leq \frac{2M}{e^{2 \pi c / h} - 1}
\end{equation*}
using
\[
I = \int_{-\infty}^\infty f(x') \, dx'.
\]
Moreover, the quantity $2 M$ in the numerator is as small as possible.
\end{theorem}

Provided the truncation error can be neglected, Theorem~\ref{theo:error_analytic} 
yields some constant $M^* > 0$ such that
\[
|\tilde{\psi}(x,t)-\psi(x,t)|
\leq \frac{2M^*}{e^{2 \pi c J / L} - 1}
\]
for every $x \in [a,b]$.
Asymptotically we have
\begin{equation*}
\|\tilde{\psi}(\cdot, t) - \psi(\cdot, t) \|_{L^\infty[a,b]}
\leq \alpha e^{-\beta J}
\end{equation*}
for some constants $\alpha, \beta > 0$ and hence
the error decrease exponentially fast (or faster) with the number of grid points $J$.

\begin{figure}[t]
\centering
\includegraphics[width=0.5\textwidth]{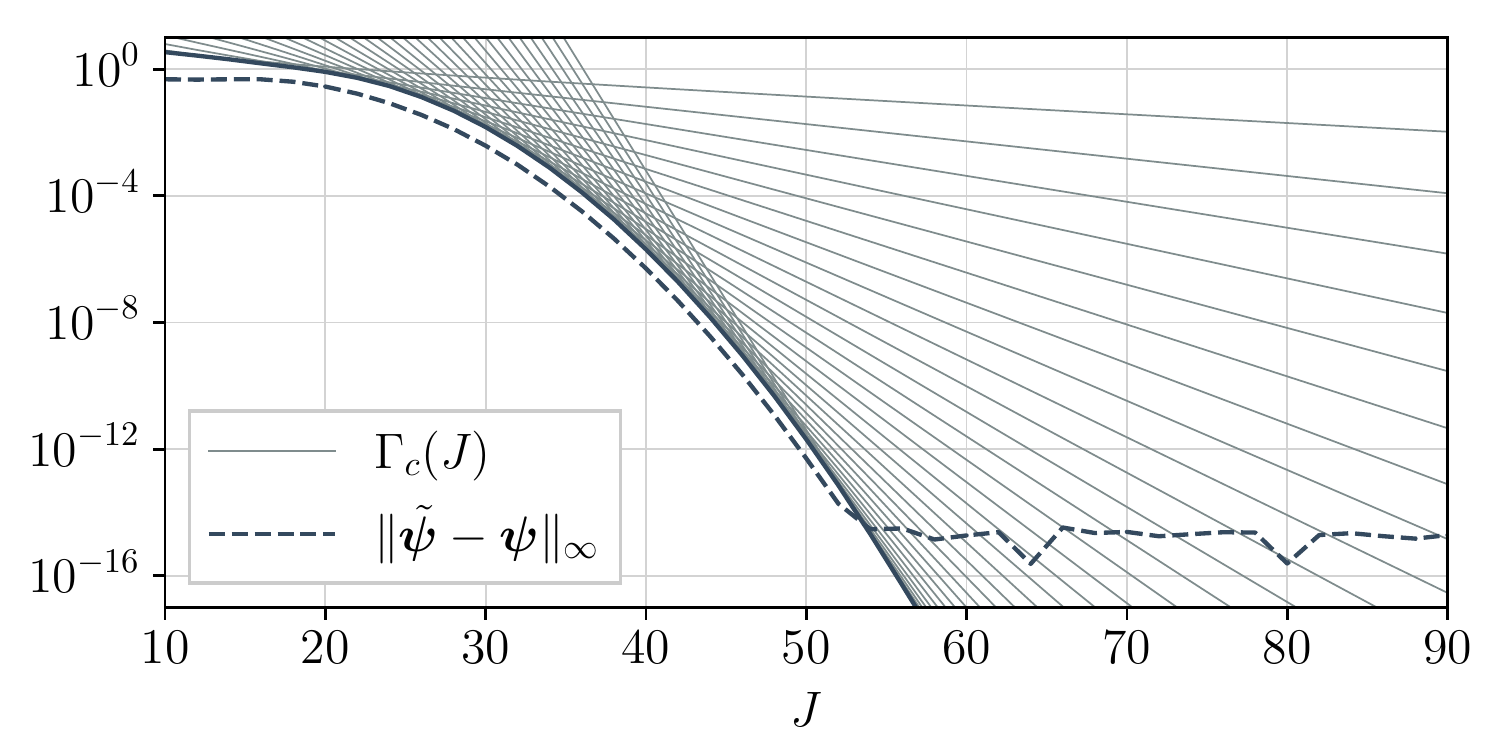}
\caption{
Convergence of the discrete Green's function approximation
for the example of two phase shifted Gaussian wave packets using the parameters
given in \eqref{eq:parameters_gaussians_1d_error_estimate}.
Error bounds $\Gamma_c(J)$ in \eqref{eq:estimate_sup_error_gaussians_1d} for 
$c=\ell \triangle c$ using $\triangle c = 0.2$ and $\ell=1,\dots,30$.
Also shown is the maximum difference between the discrete Green's function approximation
$\tilde{\bm{\psi}}$ and the exact solution $\bm{\psi}$ evaluated at $K=1024$ grid points
in the final domain $\Omega=[a,b]$.
}
\label{fig:gaussians_1d_error_analysis}
\end{figure}

For illustrative purposes, we consider the initial wave function $\psi_0$ in~\eqref{eq:superposition_1d} again.
In particular, we demonstrate the calculation of error bounds 
for the approximation of the expanded wave function at the final time $t=8$.
We use the same set of parameters
\begin{equation}
\label{eq:parameters_gaussians_1d_error_estimate}
\sigma=1/2, \;\; 
\delta=2.5,\;\; L = 20, \;\; a=-40, \;\; b=40
\end{equation}
that was used to compute the approximation depicted in the third column
of Fig.~\ref{fig:gaussians_1d_snapshots}.
Estimating the truncation error using~\eqref{eq:error_trunc} 
as well as calculating $M^*$ via 
Theorem~\ref{theo:error_analytic} is a relatively simple but tedious exercise
that is shown in~\ref{sec:appendix_error_estimate_gaussians_1d}.
The final result of these calculations is given as follows:
\begin{equation}
\label{eq:estimate_sup_error_gaussians_1d}
\|\tilde{\psi}(\cdot, t) - \psi(\cdot, t) \|_{L^\infty[a,b]}
\leq 
\Gamma_c(J)
\quad
\textrm{with}
\quad
\Gamma_c(J)
=
\frac{
e^{
\frac{257}{256} c^2 + \frac{85}{16} c 
}}
{e^{\pi c J / 10} - 1}
+ 4.22 \times 10^{-24}.
\end{equation}

We note that the initial wave function in the example is an entire function,
meaning that it has an analytic extension to the whole complex plane $\C$.
Since $G^{(1)}$ in \eqref{eq:G_1d} for fixed $t > 0$ is an entire function as well, the integrand $f$
in \eqref{eq:f_of_x_prime_real_line} has also an analytic extension to the whole complex
plane $\C$.
In particular it has an analytic extension to the strip $\mathcal{Z}_c$ in \eqref{eq:strip_Z} using any positive number $c$.
This is the reason why the error bound in
\eqref{eq:estimate_sup_error_gaussians_1d} 
is in fact true for any $c>0$.

In Fig.~\ref{fig:gaussians_1d_error_analysis} we show the
error bounds $\Gamma_c(J)$
for $c=\ell \triangle c$ using $\triangle c = 0.2$ and $\ell=1,\dots,30$.
We also show the maximum difference 
of the numerical approximation $\tilde{\bm{\psi}} \in \R^K$
and the exact solution $\bm{\psi} \in \R^K$ 
using $K=1024$ grid points in the final domain $\Omega=[-40,40]$.
It is immediately apparent that the envelope of the calculated error bounds is only slightly larger than the errors of the actual approximations.

\subsection{Error analysis for compactly supported initial wave functions}

We now consider initial wave functions of the form
\begin{equation}
\label{eq:psi_0_compact}
\psi_0(x)
=
\begin{cases}
u(x), 
\;\; &\textrm{for} \; x \in [-L/2, L/2],\\
0, \;\; &\textrm{otherwise},
\end{cases}
\end{equation}
where $u \in \C_0^{2 m + 2}([-L/2, L/2])$ for some $m \geq 0$.

Like in the previous section our aim is to estimate the error of the discrete Green's 
function approximation~\eqref{eq:discr_conv_1d} at a fixed final time $t>0$.
Using
\begin{equation*}
\tilde{\psi}(x,t)
=
\triangle x'
\sum_{j=0}^{J - 1} G^{(1)}(x - x_j', t) u(x_j')
\end{equation*}
and
\begin{equation*}
\psi(x,t)
=
\int_{-L/2}^{L/2} G^{(1)}(x-x',t) u(x') \,dx',
\end{equation*}
we consider the error
\begin{equation}
\label{eq:E_compact}
\mathcal{E} = |\tilde{\psi}(x,t) - \psi(x,t)|
\end{equation}
for fixed $x \in [a,b]$.
To simplify the notation, we introduce the function
\begin{equation}
\label{eq:f_of_x_prime_compact}
f:[-L/2, L/2] \rightarrow \C \;\;  \textrm{with} \;\; x' \mapsto G^{(1)}(x-x', t) u(x')
\end{equation}
and set $h = \triangle x'$.
Moreover, we include the rightmost grid point $x_J' = L/2$ into the set
of grid points~\eqref{eq:grid_points_x_prime_1d}.
Since $u(x_0') = u(-L/2) = 0$ and $u(x_J') = u(L/2) = 0$ we also find
$f(x_0') = f(-L/2) = 0$ and $f(x_J') = f(L/2) = 0$.
Consequently, we may write
\[
\tilde{\psi}(x, t)
=
h
\bigg[
f(x_0') / 2
+
\sum_{j=1}^{J - 1} f(x_j')
+
f(x_J') / 2
\bigg]
\]
which coincides exactly with the trapezoidal rule approximation of the integral
\begin{equation*}
\psi(x, t) = \int_{-L/2}^{L/2} f(x') \,dx'.
\end{equation*}

Estimates for the error of the trapezoidal rule approximation are readily available.
Based on the Euler-Maclaurin formula \cite{atkinson_1989, ralston_2001, weideman_2002} 
they are typically formulated for real-valued functions.
The case of a complex-valued function requires only a minor modification
and the corresponding result is given as follows:
\begin{theorem}
\label{theo:euler_maclaurin}
\begin{subequations}
Let $f:[-L/2, L/2] \to \C$  be $2m+2$ times continuously differentiable on $[-L/2, L/2]$ for some $m \geq 0$.
Further define $h = L / J$, $x_j' = -L/2 + j h$, $j=0, 1, \dots, J$
for some $J \geq 1$ and let 
\begin{equation}
\label{eq:I_J}
I_h
=
h
\bigg[
f(x_0') / 2
+
\sum_{j=1}^{J - 1} f(x_j')
+
f(x_J') / 2
\bigg]
\end{equation}
denote the trapezoidal rule approximation of the integral
\begin{equation}
\label{eq:I}
I = \int_{-L/2}^{L/2} f(x') \,dx'.
\end{equation}
Then, the error is bounded by
\begin{equation}
\label{eq:euler_maclaurin_error}
|I_h - I| 
\leq
\sum_{\ell=1}^m 
\lambda_\ell h^{2\ell}
+
\nu_m h^{2m+2},
\end{equation}
where
\begin{equation}
\label{eq:lambda_ell}
\lambda_\ell 
= 
\frac{1}{(2 \ell)!} |B_{2 \ell}| \, \big| f^{(2\ell-1)}(L/2) - f^{(2\ell-1)}(-L/2) \big|,
\quad \ell = 1, \dots, m
\end{equation}
and
\begin{equation}
\label{eq:nue_m}
\nu_m = \frac{2 L }{(2m+2)!} |B_{2m+2}| \, \big\| f^{(2m+2)} \big\|_{L^\infty[-L/2,L/2]}.
\end{equation}
\end{subequations}
The factors $B_\ell$ denote
the Bernoulli numbers for $\ell \in \N_0$.
\end{theorem}

Using Theorem~\eqref{theo:euler_maclaurin} we can estimate the error \eqref{eq:E_compact} for a fixed $x \in [a,b]$.
Moreover, it is possible to compute error bounds $C_m(J)$ such that
\begin{equation}
\label{eq:error_estimate_compact_1d}
\|\tilde{\psi}(\cdot, t) - \psi(\cdot, t) \|_{L^\infty[a,b]}
\leq C_m(J)
\end{equation}
for every integer $J>0$.
Since the calculations are of very technical nature we leave the details of our approach to the interested reader, see~\ref{sec:appendix_compactly_supported}.

\begin{figure}[!t]
\centering
\includegraphics[width=0.24\textwidth]{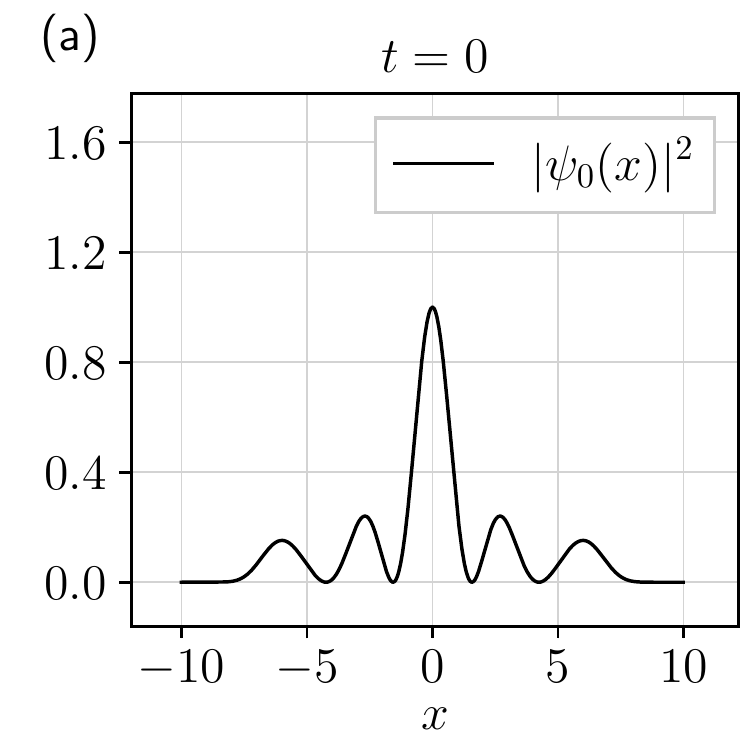}
\includegraphics[width=0.24\textwidth]{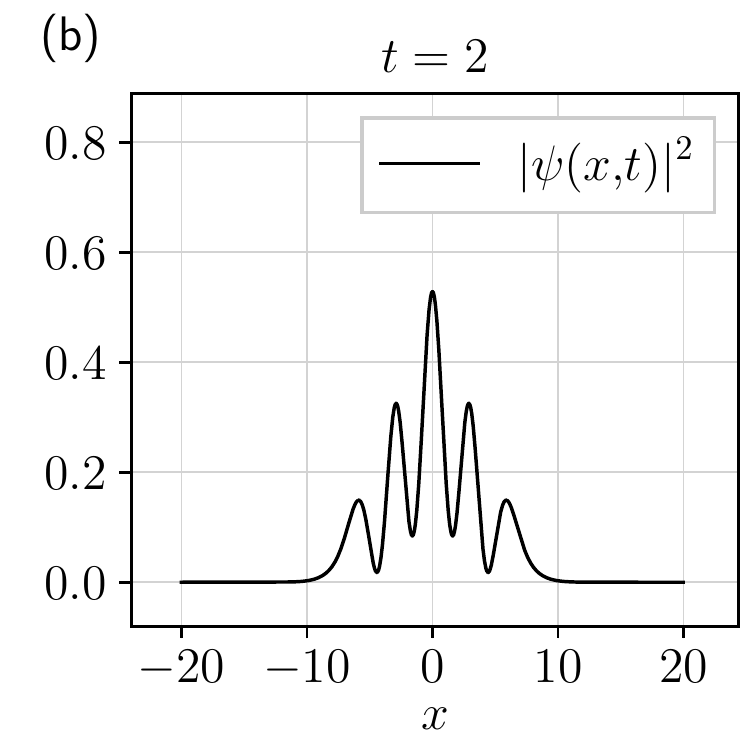}
\includegraphics[width=0.24\textwidth]{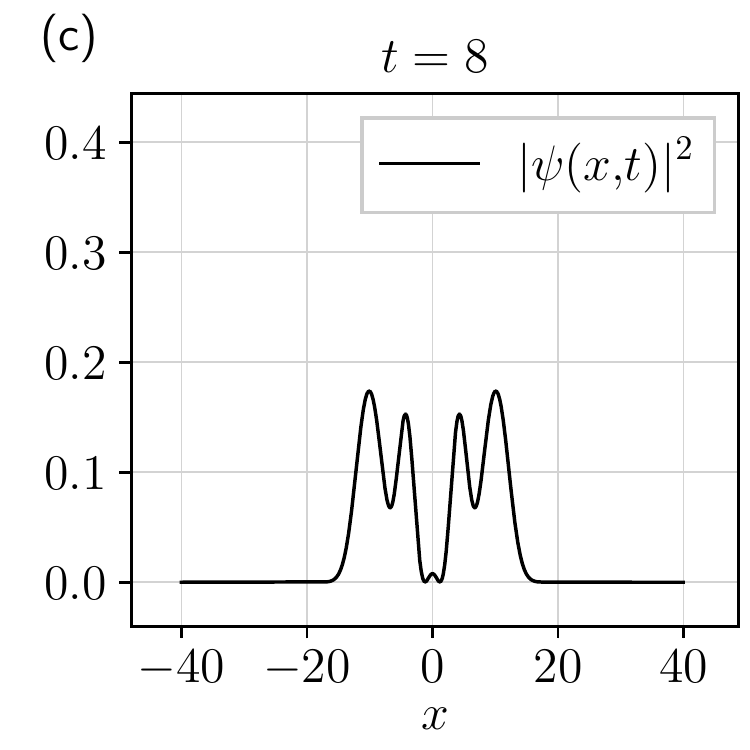}
\includegraphics[width=0.24\textwidth]{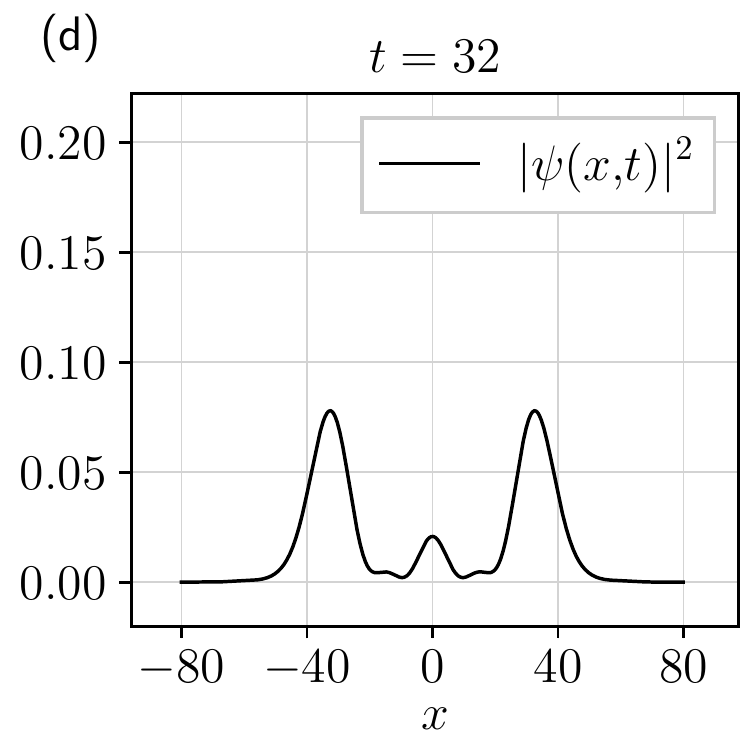}
\\
\includegraphics[width=0.24\textwidth]{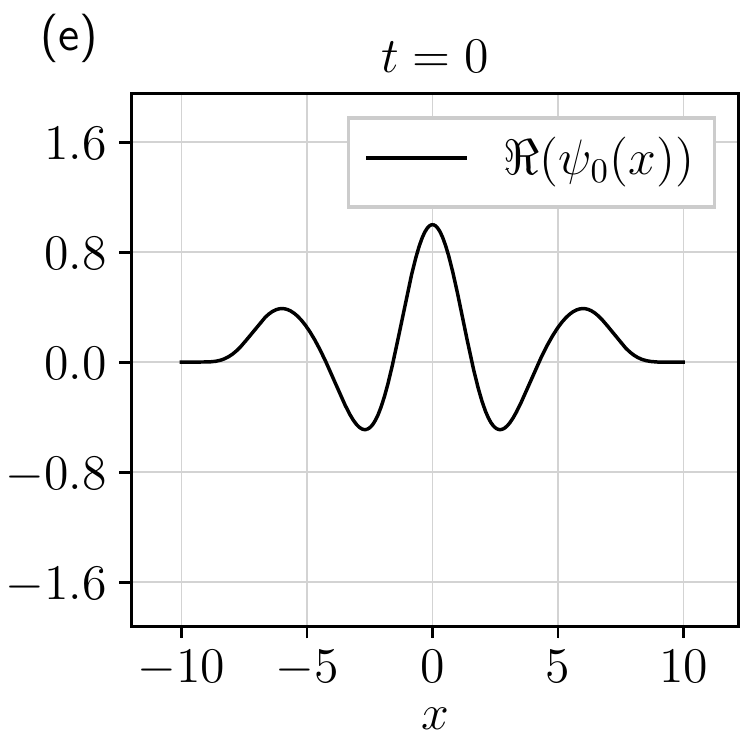}
\includegraphics[width=0.24\textwidth]{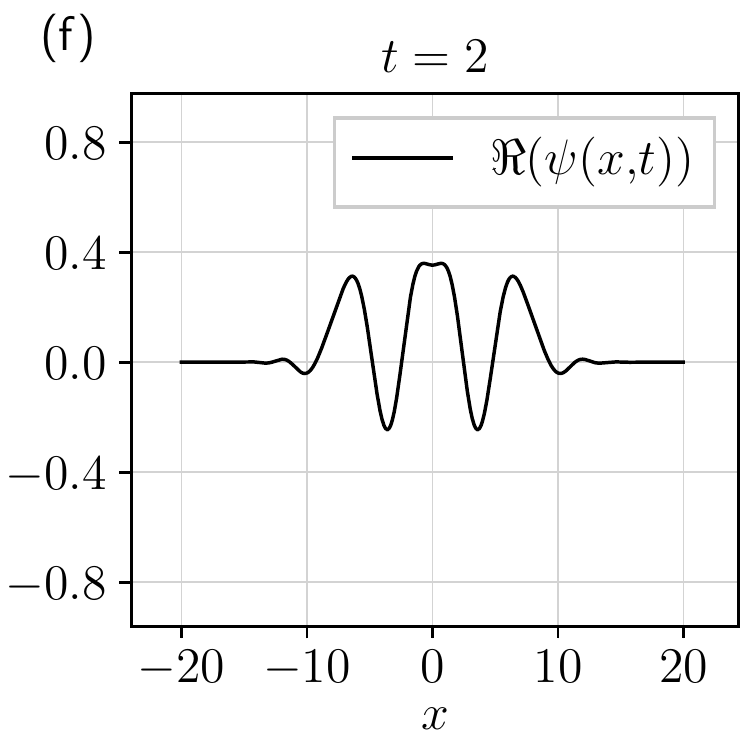}
\includegraphics[width=0.24\textwidth]{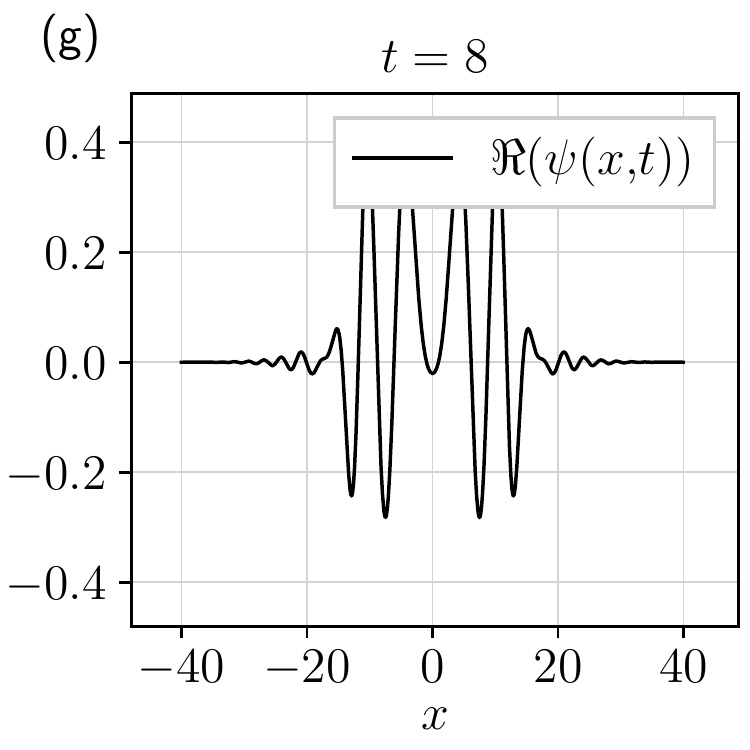}
\includegraphics[width=0.24\textwidth]{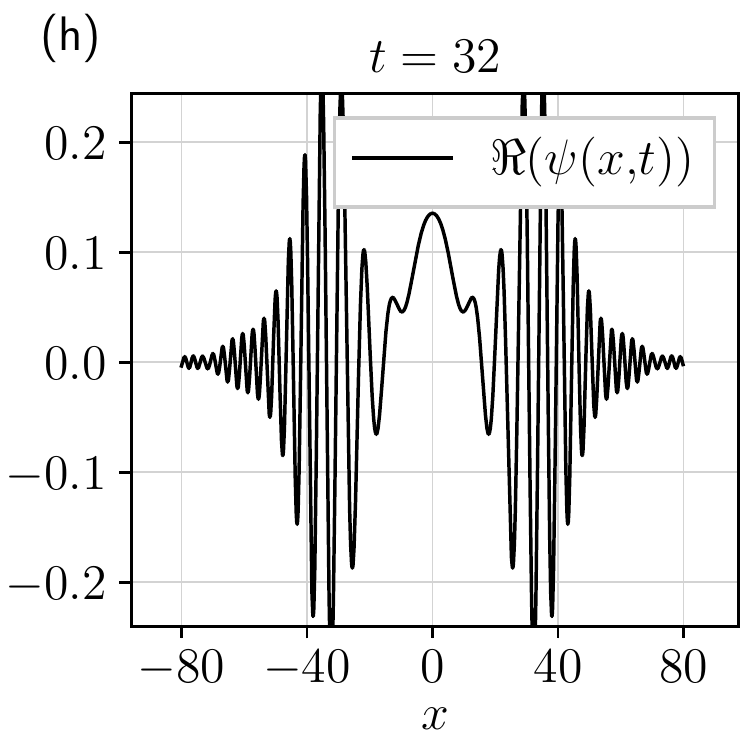}
\\
\includegraphics[width=0.24\textwidth]{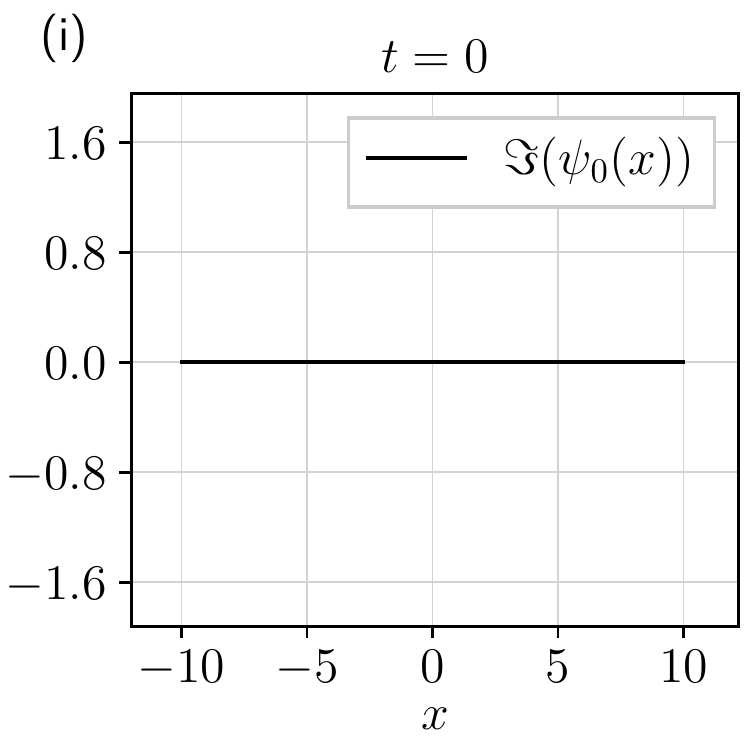}
\includegraphics[width=0.24\textwidth]{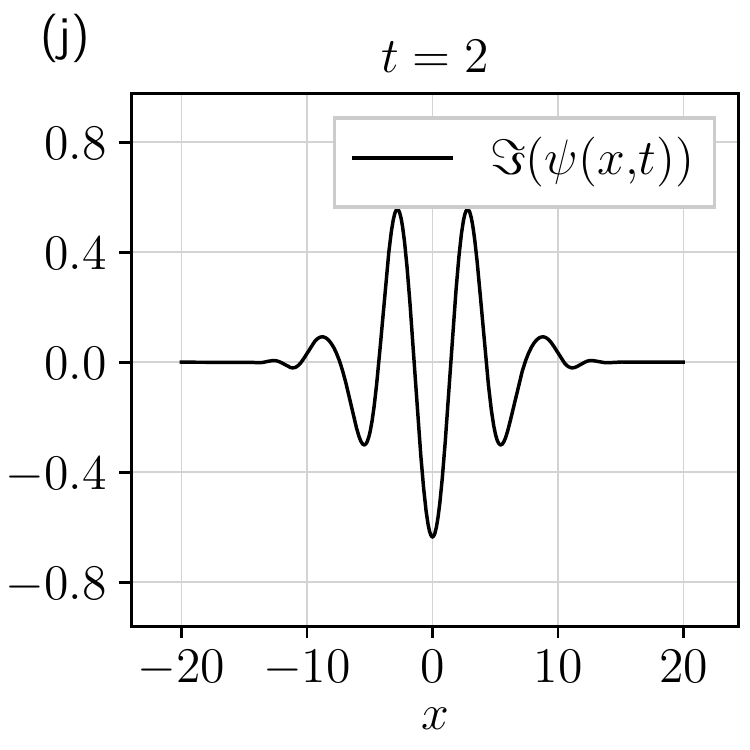}
\includegraphics[width=0.24\textwidth]{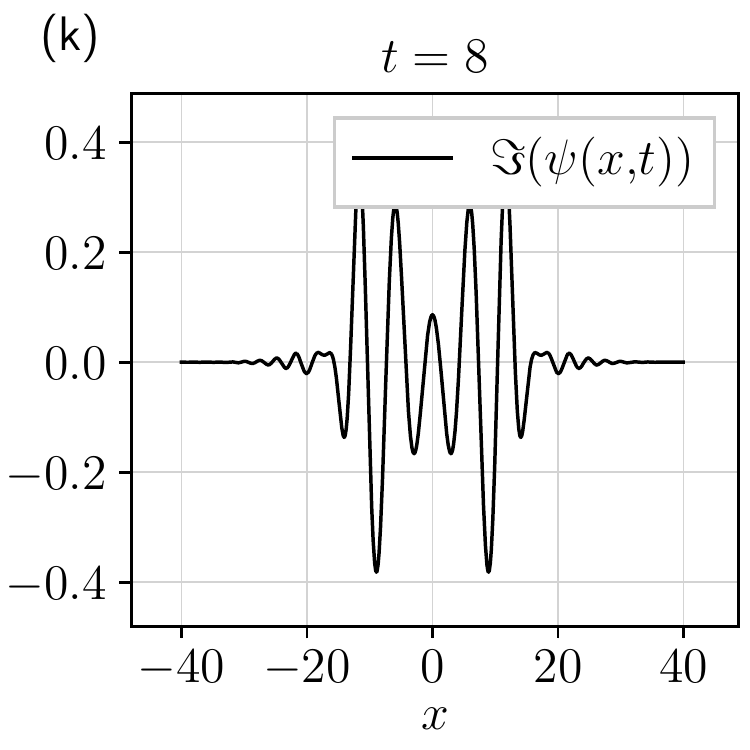}
\includegraphics[width=0.24\textwidth]{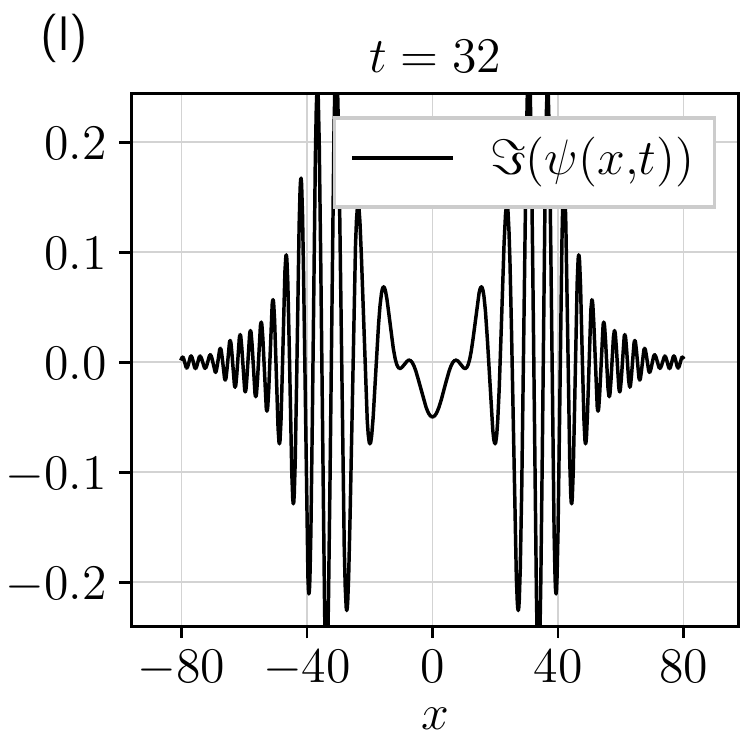}
\caption{
Wave packet expansion of a compactly supported initial wave function.
The initial wave function $\psi_0$ is given in~\eqref{eq:psi_0_compact} using the
high-order polynomial $u$ in~\eqref{eq:example_compact_1d_u_of_x}.
}
\label{fig_compact_1d_snapshots}
\end{figure}

As an example, we consider the compactly supported initial wave function $\psi_0$ 
in~\eqref{eq:psi_0_compact} using
\begin{subequations}
\label{eq:example_compact_1d_u_of_x}
\begin{equation}
u(x)
=
p_1(2 x / L) \, p_2(20 x / L), 
\;\; x \in [-L/2, L/2]
\end{equation}
with the polynomials
\begin{equation}
p_1(x) = (1 - x^2)^8,
\quad
p_2(x) = 1 - \frac{x^2}{2!} + \frac{x^4}{4!} - \frac{x^6}{6!} + \frac{x^8}{8!},
\quad x \in \R
\end{equation}
\end{subequations}
and $L=20$.
Starting from $\psi_0$ on $\Omega_0=[-10, 10]$
we compute approximations of $\psi$ on
$\Omega = [-20, 20]$, $\Omega = [-40, 40]$ and $\Omega = [-80, 80]$
corresponding to the final times $t=2$, $t=8$ and $t=32$, respectively.
Densities, real and imaginary parts of these approximations as well as the initial condition
are depicted in Fig.~\ref{fig_compact_1d_snapshots}.

The computation of an exact reference solution should, 
in principle, be possible but the final expression would look incredibly complicated. 
Nonetheless, by means of Theorem~\ref{theo:euler_maclaurin} we are able to demonstrate 
that the numerical approximations shown in~Fig.~\ref{fig_compact_1d_snapshots} 
have been computed with utmost precision.

\begin{figure}[!t]
\centering
\includegraphics[width=0.5\textwidth]{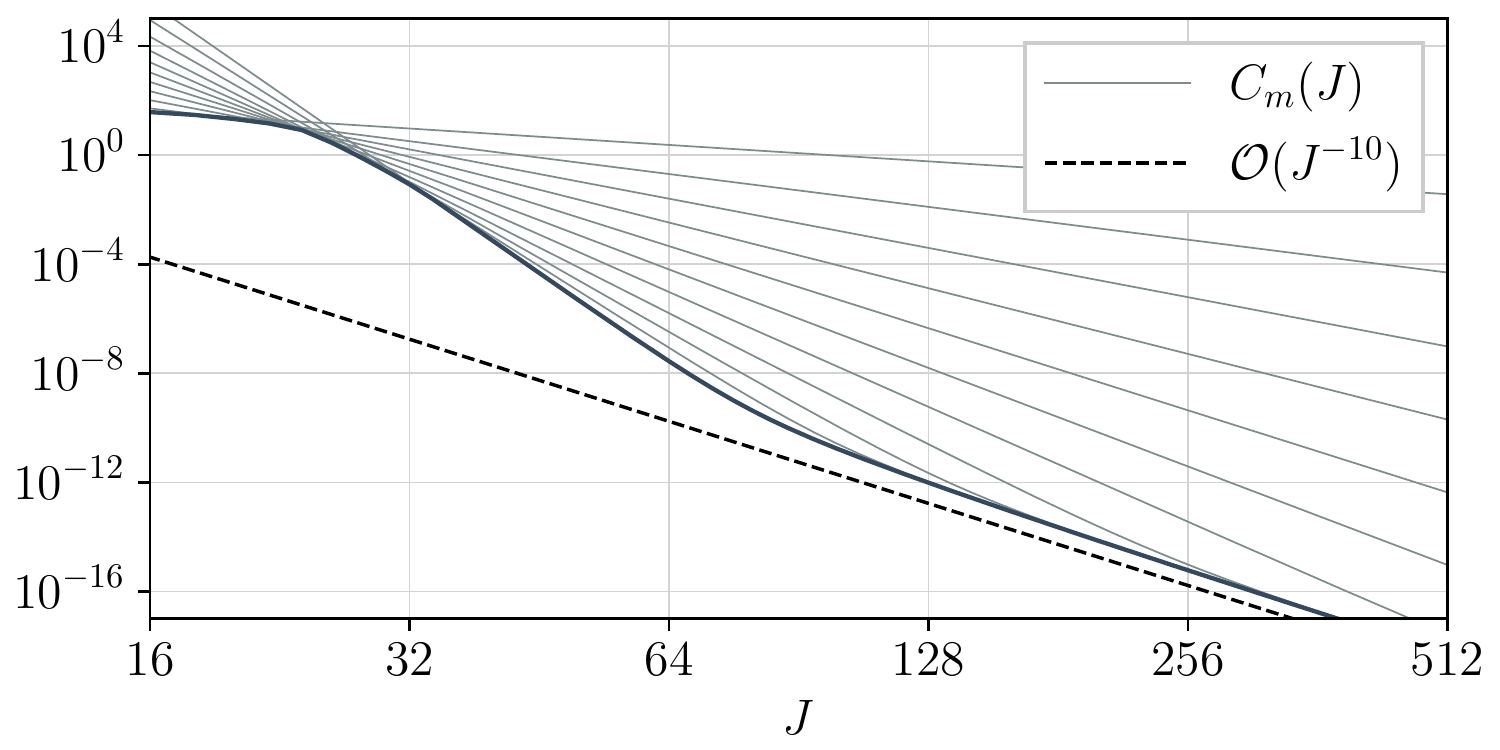}
\caption{
Error bounds $C_m(J)$, $m=0,\dots,10$ in \eqref{eq:error_estimate_compact_1d} for 
the discrete Green's function approximation in case of a compactly supported 
initial wave function~\eqref{eq:psi_0_compact} using the high-order polynomial $u$
in~\eqref{eq:example_compact_1d_u_of_x}.
}
\label{fig:fig_compact_1d_error_bounds}
\end{figure}

The fact that $u$ is a polynomial with vanishing boundary values implies
$f \in \C_0^\infty([-L/2, L/2])$.
We are therefore free to choose any $m \in \N_0$ in Theorem~\ref{theo:euler_maclaurin}.
For $t=8$ the obtained error bounds $C_m(J)$ are shown in 
Fig.~\ref{fig:fig_compact_1d_error_bounds} using $J=16, \dots, 512$ and $m=0,\dots,10$.
Quite obviously, the guaranteed speed of convergence is not as fast as in the previous examples.
It should however be noted that our estimates are comparatively coarse
as they are based on a simple expansion of the derivatives of the integrand.
Nonetheless, using $J \approx 260$ grid points the error is guaranteed to be smaller 
than the machine precision which still represents a remarkable result.
Furthermore, we find that $C_m(J)$ decrease like $\O(J^{-10})$ if $m$ and $J$ 
are sufficiently large.
This convergence behavior can be explained by the fact that
$u^{(n)}(\pm L/2)=0$ for $n=1,3,5,7$ but $u^{(9)}(\pm L/2) \neq 0$.
The same applies to the integrand $f$ in \eqref{eq:f_of_x_prime_compact}
which implies $\lambda_0 = \lambda_1 = \dots = \lambda_4 = 0$ and $\lambda_5 \neq 0$.
Consequently, the lowest order term in Theorem~\ref{theo:euler_maclaurin} 
is $\lambda_5 (L/J)^{10}$.

Finally, let us consider the general scenario where
$u \in C_0^\infty([-L/2, L/2])$ with $u^{(n)}(\pm L/2) = 0$ for $n=0,\dots, N$
using some $N \in \N_0$.
Then, according to the definition of the integrand $f$ in \eqref{eq:f_of_x_prime_compact},
we also have $f \in \C_0^\infty([-L/2, L/2])$ and $f^{(n)}(\pm L/2) = 0$ for $n=0,\dots, N$.
Using Theorem~\ref{theo:euler_maclaurin},
we see that the error of the discrete Green's function
approximation decreases like $\O(J^{-(N+2)})$ if $N$ is even
and like $\O(J^{-(N+3)})$ if $N$ is odd.

\subsection{Numerical effort of the one-dimensional approximation}

The numerical effort involved in the evaluation and application of $\bm{G}$ 
in~\eqref{eq:discr_conv_1d}
is in $\O(JK)$ or $\O(J^2)$ if $J$ and $K$ are comparable in size.
If $\Omega_0 = \Omega$, or more precisely, if $J=K$ and
\begin{equation}
\label{eq:cond_same_grid}
x_k = x_j', \quad k = 0,\dots, J-1,
\end{equation}
the same operation requires only $\O(J \log(J))$ elementary numerical operations.
In that case the discrete free particle propagator $\bm{G}$ is a square matrix
\begin{equation}
\label{eq:discr_conv_direct_1d_equal_domain_2}
\bm{G} 
= 
(G_{kj}) \in \C^{J \times J}, 
\quad 
G_{kj} = G^{(1)}(x_k - x_j, t),
\quad k,j \in \{0, \dots, J-1\},
\end{equation}
where each descending diagonal of $\bm{G}$ from left to right is constant.
Hence, $\bm{G}$ is a Toeplitz matrix and by embedding $\bm{G}$ 
in a circulant matrix of size $2J$,
the matrix-vector product in~\eqref{eq:discr_conv_1d_1} 
can be computed as follows~\cite{bjoerck_2014}: 

First, we define the two vectors
\begin{subequations}
\label{eq:discr_conv_fft_1d_equal_domain}
\begin{equation}
\label{eq:discr_conv_fft_1d_equal_domain_1}
\bm{\psi}_0
=
\begin{bmatrix}
\psi_0(x_0), \dots, \psi_0(x_{J-1}), 0, 0, \dots, 0, 0
\end{bmatrix}^\top \in \C^{2J}
\end{equation}
and
\begin{equation}
\label{eq:discr_conv_fft_1d_equal_domain_2}
\bv
=
\begin{bmatrix}
G(x_0, t), \dots, G(x_{J-1}, t), G(x_{-J}, t), \dots, G(x_{-1}, t)
\end{bmatrix}^\top \in \C^{2J}.
\end{equation}
Next, we compute
\begin{equation}
\label{eq:discr_conv_fft_1d_equal_domain_3}
\bw
=
\Big\{
\DFT^{-1}
\big\{
\DFT
\{ 
\bv
\}
\odot
\DFT\{
\bm{\psi}_0
\}
\big\}
\Big\}.
\end{equation}
Finally, the last $J$ components of $\bm{w}$ are discarded which yields
\begin{equation}
\label{eq:discr_conv_fft_1d_equal_domain_4}
\tilde{\bm{\psi}}
=
\triangle x'
\begin{bmatrix}
w_0, \dots, w_{J-1}
\end{bmatrix}^\top,
\end{equation}
\end{subequations}
where $\triangle x' = \triangle x = L/J$.

The complexity of the above method is in $\O(J \log(J))$ provided
the DFTs are computed using the FFT algorithm.
Realistically, however, the condition in~\eqref{eq:cond_same_grid} represents a 
major restriction.
Like in the Fourier collocation method, the same set of grid points 
is used to approximate the initial as well as the expanded wave function
and therefore the number of required grid points $J$ in a typical wave packet expansion
problem is very large.
This is in strong contrast to the discrete Green's function approximation 
in its most simple form~\eqref{eq:discr_conv_1d} which allows to compute an approximation 
of the expanded wave function on an arbitrary interval $\Omega = [a,b] \subset \R$ 
using a customized number of grid points $K$.
Taking into account this additional flexibility, the method in~\eqref{eq:discr_conv_1d} 
appears to be highly superior to the procedure in~\eqref{eq:discr_conv_fft_1d_equal_domain}.
This applies even more in light of the fact that the computing times to solve 
a one-dimensional wave packet expansion problem are in any case very short.

In a two- or three-dimensional wave packet expansion problem the number of grid points
is so large that a quadratic numerical effort is unacceptably high.
However, we will see shortly that the numerical effort of the multi-dimensional 
discrete Green's function approximation is not quadratic in the number of grid points 
but in fact much lower.

\section{Discrete Green's function approximation for the multi-dimensional problem}

\subsection{Exploiting the separability of the Green's function}

The Green's function formalism for the $d$-dimensional free Schrödinger equation reads~\cite{sakurai_2011}
\begin{equation}
\label{eq:time_evol_by_conv_multi_d}
\psi(x_1, \dots x_d, t)
=
\int_{\R^d} G^{(d)}(x_1-x_1',\dots, x_d - x_d', t) \, \psi_0(x_1', \dots, x_d') \,
dx_1' \dots dx_d'
\end{equation}
using the free-particle propagator
\begin{equation}
\label{eq:G_d}
G^{(d)}(x_1, \dots x_d, t) 
=
\Big( \frac{1}{2 \pi i t} \Big)^{d/2} e^{i (x_1^2 + \dots + x_d^2) / (2 t)}
\end{equation}
for $t > 0$.

Analogously to the one-dimensional case we consider the initial and final computational domains
\[
\Omega_0 = [-L_1/2, L_1/2] \times \dots \times [-L_d/2, L_d/2] \subset \R^d
\]
and
\[
\Omega = [a_1, b_1] \times \dots \times [a_d, b_d] \subset \R^d,
\]
respectively.
Moreover, we define a discrete representation
\begin{equation*}
\bm{\psi}_0 \in \C^{J_1 \times \dots \times J_d},
\quad
( \bm{\psi}_0 )_{j_1, \dots, j_d}
=
\psi_0(x_{1,j_1}', \dots, x_{d, j_d}')
\end{equation*}
of the initial wave function on $\Omega_0$ using the grid points
\begin{align*}
x_{\ell, j_\ell}'
&=
(j_\ell-J_\ell/2) \triangle x_\ell', 
\quad \triangle x_\ell' = L_\ell/J_\ell, 
\quad j_\ell=0,\dots, J_\ell-1, 
\quad \ell = 1,\dots,d,
\end{align*}
where $J_1, \dots, J_d \in \N$ are assumed to be even integers.
Likewise, we define a numerical approximation of $\psi(\cdot,t)$ on $\Omega$.
To this end, we introduce the grid points
\begin{equation*}
x_{\ell, k_\ell}
=
a_\ell + k_\ell \triangle x_\ell, 
\quad \triangle x_\ell = (b_\ell-a_\ell)/K_\ell, 
\quad k_\ell=0,\dots, K_\ell-1, 
\quad \ell=1,\dots,d
\end{equation*}
and the approximation
\[
\tilde{\bm{\psi}} \in \C^{K_1 \times \dots \times K_d},
\quad
( \tilde{\bm{\psi}} )_{k_1, \dots, k_d}
\approx
\psi(x_{1, k_1}, \dots, x_{d, k_d}, t),
\]
where $K_1, \dots, K_d \in \N$.
With these definitions and in analogy to the one-dimensional problem the approximation
of the exact solution~\eqref{eq:time_evol_by_conv_multi_d} is given by
\begin{equation}
\label{eq:approx_psi_tilde_multi_d_simple}
\begin{aligned}
( \tilde{\bm{\psi}} )_{k_1, k_2, \dots, k_{d-1}, k_d}
&=
\triangle x_d' \triangle x_{d-1}' \dots \triangle x_2' \triangle x_1' 
\sum_{j_d=0}^{J_d-1}
\sum_{j_{d-1}=0}^{J_{d-1}-1} 
\dots \\
&\qquad \dots \, 
\sum_{j_2=0}^{J_2-1} \sum_{j_1=0}^{J_1-1} 
G^{(d)}(x_{1,k_1} - x_{1,j_1}', \dots, x_{d,k_d} - x_{d,j_d}', t) 
\, (\bm{\psi}_0)_{j_1, j_2, \dots, j_{d-1}, j_d}.
\end{aligned}
\end{equation}
The numerical effort of this approximation is in 
\[
\O(J_1 \dots J_d K_1 \dots K_d).
\]
In a typical wave packet expansion problem the number of grid points 
$J_1, \dots, J_d$ and $K_1, \dots, K_d$ are of the same order of magnitude
\begin{equation}
\label{eq:assumption_J_n_K_n}
J_\ell, K_\ell \in \O(J_1), \;\; \ell = 1, \dots, d
\end{equation}
such that the numerical effort 
$\O(J^2)$ 
increases quadratically with the number of grid points $J = J_1 J_2 \dots J_{d-1} J_d$.

Noting that the free-particle propagator~\eqref{eq:G_d} is the product 
\begin{equation*}
G^{(d)}(x_1, \dots, x_d, t)
= 
\prod_{i=1}^d G^{(1)}(x_i, t)
\end{equation*}
of $d$ one-dimensional free-particle propagators $G^{(1)}$ in~\eqref{eq:G_1d}, 
the integral~\eqref{eq:time_evol_by_conv_multi_d} can also be written as
\begin{align*}
\psi(x_1, \dots x_d, t)
&=
\int_{-\infty}^\infty
G^{(1)}(x_d - x_d', t)
\int_{-\infty}^\infty
G^{(1)}(x_{d-1} - x_{d-1}', t)
\, \dots 
\\
&\qquad \dots \, 
\int_{-\infty}^\infty
G^{(1)}(x_2 - x_2', t)
\int_{-\infty}^\infty
G^{(1)}(x_1 - x_1', t) \, \psi_0(x_1', \dots, x_d') \,
dx_1' \dots dx_d'
\end{align*}
and the approximation~\eqref{eq:approx_psi_tilde_multi_d_simple} becomes
\begin{subequations}
\label{eq:approx_psi_tilde_multi_d}
\begin{equation}
\label{eq:approx_psi_tilde_multi_d_1st}
\begin{aligned}
(\tilde{\bm{\psi}})_{k_1, k_2, \dots, k_{d-1}, k_d} 
&=
\triangle x_d' \triangle x_{d-1}' \dots \triangle x_2' \triangle x_1'
\sum_{j_d=0}^{J_d-1} (\bm{G}_d)_{k_d, j_d}
\sum_{j_{d-1}=0}^{J_{d-1}-1} (\bm{G}_{d-1})_{k_{d-1}, j_{d-1}} \, \dots \, \\
&\qquad \dots \,
\sum_{j_2=0}^{J_2-1} (\bm{G}_2)_{k_2, j_2}
\sum_{j_1=0}^{J_1-1} (\bm{G}_1)_{k_1, j_1}(\bm{\psi}_0)_{j_1, j_2, \dots, j_{d-1}, j_d}
\end{aligned}
\end{equation}
using
\begin{equation}
\label{eq:approx_psi_tilde_multi_d_2nd}
\bm{G}_\ell \in \C^{K_\ell \times J_\ell}, 
\;\;
(\bm{G}_\ell)_{k_\ell, j_\ell} = G^{(1)}(x_{\ell,k_\ell}-x_{\ell,j_\ell}', t),
\;\;
k_\ell = 0, \dots, K_\ell,
\;\;
j_\ell = 0, \dots, J_\ell
\end{equation}
\end{subequations}
for $\ell = 1, \dots, d$.
In $d$ steps, the initial wave function is transformed along the different spatial directions.
By counting all elementary operations in \eqref{eq:approx_psi_tilde_multi_d_1st}
we immediately see that the multi-dimensional discrete Green's function approximation
can be evaluated in
\begin{equation*}
\O(
K_1 J_1 J_2 \dots J_{d-1} J_d 
+ 
K_1 K_2 J_2 \dots J_{d-1} J_d
+
\dots
+
K_1 K_2 K_3 \dots K_{d-1} J_d)
\end{equation*}
computational time.
Consequently, under the assumptions in~\eqref{eq:assumption_J_n_K_n}, 
the numerical effort is in $\O(d J_1^{d+1})$.
In terms of the total number of grid points $J=J_1 J_2 \dots, J_{d-1} J_d$
we have $J_1 \approx J^{1/d}$ and hence the numerical effort is in $\O(d J^{(d+1)/d})$.
Our main interest is in the spatial dimensions $d=1, 2, 3$,
in which case the numerical effort 
is in $\O(J^2)$, $\O(J^{3/2})$ and $\O(J^{4/3})$, respectively.

We refrain from deriving error estimates of the multi-dimensional approximation
but would like to point out that the multi-dimensional Green's function method is based
on the same integral approximation as the one-dimensional method.
Therefore, we expect that the observations from the previous section are largely transferable to the multidimensional case.

\subsection{Implementation}

\begin{algorithm}[!ht]
\small
\caption{Wave packet expansion in three spatial dimensions}
\begin{algorithmic}[1]
\Require $\bm{\psi}_0 \in \C^{J_1 \times J_2 \times J_3}$
\State Compute $\bm{G}_1 \in \C^{K_1 \times J_1}$, $\bm{G}_2 \in \C^{K_2 \times J_2}$ and
$\bm{G}_3 \in \C^{K_3 \times J_3}$ using~\eqref{eq:approx_psi_tilde_multi_d_2nd}
\State $\bm{\psi} \gets$ \Call{ExpandAxis}{$\bm{\psi}_0$, $\bm{G}_1$, $1$}
\State $\bm{\psi} \gets$ \Call{ExpandAxis}{$\bm{\psi}$, $\bm{G}_2$, $2$}
\State $\bm{\psi} \gets$ \Call{ExpandAxis}{$\bm{\psi}$, $\bm{G}_3$, $3$}
\State \Return $\triangle x_1' \triangle x_2' \triangle x_3' \bm{\psi}$ 
\newline
\Function{ExpandAxis}{$\bm{\psi}$, $\bm{G}$, $\ell$}
\If{$\ell=2$}
\State $\bm{\psi} \gets \mathrm{swapaxes}(\bm{\psi}, 1, 2)$
\ElsIf{$\ell=3$}
\State $\bm{\psi} \gets \mathrm{swapaxes}(\bm{\psi}, 1, 3)$
\EndIf
\State $J_1, J_2, J_3 \gets \mathrm{shape}(\bm{\psi})$
\State $K_1 \gets \mathrm{shape}(\bm{G}, 1)$
\State $\bm{\psi} \gets \mathrm{reshape}(\bm{\psi}, J_1, J_2 J_3)$ \Comment{Create $J_1 \times (J_2 J_3)$ matrix}
\State $\bm{\psi} \gets \bm{G} \, \bm{\psi}$ \Comment{Dense matrix-matrix multiplication}
\State $\bm{\psi} \gets \mathrm{reshape}(\bm{\psi}, K_1, J_2, J_3)$ \Comment{Restore dimensions of tensor}
\If{$\ell=2$}
\State $\bm{\psi} \gets \mathrm{swapaxes}(\bm{\psi}, 2, 1)$ \Comment{Restore original order of tensor}
\ElsIf{$\ell=3$}
\State $\bm{\psi} \gets \mathrm{swapaxes}(\bm{\psi}, 3, 1)$
\EndIf
\State \Return $\bm{\psi}$ 
\EndFunction
\end{algorithmic}
\label{alg:expansion_3d}
\end{algorithm}

The discrete Green's function approximation of the one-dimensional wave packet expansion problem is computed using a single dense matrix-vector multiplication 
for which highly optimized routines are available in practically 
any programming language.
The multi-dimensional approximation~\eqref{eq:approx_psi_tilde_multi_d}
can be realized using a series of simple for-loops which, however, is very slow 
in many scripting languages like \textit{Matlab} or \textit{Python}.

An alternative approach is to combine all matrix-vector multiplications in the $\ell$th step 
of the calculation in a single matrix-matrix multiplication. 
To do this, the tensor to be transformed must first be arranged in the form of an extremely large
matrix using a reshape operation.
This matrix is then multiplied from left by the matrix $\bm{G}_\ell$ 
in~\eqref{eq:approx_psi_tilde_multi_d_2nd}.
Finally, the result matrix is reshaped back into the form of a tensor.
This process is described for the three-dimensional wave packet expansion problem 
in Algorithm~\ref{alg:expansion_3d}.
Effectively, the entire calculation is carried out using three large scale 
matrix-matrix multiplications, for which very efficient routines are available 
in any practically relevant programming language.
The calculations in this work were implemented in \textit{Python} 
using the packages Numpy~\cite{harris_2020_numpy} and PyTorch~\cite{paszke_2019_pytorch}.

\subsection{Example: Interference of three Gaussian wave packets}

\begin{figure}[!ht]
\centering
\includegraphics[width=0.49\textwidth]{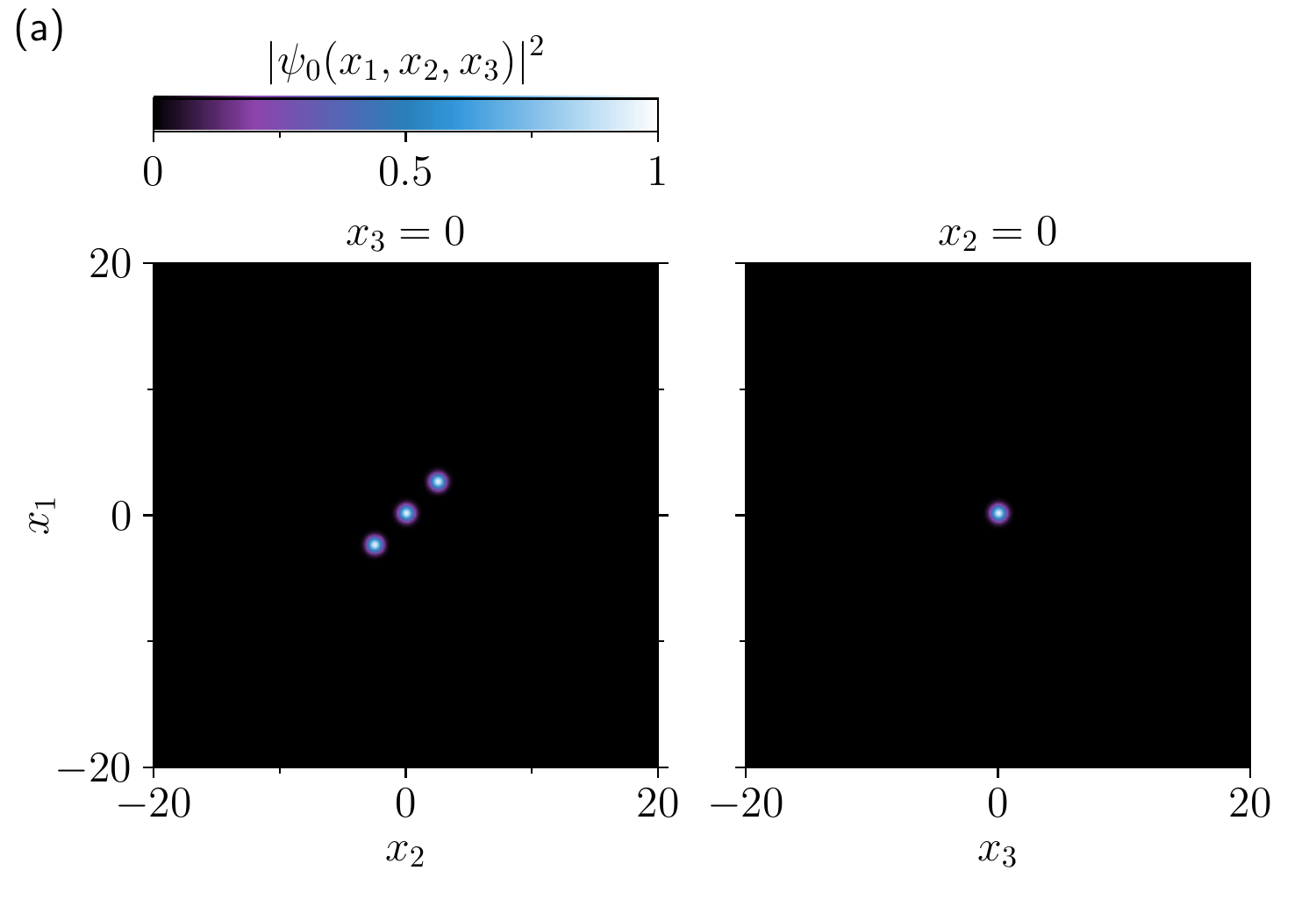}
\includegraphics[width=0.49\textwidth]{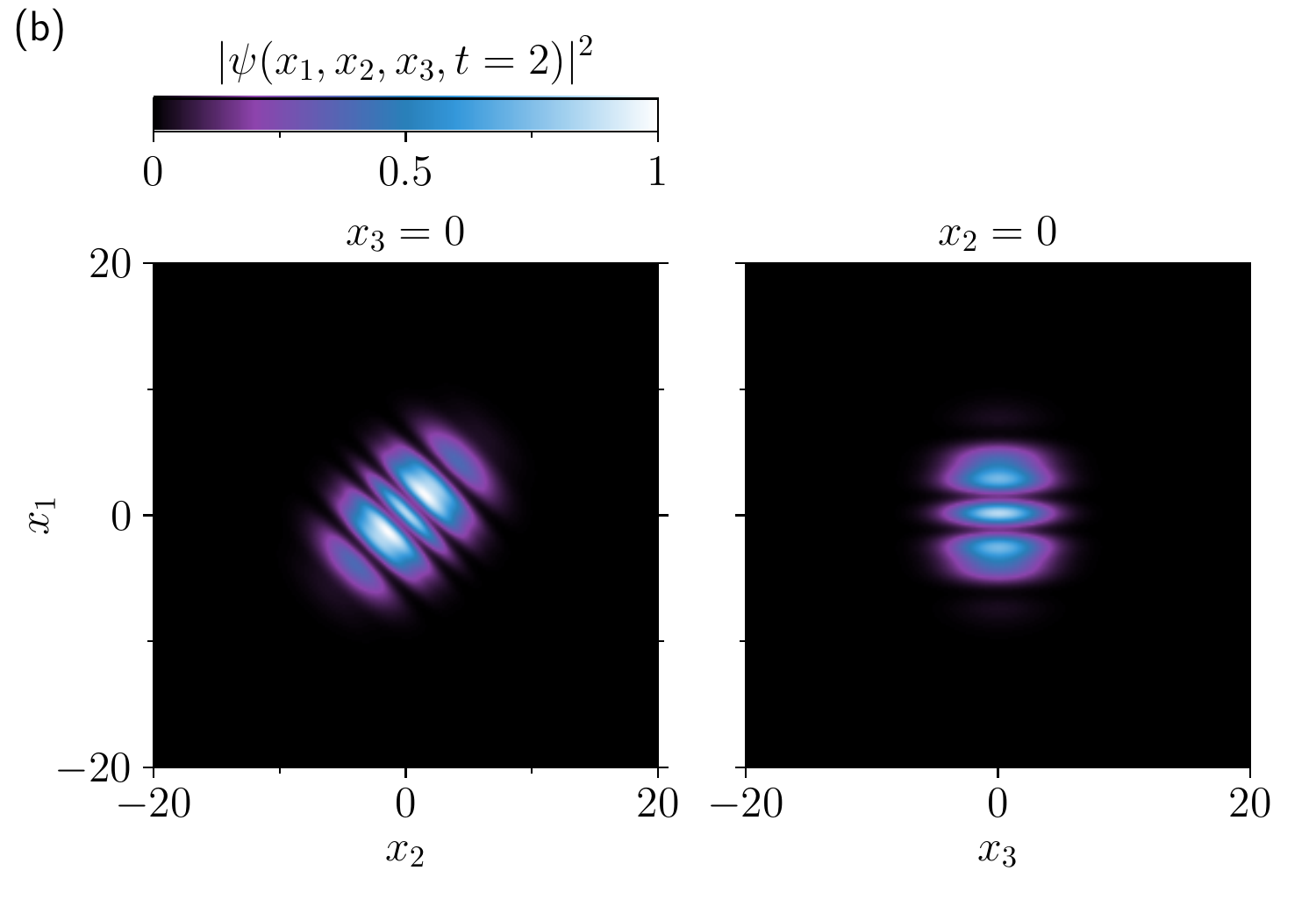} \\
\includegraphics[width=0.49\textwidth]{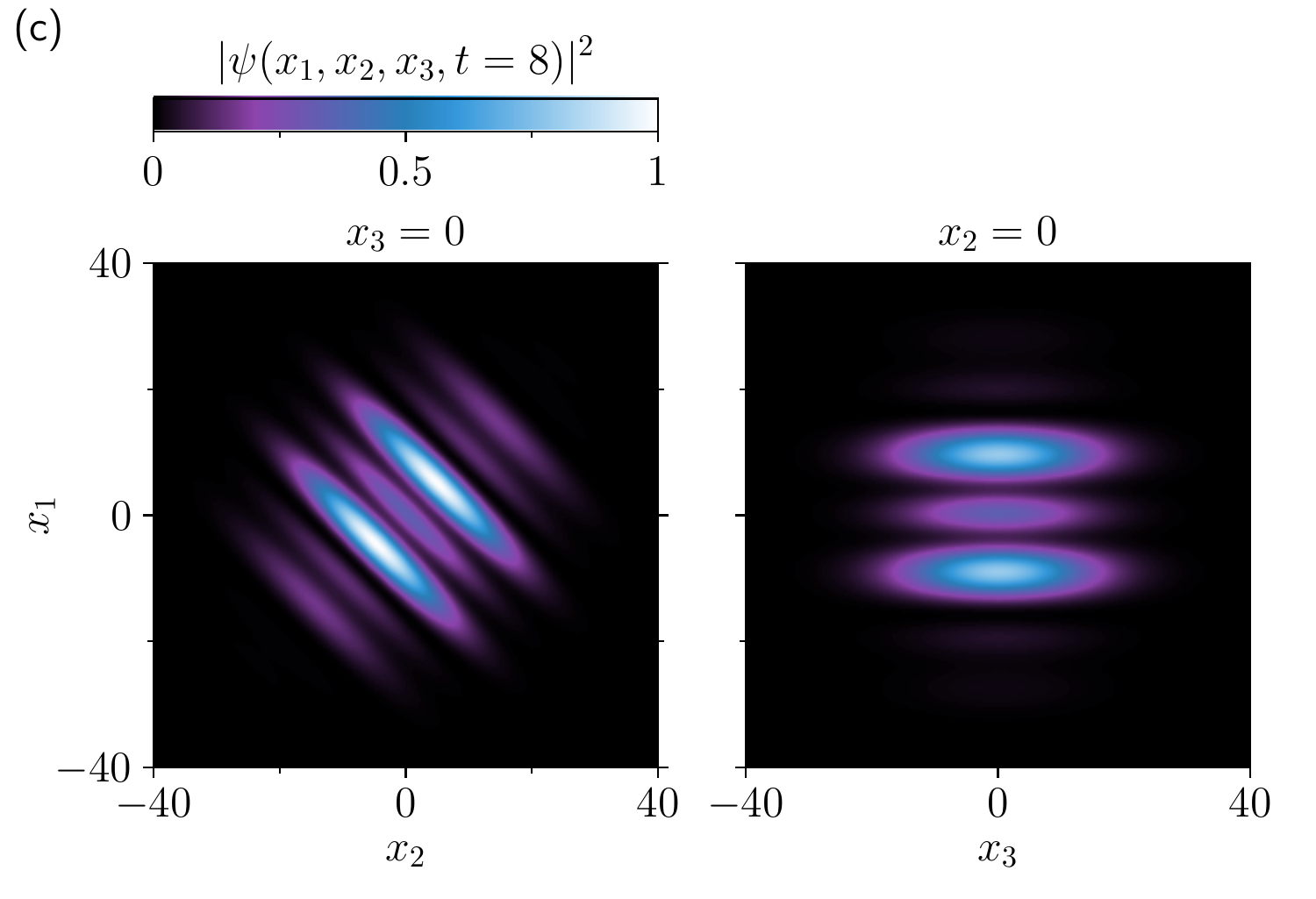}
\includegraphics[width=0.49\textwidth]{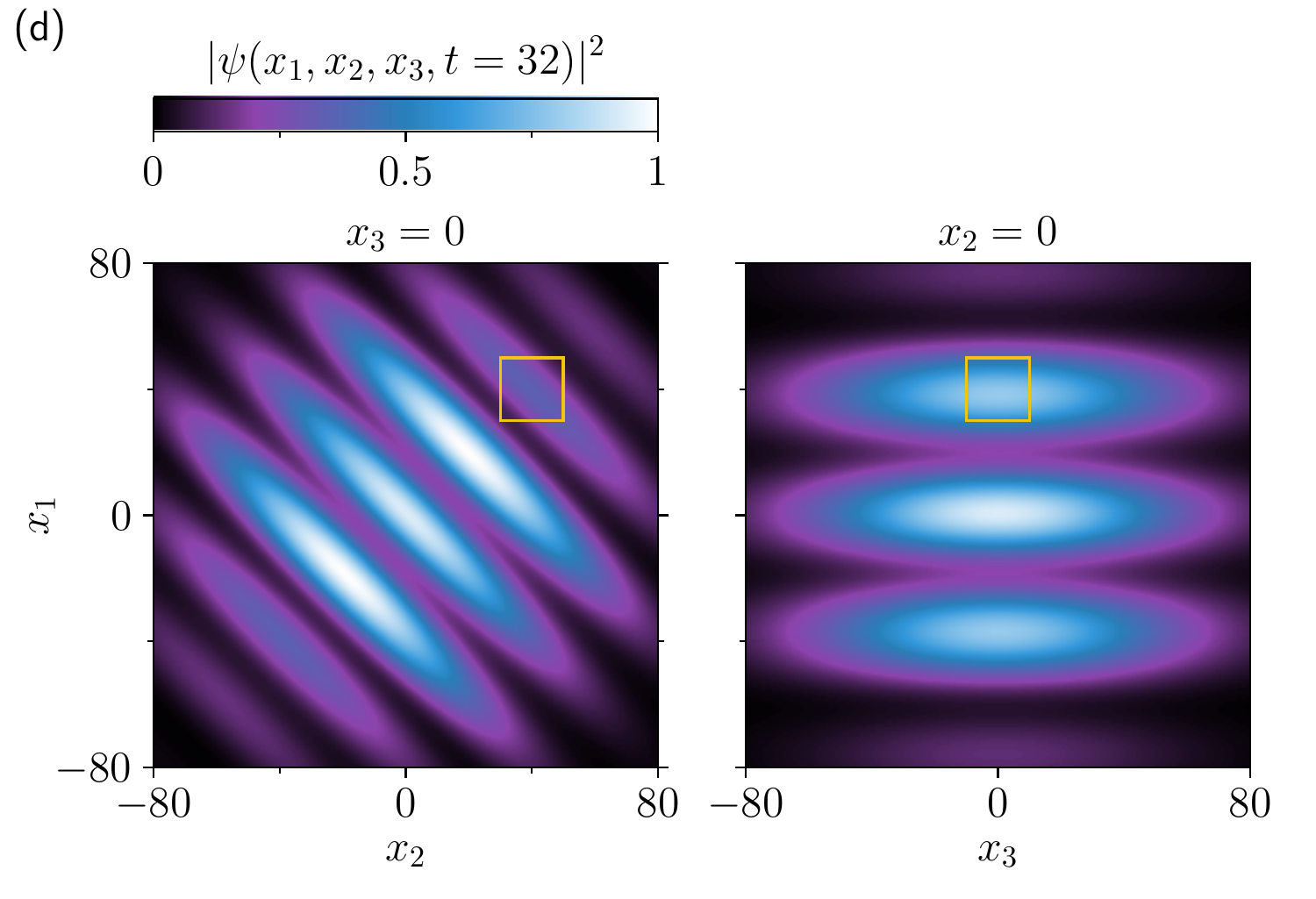}
\caption{
Expansion and interference of three phase shifted Gaussian wave packets 
in three spatial dimensions at different times.
The initial wave function is given in~\eqref{eq:three_gaussians_3d}
using $t=0$.
(a) Density of the initial wave function.
(b)-(d) Density of the approximations of the expanded wave function 
at $t=2$, $t=8$ and $t=32$.
}
\label{fig:gaussians_3d_density}
\end{figure}

\begin{figure}[!ht]
\centering
\includegraphics[width=0.49\textwidth]{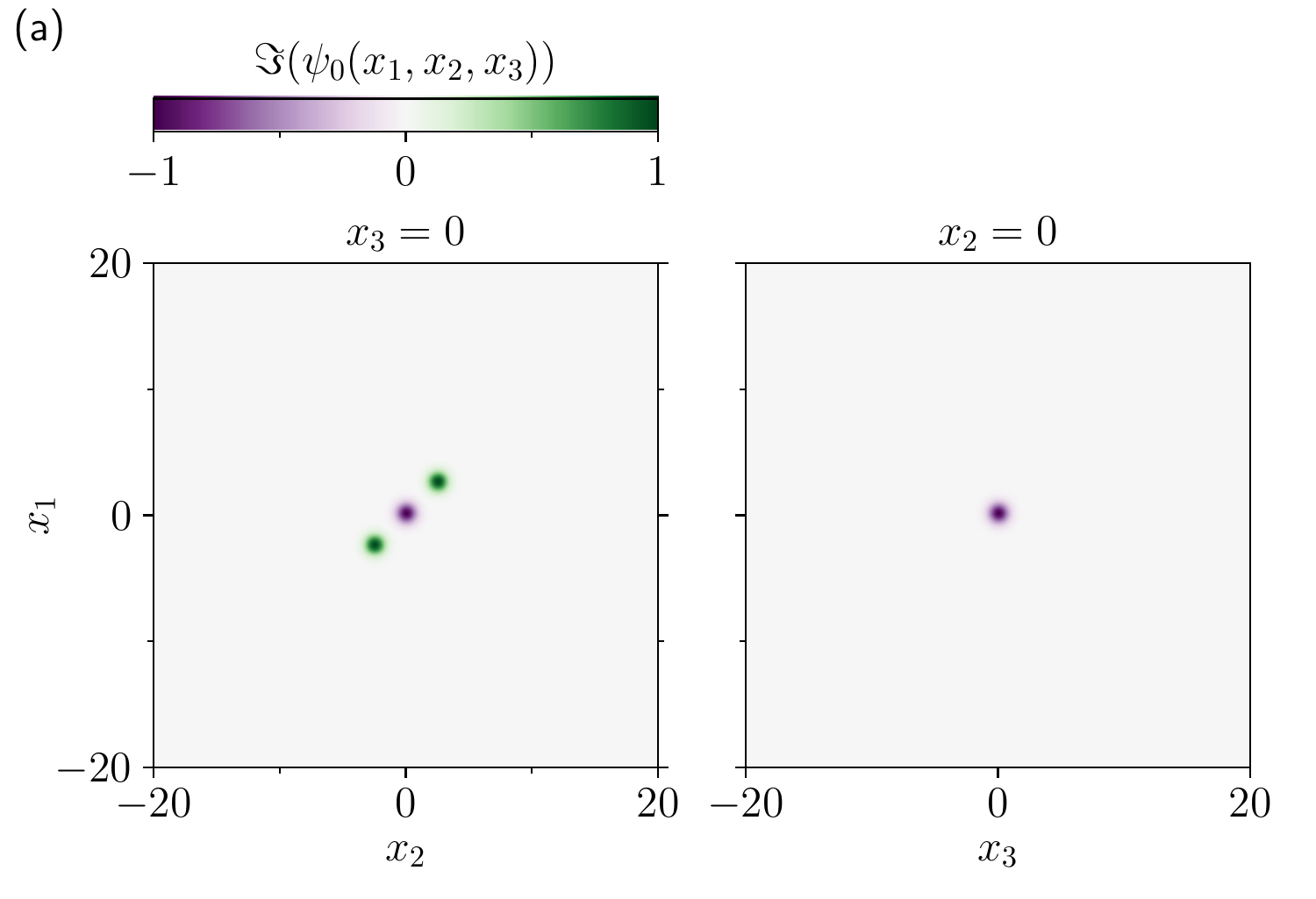}
\includegraphics[width=0.49\textwidth]{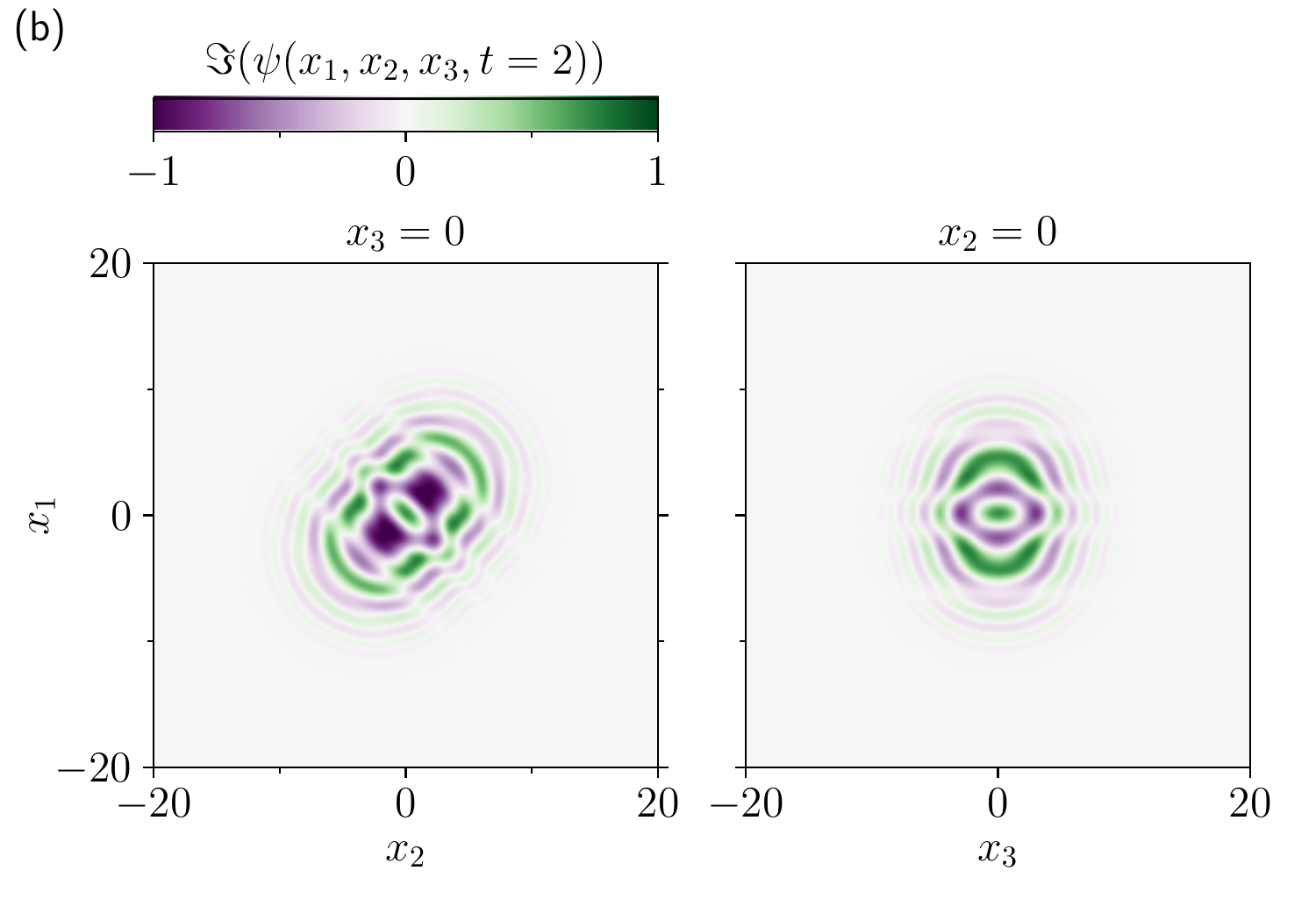} \\
\includegraphics[width=0.49\textwidth]{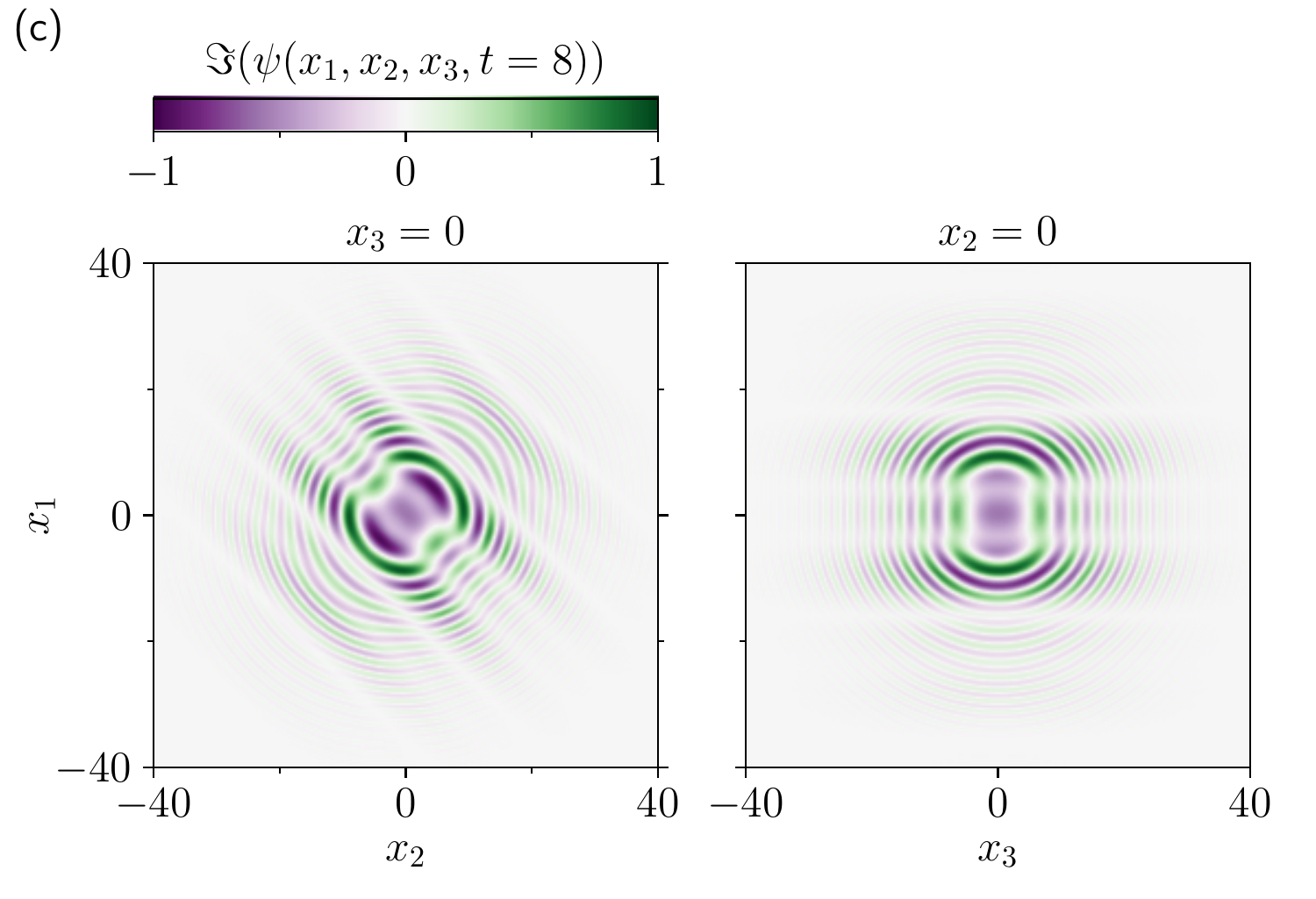}
\includegraphics[width=0.49\textwidth]{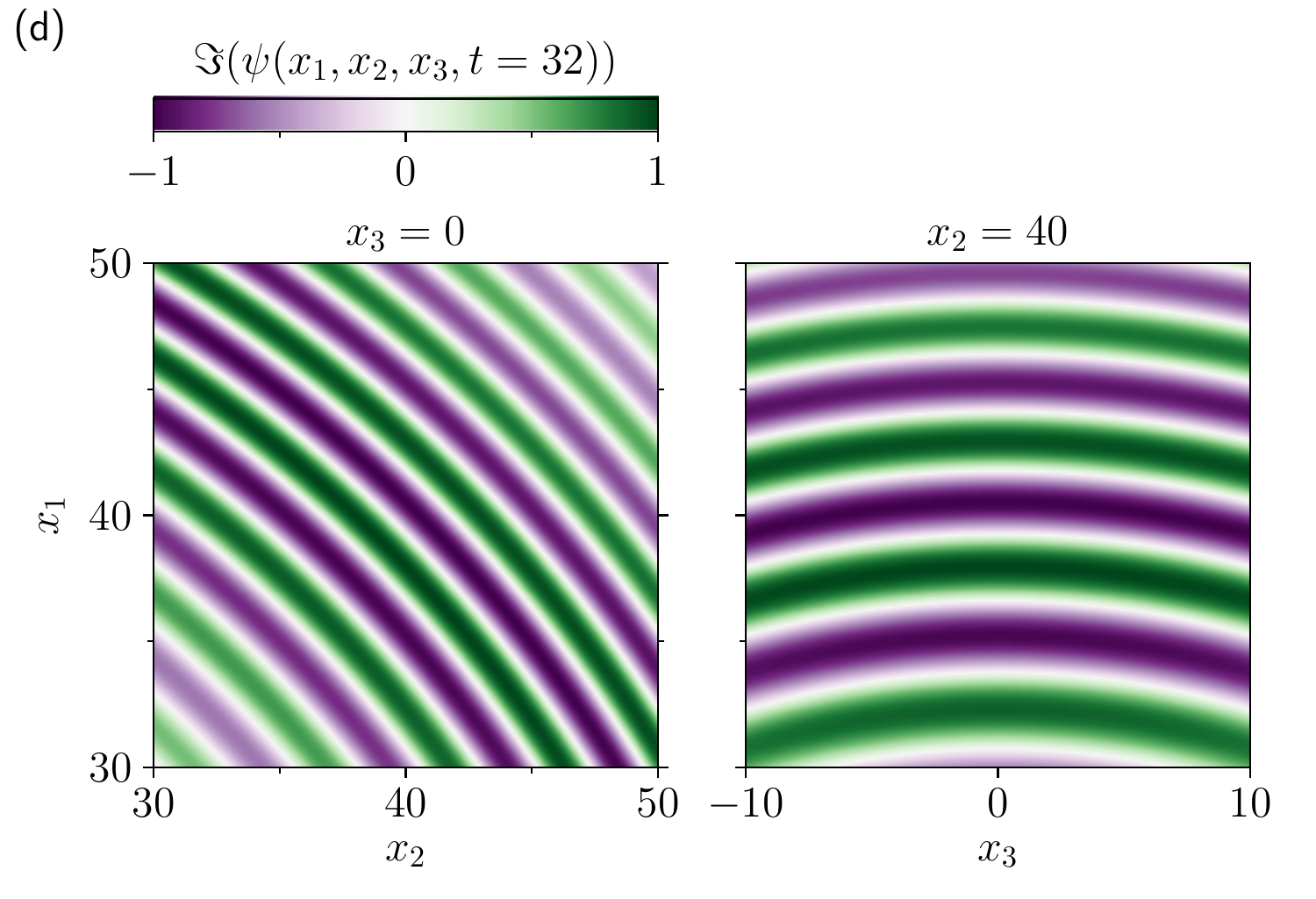}
\caption{
Expansion and interference of three phase shifted Gaussian wave packets 
in three spatial dimensions at different times.
The initial wave function is given in~\eqref{eq:three_gaussians_3d}
using $t=0$.
(a) Imaginary part of the initial wave function.
(b)-(d) Imaginary part of the approximations of the expanded wave function 
at $t=2$, $t=8$ and $t=32$.
In (d) the approximation to the expanded wave function is computed on the region marked by 
a yellow square depicted in Fig.~\ref{fig:gaussians_3d_density}~(d).
}
\label{fig:gaussians_3d_imag}
\end{figure}

\begin{figure}[!ht]
\centering
\includegraphics[width=0.49\textwidth]{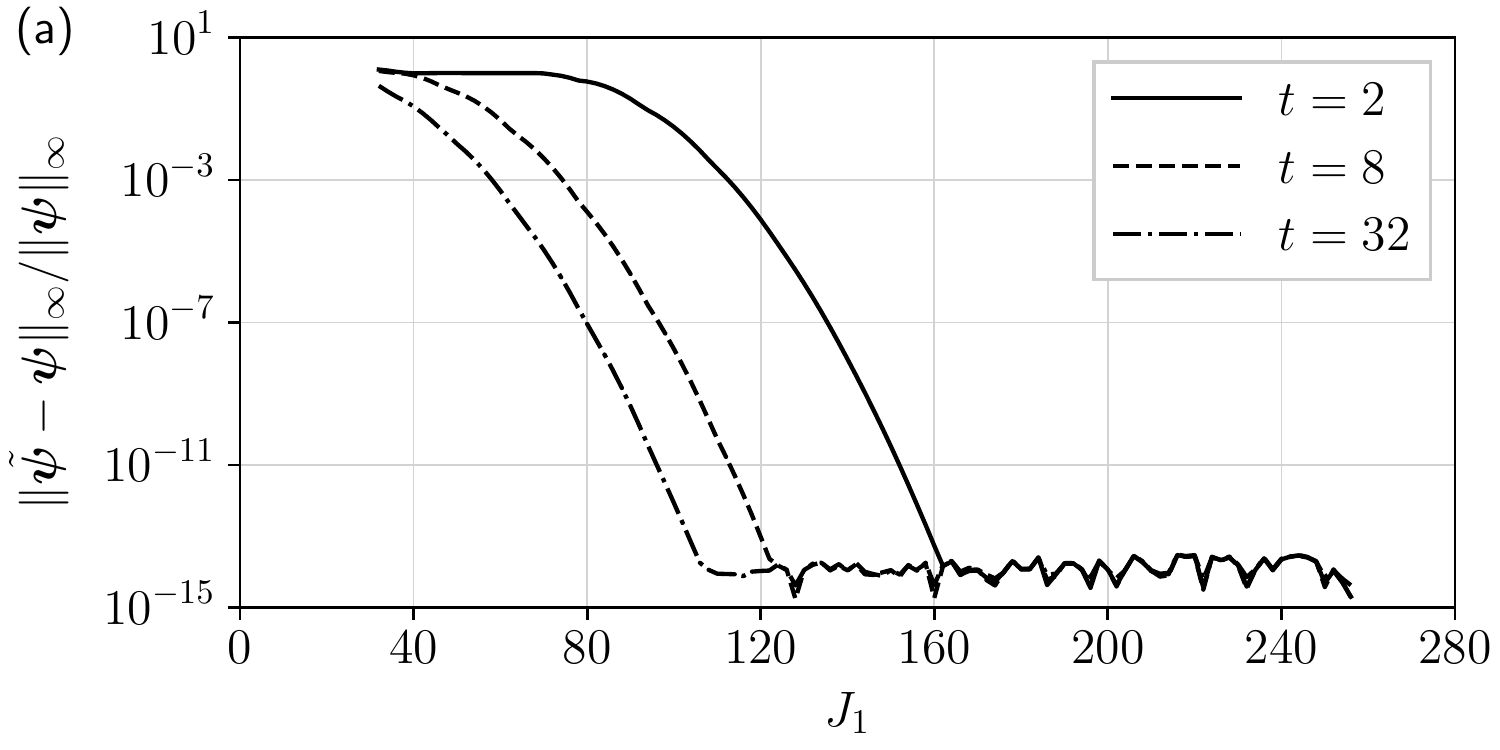}
\includegraphics[width=0.49\textwidth]{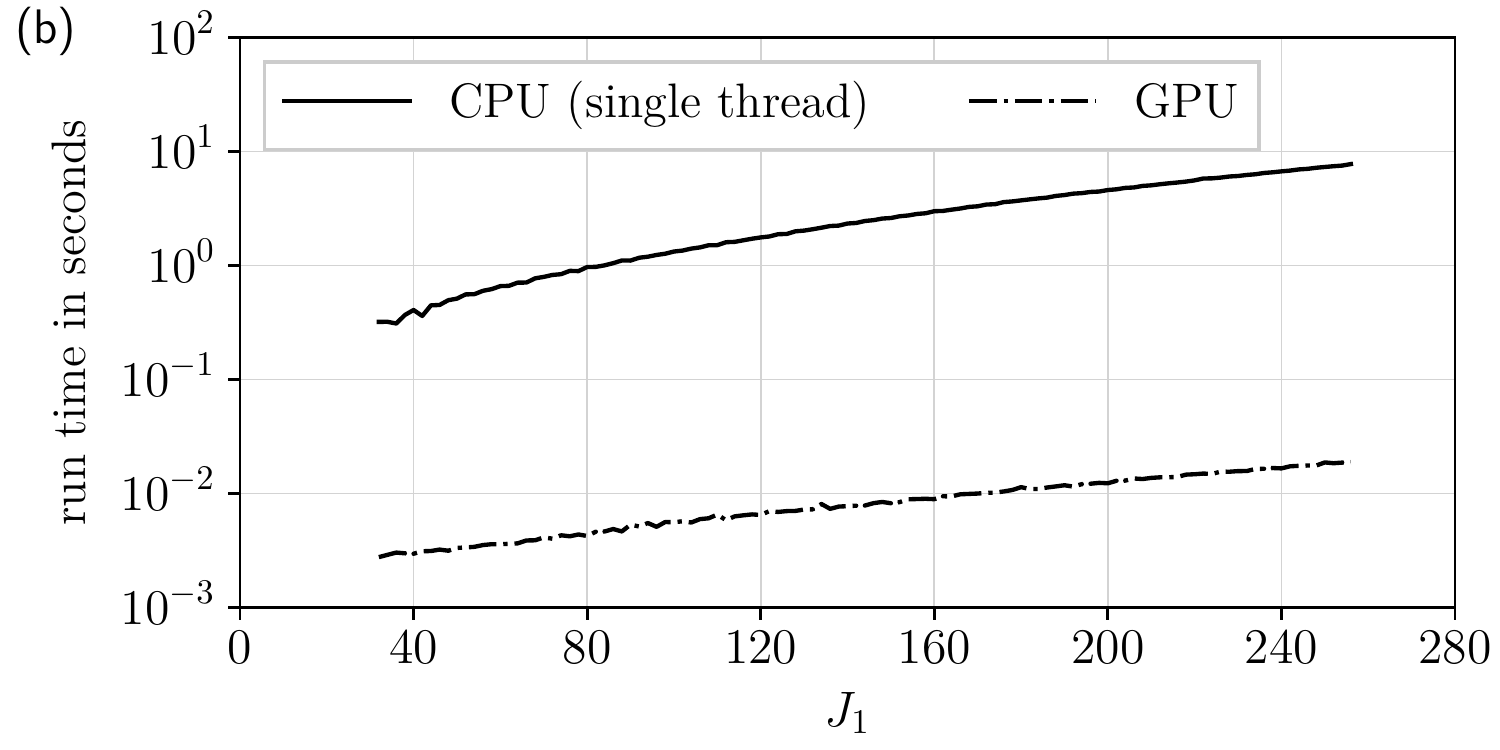}
\caption{
Relative error (a) and run time (b) as a function of the parameter $J_1 = J_2 = J_3$
for the example showing the expansion and interference of three phase-shifted 
Gaussian wave packets.
}
\label{fig:gaussians_3d_convergence_runtime}
\end{figure}

Let us first consider the interference of three Gaussian wave packets
\begin{subequations}
\label{eq:three_gaussians_3d}
\begin{equation}
\psi(x_1, x_2, x_3, t)
=
e^{i \pi / 4} \chi_1(x_1, x_2, x_3, t)
+
e^{-i \pi / 4} \chi_2(x_1, x_2, x_3, t)
+
e^{i \pi / 4} \chi_3(x_1, x_2, x_3, t)
\end{equation}
for $x_1, x_2, x_3 \in \R$ and $t \geq 0$ using
\begin{equation}
\chi_n(x_1, x_2, x_3, t)
=
\bigg[ 
\frac{1}{1 + i (t / \tau_n)} 
\bigg]^{3/2}
\exp 
\bigg[ 
-\frac{(x_1-\delta_{n,1})^2 + (x_2-\delta_{n,2})^2 + (x_3-\delta_{n,3})^2}{4 \sigma_n^2 [1 + i (t/\tau_n)]} 
\bigg],
\end{equation}
\end{subequations}
$\tau_n = 2 \sigma_n^2$ and $n=1,2,3$.
The wave function in \eqref{eq:three_gaussians_3d} defines our initial condition at $t=0$ and serves as a reference solution for $t>0$.
The parameters in the example are given by
$\sigma_1 = \sigma_2 = \sigma_3 = 0.4$ and
\begin{equation*}
\delta_{1,1} = \delta_{1,2} = 2.5,\, \delta_{1,3} = 0,
\qquad
\delta_{2,1} = \delta_{2,2} = \delta_{2,3} = 0,
\qquad
\delta_{3,1} = \delta_{3,2} = -2.5,\, \delta_{3,3} = 0.
\end{equation*}
Furthermore, we choose
\[
\Omega_0 
=
[-20, 20] \times [-20, 20] \times [-20, 20]
\]
as the domain where we evaluate the initial wave function $\psi_0$ and
\begin{align*}
\Omega &= [-20, 20] \times [-20, 20] \times [-20, 20], \\
\Omega &= [-40, 40] \times [-40, 40] \times [-40, 40], \\
\Omega &= [-80, 80] \times [-80, 80] \times [-80, 80]
\end{align*}
as the domains of the approximations to the expanded wave function $\psi$ at the final times $t=2$, $t=8$ and $t=32$, respectively.

The results of the discrete Green's function approximation
using $J_1 = J_2 = J_3 = 256$ and $K_1 = K_2 = K_3 = 256$ grid points
are shown in Fig.~\ref{fig:gaussians_3d_density} and Fig.~\ref{fig:gaussians_3d_imag}
in form of false-color plots at $x_2 \equiv 0$ and $x_3 \equiv 0$.
The first figure shows the density and the second figure 
shows the imaginary part of the wave function.
Here, and in all false-color plots below, the densities and the imaginary parts
of the three-dimensional wave functions have been normalized 
to their individual maximum (absolute) value.
At the given resolution, the countless oscillations in the 
imaginary part of the wave function at $t=32$ 
cause undesirable aliasing effects in the graphical representation.
In order to make the high-frequency character of the expanded wave function visible, 
we have computed an additional approximation on the domain
$\Omega = [30, 50] \times [30, 50] \times [-10, 10]$ (at the same
resolution $K_1 = K_2 = K_3 = 256$).
This region is indicated by a yellow box in Fig.~\ref{fig:gaussians_3d_density}\,(d).
The corresponding imaginary part of the computed approximation 
can be seen in Fig.~\ref{fig:gaussians_3d_imag}\,(d).

The approximations depicted in Fig.~\ref{fig:gaussians_3d_density} 
and Fig.~\ref{fig:gaussians_3d_imag} are practically indistinguishable
from the exact solution~\eqref{eq:three_gaussians_3d}.
This is illustrated by Fig.~\ref{fig:gaussians_3d_convergence_runtime}\,(a)
in which we plot the relative error as a function 
of the parameter $J_1=32, 34, \dots, 256$ using $J_1=J_2=J_3$
and $K_1=K_2=K_3=256$.

Additionally, Fig.~\ref{fig:gaussians_3d_convergence_runtime}\,(b) shows the runtime 
of the three-dimensional discrete Green's function method as a function 
of the parameter $J_1$.
We compare two implementations of Algorithm~\ref{alg:expansion_3d}.
In the first case, the algorithm is implemented on single core of a 
CPU\footnote{Intel\textregistered\,Xeon(R)\,W-2145\,CPU\,@\,3.70GHz\,$\times$\,16} (Numpy),
while in the second case, the algorithm is implemented on a 
GPU\footnote{NVIDIA\,Quadro\,GV100} (PyTorch).
The GPU computations are incredibly fast.
In fact, they are several orders of magnitude faster 
than the computations on the CPU.
For example, using $J_1 = 256$ ($J=J_1 J_2 J_3=16\,777\,216$ grid points), 
the GPU calculation is $411$ times faster than the corresponding calculation 
on the CPU.

\subsection{Example: Interference of two ring-shaped wave packets}

\begin{figure}[!ht]
\centering
\includegraphics[width=0.45\textwidth]{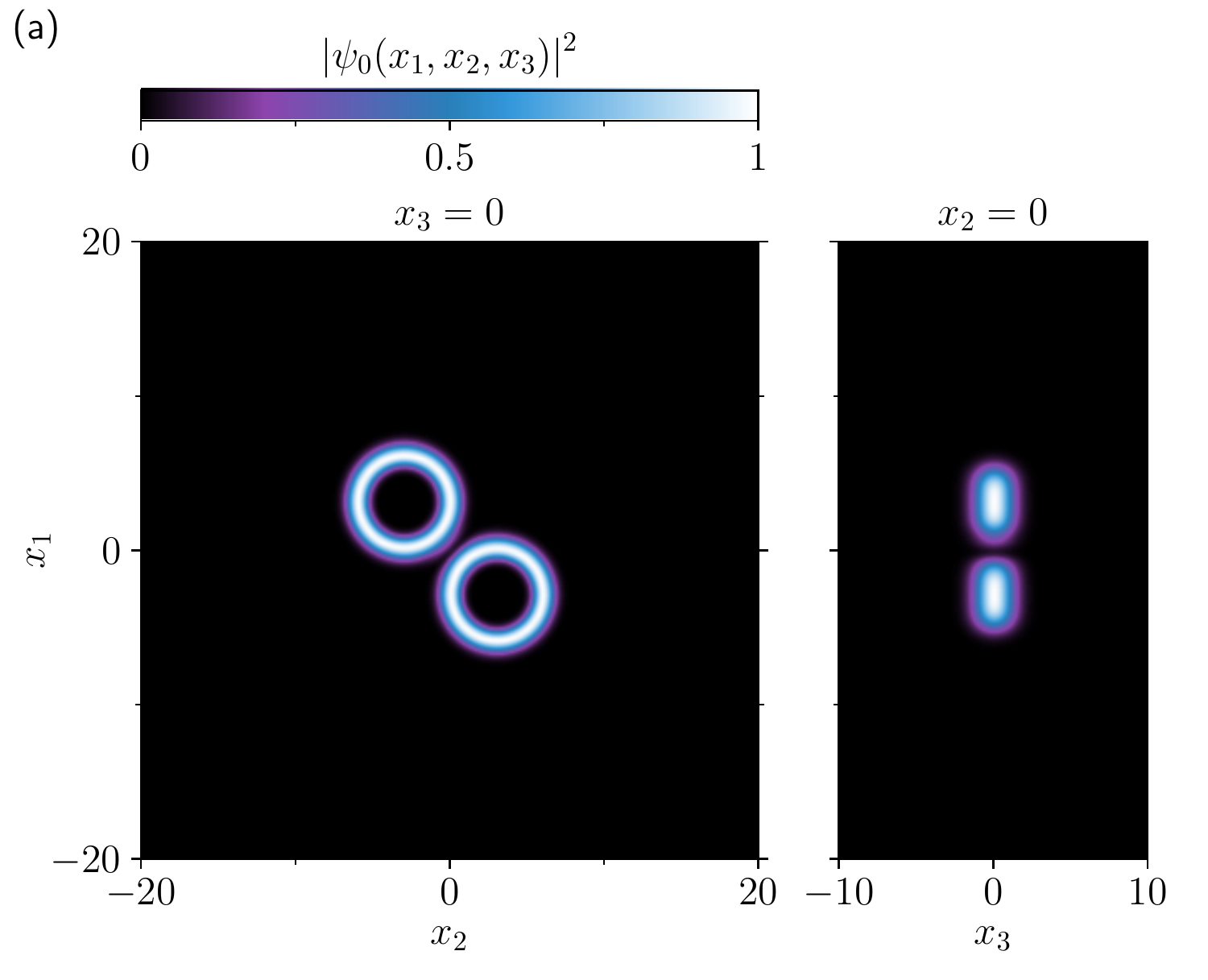}
\includegraphics[width=0.45\textwidth]{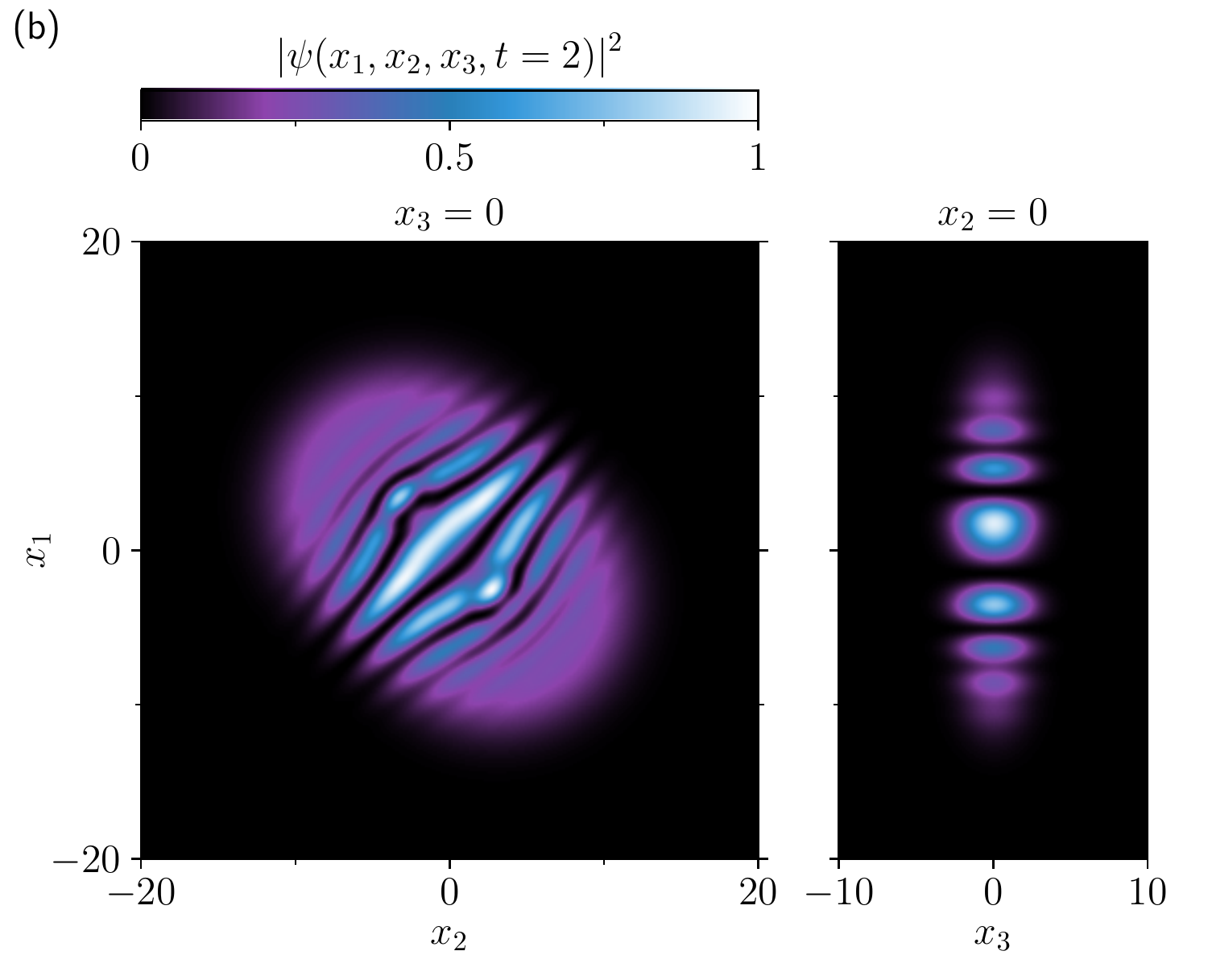} \\
\includegraphics[width=0.45\textwidth]{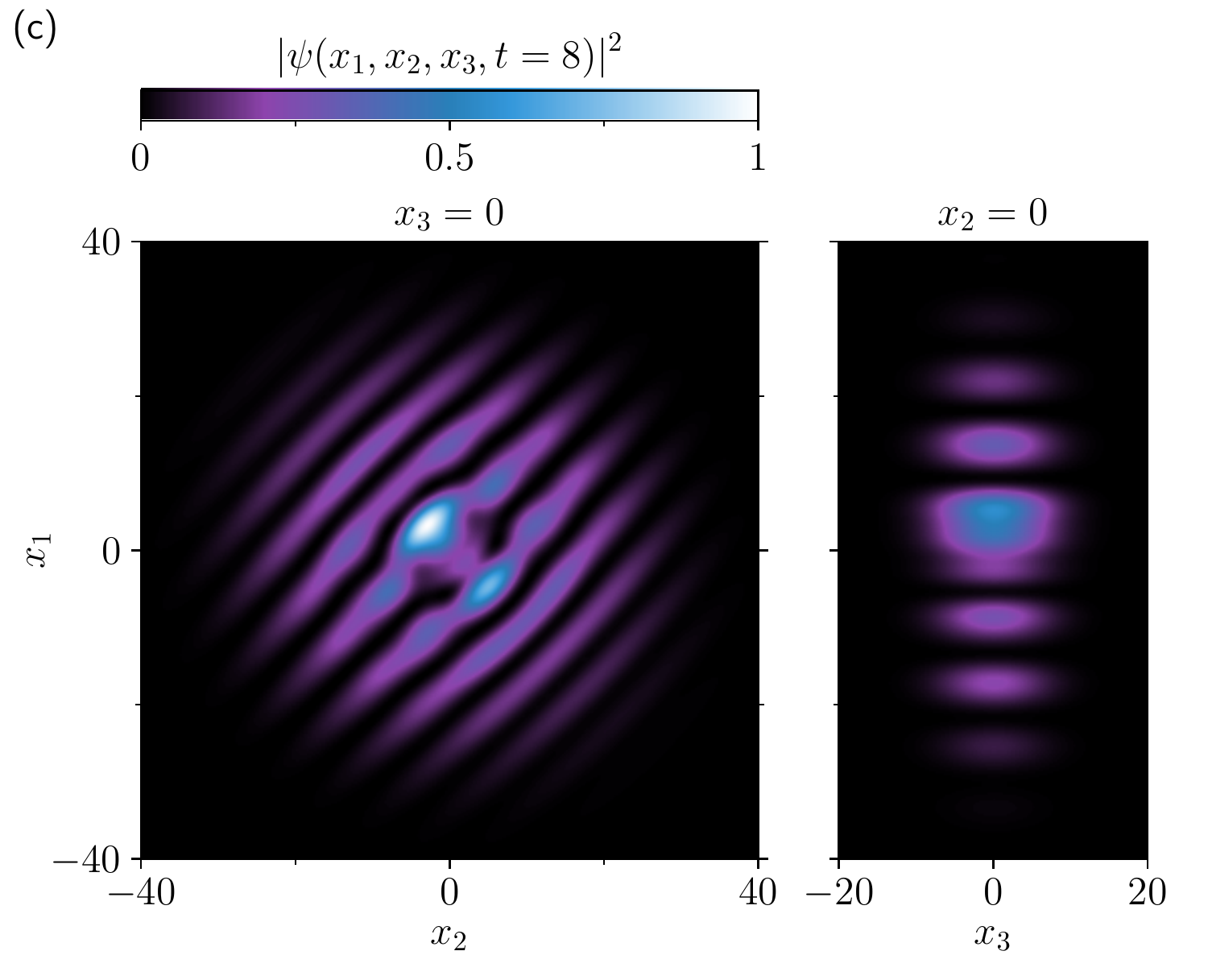}
\includegraphics[width=0.45\textwidth]{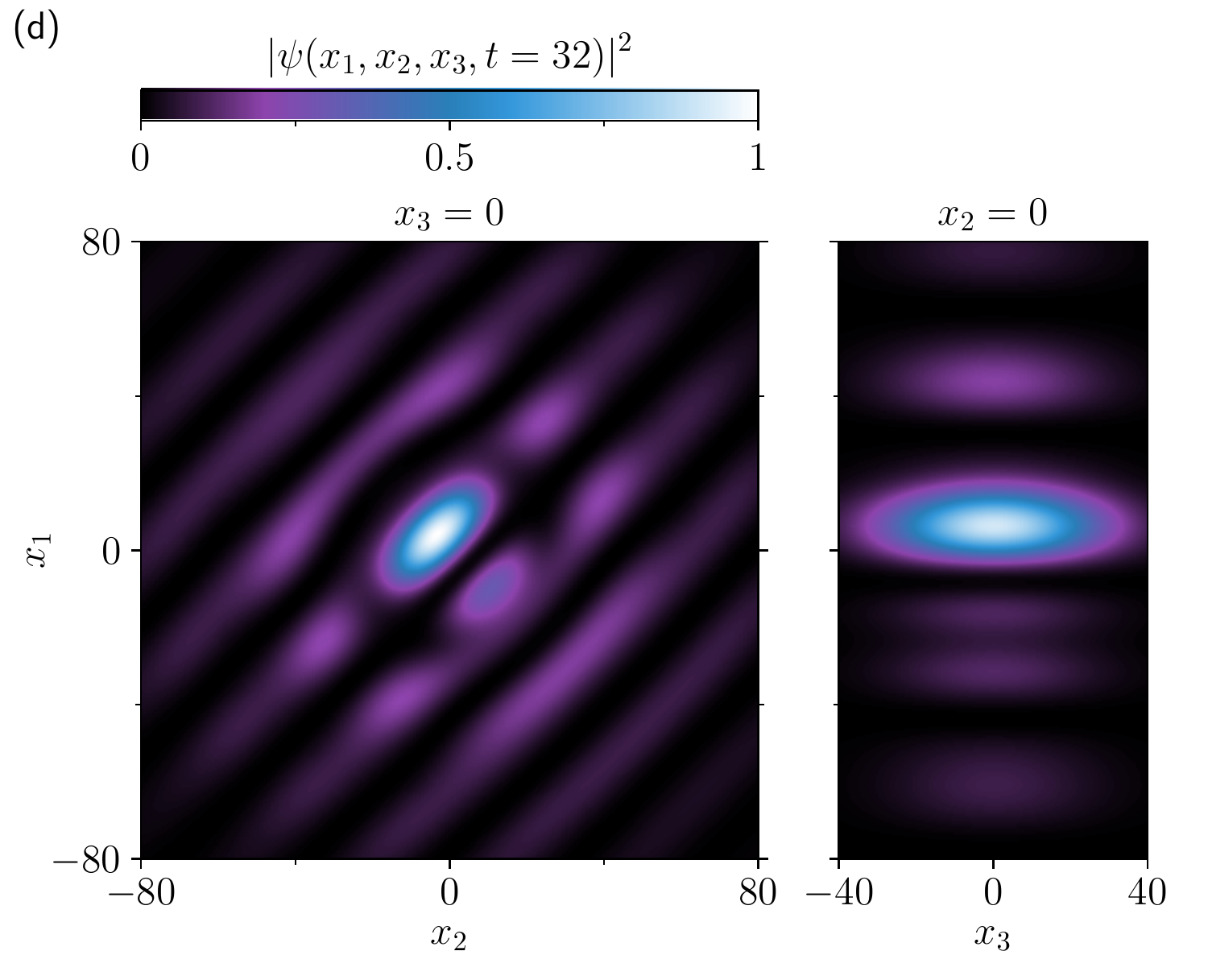}
\caption{
Expansion and interference of two phase shifted ring-shaped wave packets 
in three spatial dimensions at different times.
The initial wave function is given in~\eqref{eq:two_rings_3d} using $t=0$.
(a) Density of the initial wave function.
(b)-(d) Density of the approximations of the expanded wave function 
at $t=2$, $t=8$ and $t=32$.
}
\label{fig:rings_3d_density}
\end{figure}

\begin{figure}[!ht]
\centering
\includegraphics[width=0.45\textwidth]{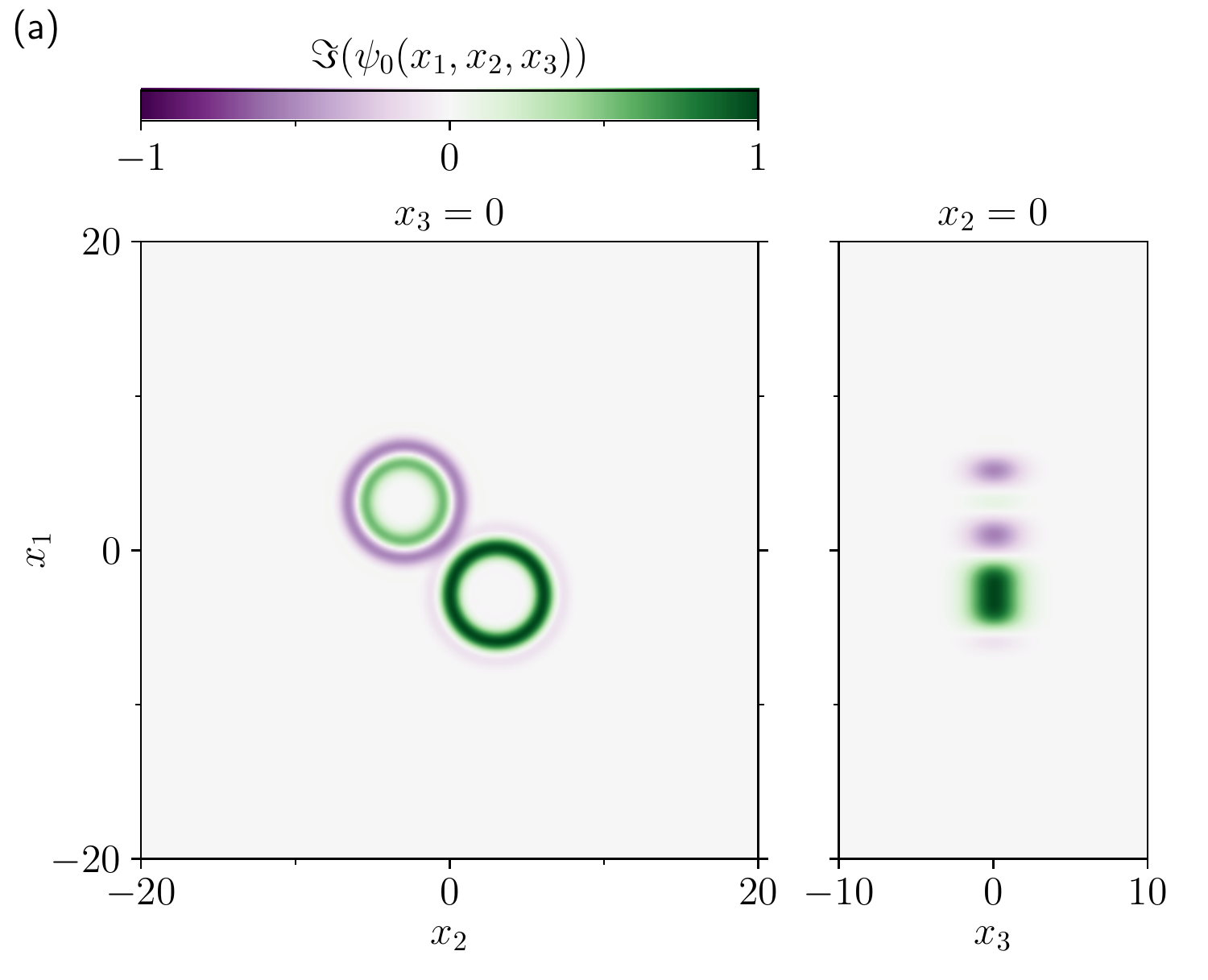}
\includegraphics[width=0.45\textwidth]{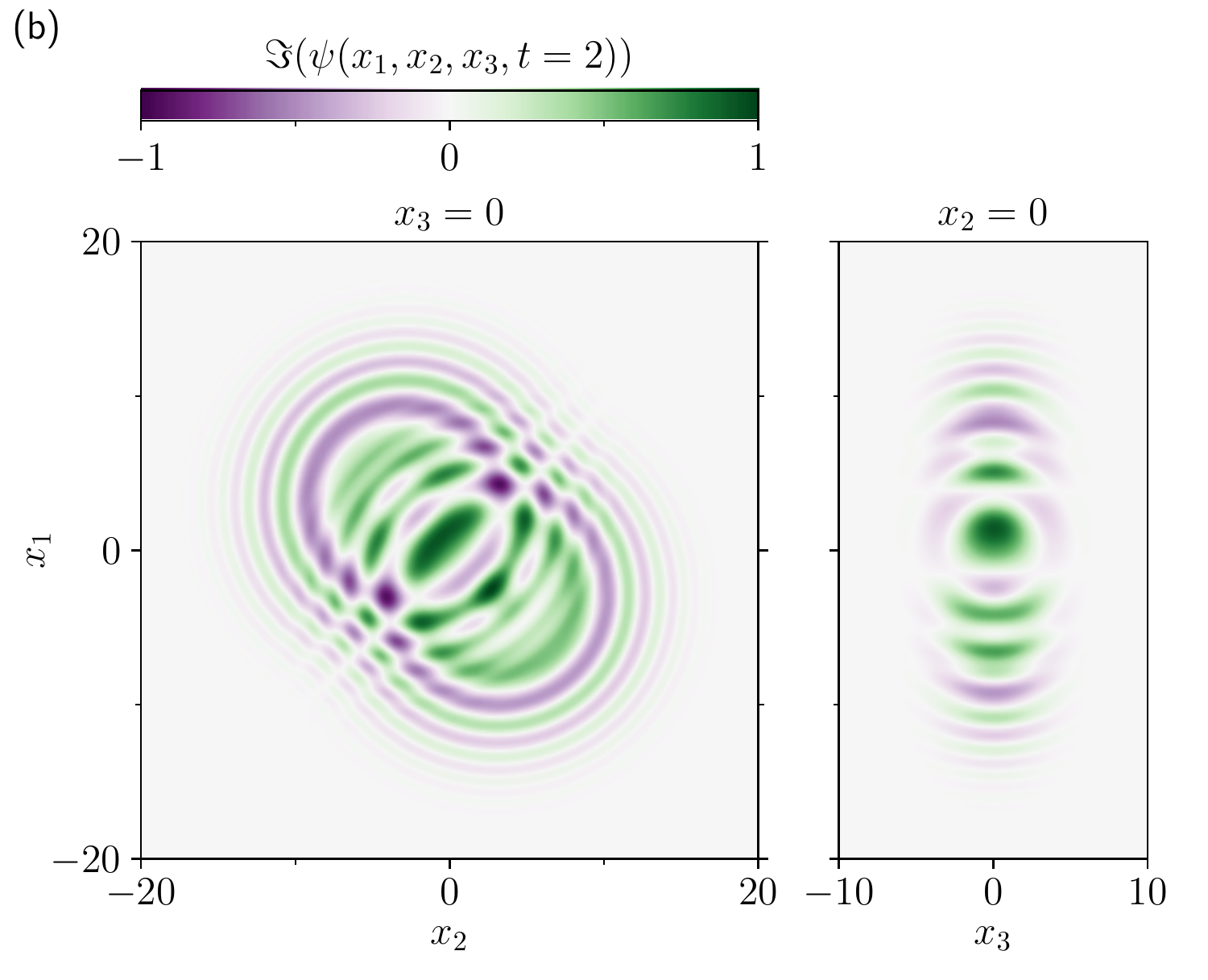} \\
\includegraphics[width=0.45\textwidth]{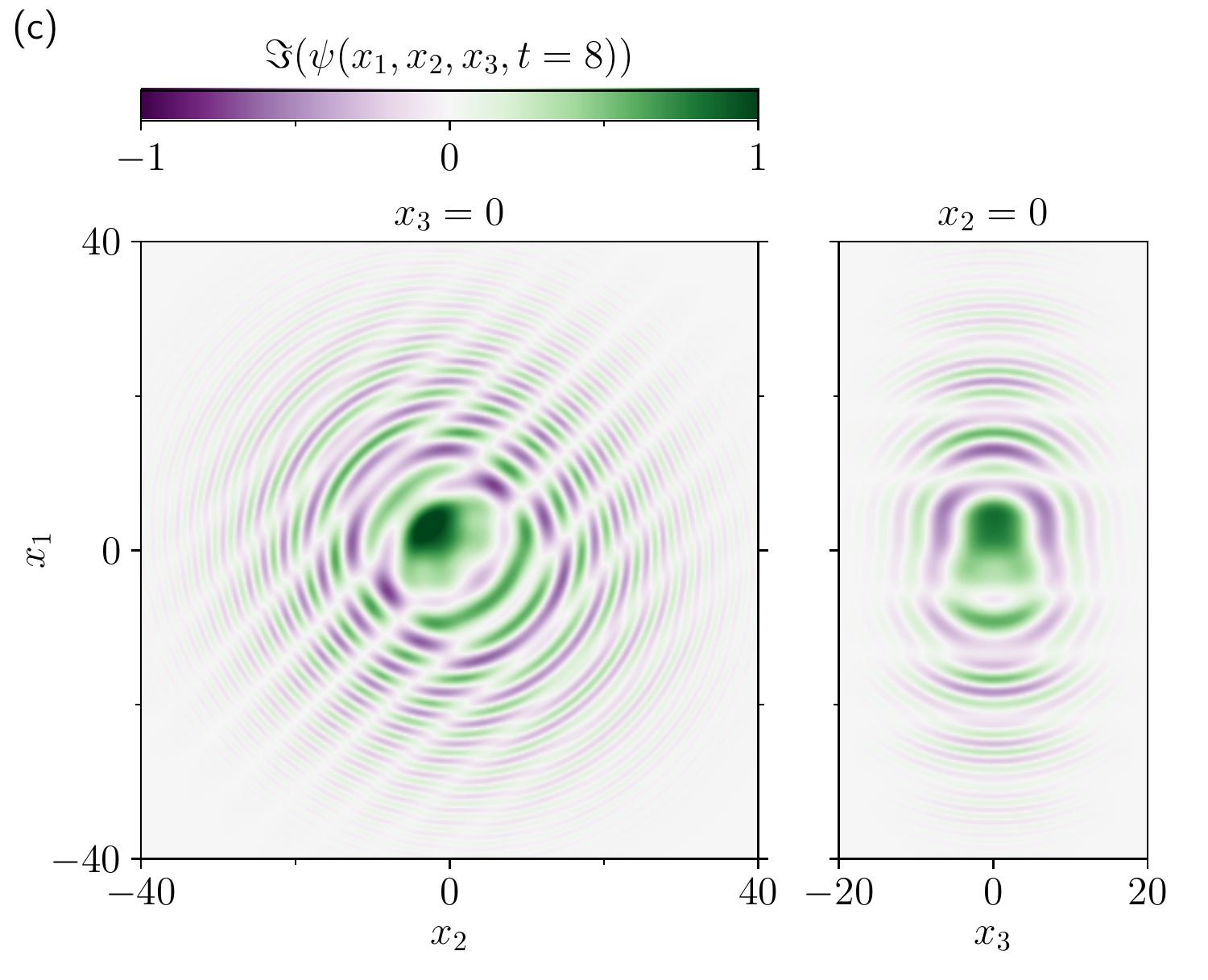}
\includegraphics[width=0.45\textwidth]{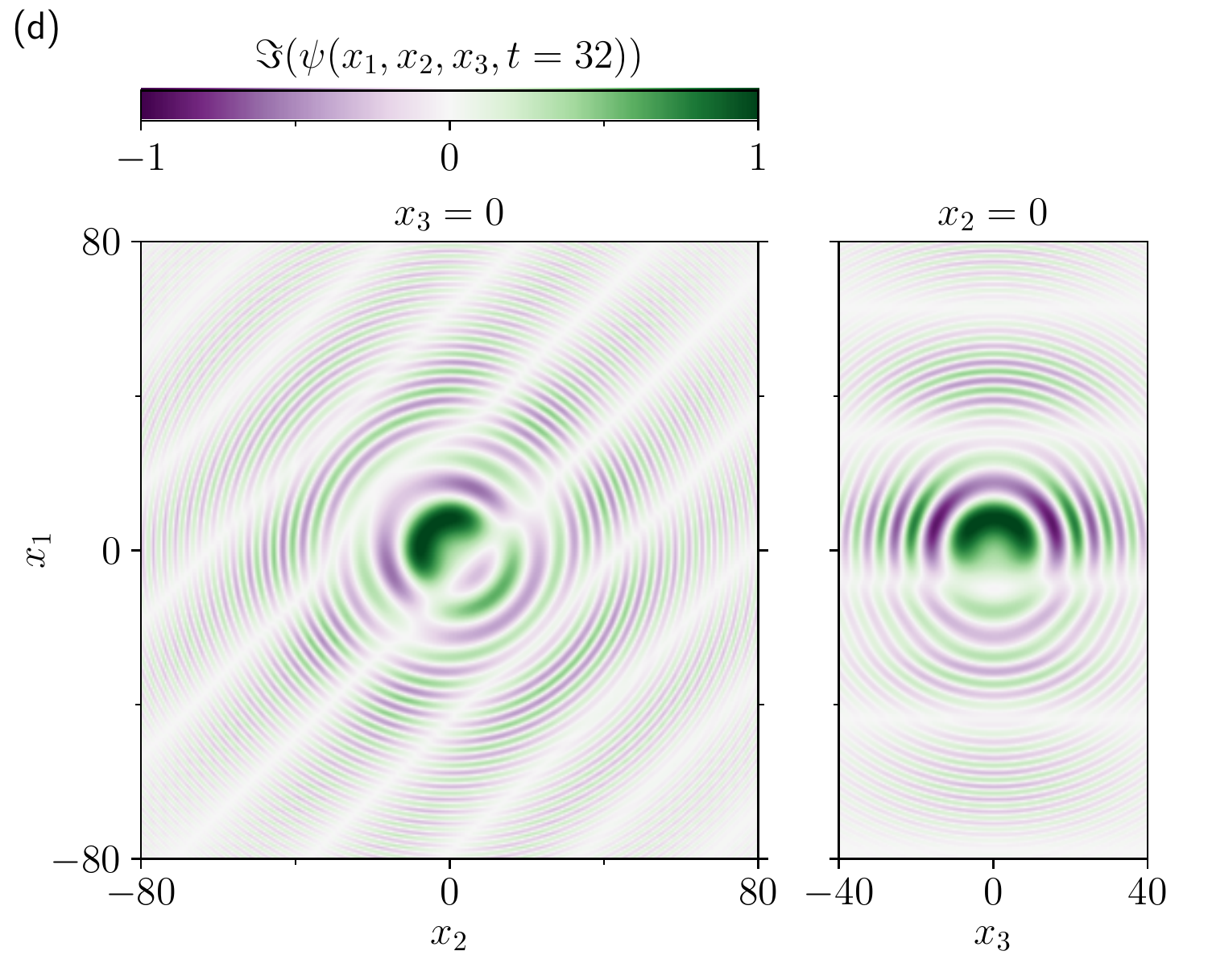}
\caption{
Expansion and interference of two phase shifted ring-shaped wave packets 
in three spatial dimensions at different times.
The initial wave function is given in~\eqref{eq:two_rings_3d} using $t=0$.
(a) Imaginary part of the initial wave function.
(b)-(d) Imaginary part of the approximations of the expanded wave function 
at $t=2$, $t=8$ and $t=32$.
}
\label{fig:rings_3d_imag}
\end{figure}

Finally, we consider the interference of two ring-shaped wave packets in three spatial dimensions.
In this context, we first construct a ring-shaped solution 
\begin{equation}
\label{eq:psi_ring_3d}
\chi(x_1, x_2, x_3, t)
=
\xi(x_1, x_2, t) \zeta(x_3, t)
\end{equation}
to the free Schrödinger equation in three spatial dimensions.
Here, the motion in the $x_3$-direction is described 
by a simple Gaussian wave packet~\cite{schiff_quantum_mechanics}
\begin{equation}
\zeta(x_3,t)
= 
\Big[ \frac{1}{1 + i (t / \tau)} \Big]^{1/2}
\exp 
\Big[ 
-\frac{x_3^2}{4 \sigma^2 [1 + i (t/\tau)]} 
\Big], 
\;\; x_3 \in \R,\;\; t \geq 0
\end{equation}
using $\tau = 2 \sigma^2$,
while the motion in the $(x_1, x_2)$-plane is given by a less trivial ring-like function
that still needs to be specified.

According to~\eqref{eq:time_evol_by_conv_multi_d} we have
\begin{equation*}
\xi(x_1, x_2, t) 
=
\frac{1}{2 \pi i t}
\int_{-\infty}^\infty dx_1^\prime \int_{-\infty}^\infty dx_2^\prime \, 
\exp 
\big[
i (x_1-x_1^\prime )^2+(x_2-x_2^\prime )^2 / (2 t)
\big] 
\xi(x_1^\prime, x_2^\prime, 0) 
\end{equation*}
which gives
\begin{equation*}
\xi(r, \theta, t) 
=
\frac {1}{2 \pi i t}
\int_0^\infty dr' \, r' \int_0^{2 \pi} d\theta' \, 
\exp 
\bigg[ 
\frac{i (r^2+r^{\prime \, 2} )}{2 t} 
\bigg] 
\exp \big(-i r r' \cos(\theta -\theta') / t \big) \,
\xi (r', \theta', 0)
\end{equation*}
using polar coordinates $r = \sqrt{x_1^2+x_2^2}$ and $\theta =\arctan (x_2/x_1)$.
From now on we 
we restrict ourselves to axially symmetric initial conditions, i.e.,
$\xi(r', \theta', 0) = \xi(r', 0)$ being independent of $\theta$.
Due to the rotational symmetry of the Schrödinger equation it remains depending 
on $r$ only also at $t>0$. 
By means of the well-known formula~\cite{abramowitz_stegun_1964} 
\[
\frac 1{2\pi }\int _0^{2\pi }d\theta \, \exp (-i\alpha \cos \theta )=J_0(\alpha ), 
\]
where $J_0 (\alpha )$ is the Bessel function of the first kind of the order 0, we obtain
\begin{equation} 
\label{ff.3} 
\xi(r,t) 
=
\frac{1}{i t}
\int_0^\infty dr' \, r' 
\exp 
\bigg[
\frac{i (r^2+r^{\prime \, 2} )}{2 t}
\bigg] 
J_0 
\bigg( 
\frac{r r^{\prime}}{t}
\bigg) \xi(r^\prime, 0). 
\end{equation}
Next, we further restrict ourselves to initial conditions of the form
\begin{equation}
\label{ff.4}
\xi(r^\prime, 0) 
= 
A (\kappa r^\prime)^s \exp [-(\kappa^2-iq^2)r^{\prime \, 2}],
\end{equation}
where $\kappa $, $q$, and $s$ are real parameters. 
If $s>0$, the wave function at $t=0$ has a ring-like structure with 
the maximum of its absolute value at $r=\sqrt{s/2} \, \kappa ^{-1}$, where, 
without loss of generality, 
we assume $\kappa >0$.  
The normalization $2\pi \int _0^\infty dr\,r|\psi_\perp(r, 0)|^2=1$ is ensured by setting 
\[
A = \sqrt{\frac {2^{s+1}}{\pi \Gamma (s+1)}}\, \kappa , 
\]
where $\Gamma (z)$ is the gamma-function \cite{abramowitz_stegun_1964}. 
We then apply the series expansion of the Bessel function~\cite{abramowitz_stegun_1964} 
\[
J_0 (m r r' / t) 
= 
\sum_{k=0}^\infty \frac {(-1)^k}{(k!)^2} 
\Big( 
\frac{m r r'}{2 t}
\Big)^{2k}
\]
and perform, after substitution of \eqref{ff.4} into \eqref{ff.3}, 
the integration over $r'$ term by term. Summation of the obtained series yields 
\begin{equation}
\label{ff.5}
\xi(r, t)
=
A
\frac{
\kappa^s \Gamma(\frac{s}{2}+1)
}
{
2 i t \big(\kappa^2 - iq^2 - i / (2 t) \big)^{s/2+1}
} 
\exp 
\Big( 
\frac {i r^2}{2 t}
\Big) 
\, 
_1F_1 
\Big[
\frac{s}{2} + 1;
\, 
1;
\, 
-\frac{r^2}{4 t^2 \big( \kappa ^2 - i q^2 - i / (2 t) \big) }
\Big], 
\end{equation} 
where $_1F_1(a;\, b;\, z)=\sum _{k=0}^\infty (a)_kz^k/[(b)_kk!]$ is the 
confluent hypergeometric (Kummer's) function~\cite{abramowitz_stegun_1964} 
and $(a)_k = \Gamma(a+k) / \Gamma(a)$. 

Using the ring-like solution in~\eqref{eq:psi_ring_3d} we define the wave function
\begin{equation}
\label{eq:two_rings_3d}
\psi(x_1, x_2, x_3,t)
=
e^{i \pi / 4} \chi(x_1-\delta, x_2+\delta, x_3, t) 
+ 
e^{-i \pi/ 4} \chi(x_1+\delta, x_2-\delta, x_3, t)
\end{equation}
which defines our initial condition at $t=0$ and serves as a reference solution for $t>0$.
In the example below we choose $\delta=3$, $\kappa = 0.75$, $s=10$, $q=0.5$ and $\sigma=0.85$. 
Moreover, the initial wave function is evaluated on the domain
\[
\Omega_0 
=
[-20, 20] \times [-20, 20] \times [-10, 10]
\]
while the approximations of the expanded wave function at $t=2$, $t=8$ and $t=32$
are computed on the domains 
\begin{align*}
\Omega &= [-20, 20] \times [-20, 20] \times [-10, 10], \\
\Omega &= [-40, 40] \times [-40, 40] \times [-20, 20],
\end{align*}
and
\begin{align*}
\Omega &= [-80, 80] \times [-80, 80] \times [-40, 40],
\end{align*}
respectively.

Densities and imaginary parts of such approximations are shown in the form 
of false-color plots at $x_2 \equiv 0$ and $x_3 \equiv 0$ 
in Fig.~\ref{fig:rings_3d_density} and Fig.~\ref{fig:rings_3d_imag}, respectively. 
In the calculations, the initial condition was evaluated using 
$J_1 = J_2 = 256$ and $J_3 = 128$ grid points.
The same number of grid points $K_1 = K_2 = 256$ and $K_3 = 128$ was used
to approximate the expanded wave functions.
Once again, it becomes evident how much the temporal development 
of the density $\rho = |\psi|^2$ differs from the temporal development of 
the imaginary part $\Im(\psi)$. 
While the density of the expanded wave function at $t=32$ appears to be very smooth, 
the short wavelengths in the imaginary part of the initial condition are
still visible in the imaginary part of the final approximation.

\begin{figure}[!ht]
\centering
\includegraphics[width=0.49\textwidth]{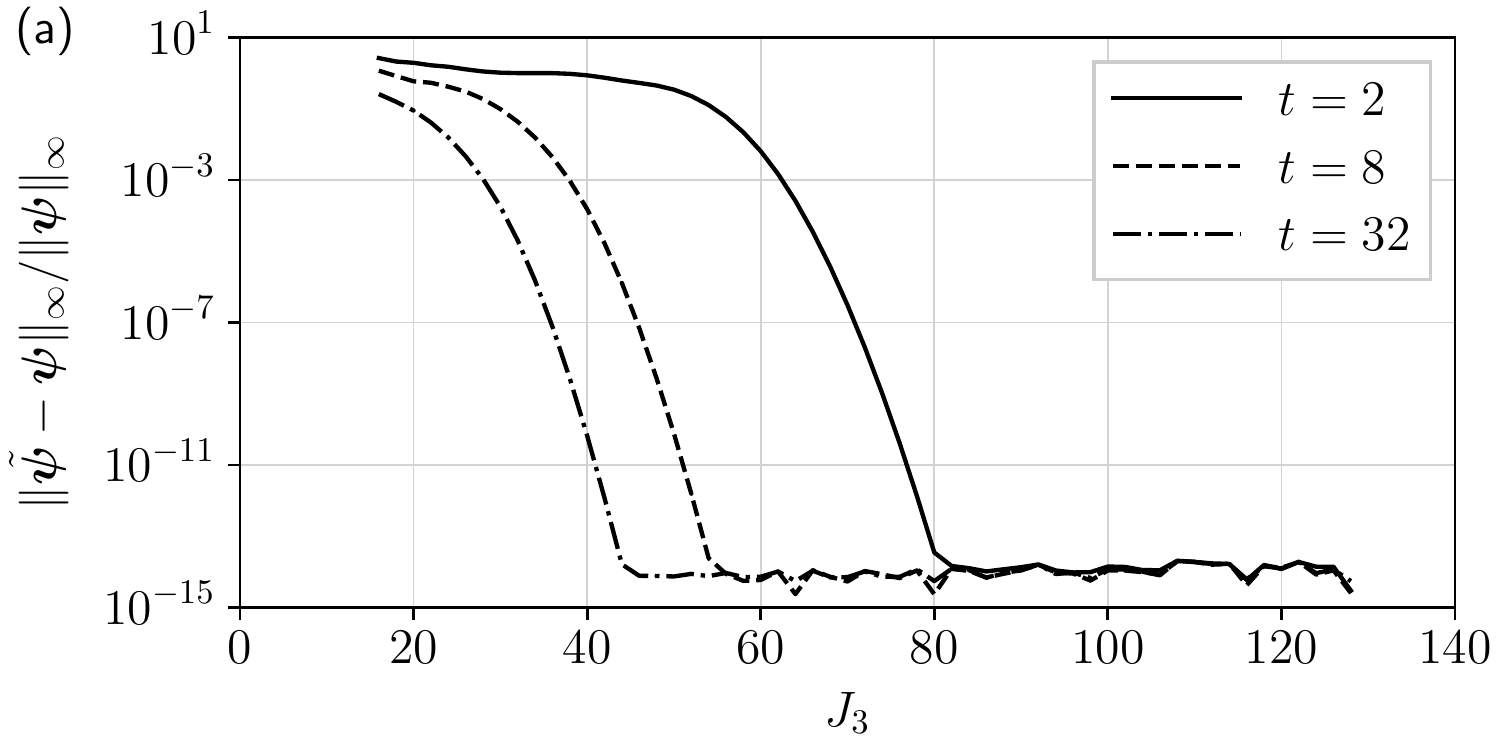}
\includegraphics[width=0.49\textwidth]{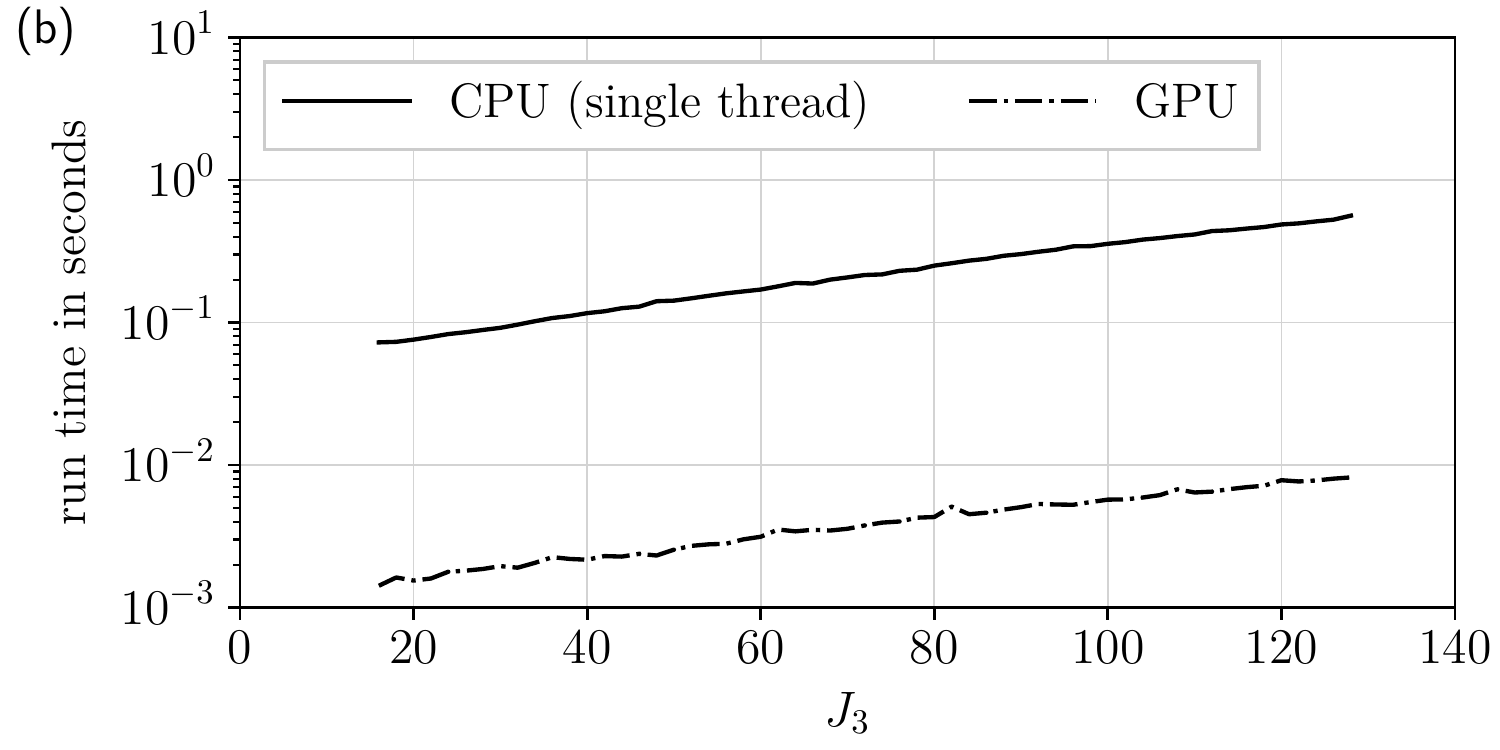}
\caption{
Relative error (a) and run time (b) as a function of the parameter $J_3$
using $J_1 = J_2 = 2 J_3$ for the example showing the expansion 
and interference of two phase-shifted ring-shaped wave packets.
}
\label{fig:rings_3d_conergence_runtime}
\end{figure}

Even this example is solved with incredible accuracy, as shown in Fig.~\ref{fig:rings_3d_conergence_runtime}\,(a) where the relative error 
is plotted as a function of the parameter $J_3=16, 18, \dots, 128$ with $J_1=J_2=2 J_3$.
The parameters $K_1 = K_2 = 256$ and $K_3 = 128$
remain constant throughout the whole simulation.
For completeness, Fig.~\ref{fig:rings_3d_conergence_runtime}\,(b) shows the
computational runtimes which, due to the smaller number of grid points, are even shorter than 
in the previous example.

\section{Conclusion}

We have shown that the discrete Green's function approximation yields highly accurate 
numerical solutions to the free wave packet expansion problem at small numerical costs.
This method will, for example, greatly simplify the simulation of expanding Bose-Einstein
condensates after they have been released from the trap and at times when
atomic interactions have become negligible.
We are therefore convinced that this work represents an important progress 
in terms of modeling, planning and interpretation of experiments in matter-wave interferometry.

\section*{Acknowledgements}

We acknowledge support by the Wiener Wissenschafts- und  TechnologieFonds (WWTF) 
via the project No. MA16-066  (``SEQUEX”) and by the Austrian Science Fund (FWF) 
via the grant SFB F65 ``Taming Complexity in PDE systems”. 
S.E. acknowledges support by the DFG/FWF Collaborative Research Centre via the project SFB 1225 (``ISOQUANT”).
I.M. acknowledges the support by the FQXI program on ``Fueling quantum field machines 
with information''.

\appendix

\section{Error estimate for the Gaussian wave packets example}
\label{sec:appendix_error_estimate_gaussians_1d}

\subsection{Truncation error}
\label{sec:appendix_gaussians_1d_truncation_error}

For $0 < \delta < L/2$, where $L = J h$ and $h = \triangle x'$,
the truncation error \eqref{eq:error_trunc} is bounded by
\begin{align*}
\mathcal{E}_\mathrm{trunc}
\leq 
\frac{h}{\sqrt{2 \pi t}}
\sum_{|j| \geq J/2} |\psi_0(j h)|
\leq
\frac{2 h}{\sqrt{2 \pi t}} 
\sum_{j = J/2}^\infty 
\Big[
e^{-\frac{(jh-\delta)^2}{4 \sigma^2}}
+
e^{-\frac{(jh+\delta)^2}{4 \sigma^2}}
\Big]
\end{align*}
which gives
\begin{align*}
\mathcal{E}_\mathrm{trunc}
\leq
\frac{4 h}{\sqrt{2 \pi t}} 
\sum_{j = J/2}^\infty
e^{-\frac{(jh-\delta)^2}{4 \sigma^2}}
\leq
\frac{4 h}{\sqrt{2 \pi t}} e^{-\frac{((J/2)h-\delta)^2}{4 \sigma^2}}
+
\frac{4}{\sqrt{2 \pi t}} \int_{L/2}^\infty e^{-\frac{(x-\delta)^2}{4 \sigma^2}} \, dx
\end{align*}
and hence
\begin{align*}
\mathcal{E}_\mathrm{trunc}
=
\frac{4}{\sqrt{2 \pi t}} \frac{L}{J}
e^{-\frac{(L/2-\delta)^2}{4 \sigma^2}}
+
\frac{4}{\sqrt{2 \pi t}} \sqrt{\pi} \sigma \operatorname{erfc} \Big( \frac{L - 2 \delta}{4 \sigma} \Big).
\end{align*}
Finally, we replace $J$ with $1$ which yields
\begin{equation*}
\mathcal{E}_\mathrm{trunc}
\leq
\frac{4}{\sqrt{2 \pi t}} L
e^{-\frac{(L/2-\delta)^2}{4 \sigma^2}}
+
\frac{4}{\sqrt{2 \pi t}} \sqrt{\pi} \sigma \operatorname{erfc} \Big( \frac{L - 2 \delta}{4 \sigma} \Big).
\end{equation*}
Using the parameters listed in~\eqref{eq:parameters_gaussians_1d_error_estimate} 
we find that the truncation error in the example
\begin{equation}
\label{eq:E_trunc_gaussians_1d}
\mathcal{E}_\mathrm{trunc} < 4.22 \times 10^{-24}
\end{equation}
is far below the machine precision.

\subsection{Discretization error}
\label{sec:appendix_gaussians_1d_discr_error}

To estimate the discretization error using Theorem~\ref{theo:error_analytic}, 
we consider the expression
\begin{equation*}
|f(x'+iy')| 
= 
|G^{(1)}(x-(x'+iy'), t)| |\psi_0(x'+iy')|
=
\frac{1}{\sqrt{2 \pi t}} e^{xy' / t} e^{-x' y' / t} |\psi_0(x'+iy')|
\end{equation*}
for $x' \in \R, \;\; y' \in (-c, c)$ and some $c>0$.
By means of the estimate
\begin{align*}
|\psi_0(x' + i y')|
\leq
\Big|e^{-\frac{[(x'+iy')-\delta]^2}{4 \sigma^2}}\Big|
+
\Big|e^{-\frac{[(x'+iy')+\delta]^2}{4 \sigma^2}}\Big|
=
e^{\frac{y'^2}{4 \sigma^2}}
\Big(
e^{-\frac{(x'-\delta)^2}{4 \sigma^2}}
+
e^{-\frac{(x'+\delta)^2}{4 \sigma^2}}
\Big)
\end{align*}
we find
\[
\int_{-\infty}^\infty |f(x'+iy')| \,dx'
\leq
\frac{e^{x y' / t}}{\sqrt{2 \pi t}} e^{y'^2 / (4 \sigma^2)}
\int_{-\infty}^\infty e^{-x' y' / t} 
\Big(
e^{-(x'-\delta)^2/(4 \sigma^2)}
+
e^{-(x'+\delta)^2/(4 \sigma^2)}
\Big) \, dx'.
\]
Furthermore,
\[
\int_{-\infty}^\infty e^{-x' y' / t} 
\Big(
e^{-(x'-\delta)^2/(4 \sigma^2)}
+
e^{-(x'+\delta)^2/(4 \sigma^2)}
\Big) \, dx'
=
2 \sqrt{\pi} \sigma e^{(\sigma y')^2 / t^2}
\big(
e^{\delta y' / t} + e^{-\delta y' / t}
\big)
\]
which yields
\begin{align*}
\int_{-\infty}^\infty |f(x'+iy')| \,dx'
&\leq
\sqrt{\frac{2}{t}} \sigma
\exp 
\big[
x y' / t + y'^2 / (4 \sigma^2) + (\sigma y')^2 / t^2
\big]
\big(
e^{\delta y' / t} + e^{-\delta y' / t}
\big).
\end{align*}
The restriction $y' \in (-c,c)$ finally gives
\begin{equation*}
\int_{-\infty}^\infty |f(x'+iy')| \,dx' 
\leq M,
\end{equation*}
where
\[
M
=
\sigma \sqrt{8/t} 
\exp
\Big(
\frac{x c}{t} + \frac{\delta c}{t} + \frac{c^2}{4 \sigma^2} + \frac{\sigma^2 c^2}{t^2}
\Big).
\]

According to Theorem~\ref{theo:error_analytic}, the discretization error is bounded by 
\[
\mathcal{E}_\mathrm{discr} \leq \frac{2M}{e^{2 \pi c / \triangle x'} - 1}.
\]
Next, we substitute $\triangle x'$ with  $L / J$ and
use the parameters listed in~\eqref{eq:parameters_gaussians_1d_error_estimate} which finally
shows that the discretization error is bounded by
\begin{equation}
\label{eq:E_discr_gaussians_1d}
\mathcal{E}_\mathrm{discr}
\leq
\frac{
e^{
\frac{257}{256} c^2 + \big(\frac{1}{8} x + \frac{5}{16}\big) c
}}
{e^{\pi c J / 10} - 1}
\end{equation}
for any $c>0$.

\subsection{Total error}
\label{sec:appendix_gaussians_1d_total_error}

According to \eqref{eq:sum_error_trunc_discr} the total error is bounded by the sum
of the truncation and discretization error.
The estimates in \eqref{eq:E_trunc_gaussians_1d} and \eqref{eq:E_discr_gaussians_1d}
using $x \in [-40, 40]$
therefore show that the error of the discrete Green's function
approximation in the example is bounded by
\begin{equation*}
\|\tilde{\psi}(\cdot, t) - \psi(\cdot, t) \|_{L^\infty[a,b]}
\leq 
\Gamma_c(J)
\quad
\textrm{with}
\quad
\Gamma_c(J)
=
\frac{
e^{
\frac{257}{256} c^2 + \frac{85}{16} c 
}}
{e^{\pi c J / 10} - 1}
+ 4.22 \times 10^{-24}
\end{equation*}
for any positive number $c$.


\section{Error estimates for compactly supported initial wave functions}
\label{sec:appendix_compactly_supported}

This section provides helpful tools to estimate the error of the discrete
Green's function approximation for compactly supported initial wave functions.

\subsection{Derivatives of the one-dimensional Green's function}
\label{sec:appendix_derivatives_of_G_1d}

Let $x \in \mathbbm{R}$ and $\sigma > 0$.
The $n$th derivative of a Gaussian wave packet is given by
\[
\frac{d^n}{d(x')^n} e^{-\frac{1}{2}\frac{(x'-x)^2}{\sigma^2}}
=
(-1)^n \Big( \frac{1}{\sigma \sqrt{2}} \Big)^n 
H_n \Big( \frac{x'-x}{\sigma \sqrt{2}} \Big) e^{-\frac{1}{2}\frac{(x'-x)^2}{\sigma^2}}, 
\;\; x' \in \mathbbm{R}.
\]
Application to the one-dimensional Green's function $G^{(1)}$ in~\eqref{eq:G_1d} using $\sigma = \sqrt{i t}$ yields
\begin{align*}
\frac{\partial^n}{\partial (x')^n} G^{(1)}(x-x', t)
&=
\frac{e^{-i \pi / 4}}{\sqrt{2 \pi t}} \frac{\partial^n}{\partial (x')^n} e^{i (x'-x)^2 / (2 t)} \\
&=
\frac{e^{-i \pi / 4}}{\sqrt{\pi}} (-1)^n e^{-i n \pi/4} q^{n+1}
H_n \big( e^{-i \pi / 4} q (x' - x) \big) e^{i q^2 (x' - x)^2},
\end{align*}
where we introduced the real-valued parameter $q = 1 / \sqrt{2 t}$.
By taking the absolute value we find
\begin{equation}
\label{eq:derivative_G_1d_abs}
\Big| \frac{\partial^n}{\partial (x')^n} G^{(1)}(x - x', t) \Big|
=
\frac{q^{n+1}}{\sqrt{\pi}}
\big| H_n \big( e^{-i \pi / 4} q  (x' - x) \big) \big|.
\end{equation}

\subsection{An unexpected property of the Hermite polynomials}

\begin{lemma}
\label{lemma:f_n_r1_r2}
Let $f_n:\mathbbm{R} \rightarrow \mathbbm{R}$ be given by 
\[f_n(r) = |H_n(r e^{-i \pi / 4})|,
\]
where
\[
H_n(z) = (-1)^n e^{z^2} \frac{d^n}{dz^n} e^{-z^2}, \;\; z \in \C,
\]
denotes the Hermite polynomial of order $n \in \mathbbm{N}_0$.
Moreover, let $r_1, r_2 \in \mathbbm{R}$ with $|r_1| \leq |r_2|$.
Then
\[
0 \leq f_n(r_1) \leq f_n(r_2).
\]
\end{lemma}

To prove  Lemma~\ref{lemma:f_n_r1_r2} we first note that $H_n(-z) = (-1)^n H_n(z)$
for every $z \in \mathbbm{C}$ and $n \in \mathbbm{N}_0$.
Consequently, it remains to proof that $f_n$ is non-decreasing on $[0, \infty)$.
To this end, we will show  that
\begin{align}
\label{eq:A}
\frac{d}{dr} |H_n(r e^{-i \pi / 4})|^2 \geq 0, \;\; r \geq 0, \;\; n \in \mathbbm{N}_0.
\end{align}

Using $|H_n(r e^{-i \pi / 4})|^2 = H_n(r e^{-i \pi /4}) H_n(r e^{i \pi /4})$ and
\begin{equation*}
\frac d{dz} H_n(z) = 2n H_{n-1}(z), \;\; n \in \mathbbm{N},
\end{equation*}
we obtain
\begin{equation}
\label{eq:B}
\begin{aligned}
\frac{d}{dr} \Big[ |H_n(r e^{-i \pi /4})|^2 \Big]
&=
2n
\Big\{
e^{-i\pi /4} H_{n-1}(r e^{-i \pi /4}) H_n(r e^{i \pi /4}) \\
&\qquad+ 
e^{i \pi /4} H_n(r e^{-i \pi /4}) H_{n-1}(r e^{i \pi /4})
\Big\}. 
\end{aligned}
\end{equation}
We then apply
\begin{equation} 
\label{eq:C} 
H_{n+1}(z) = 2 z H_n(z)-2nH_{n-1}(z)     
\end{equation}
to the terms $H_n(r e^{i \pi / 4})$ and $H_n(r e^{-i \pi / 4})$, which yields
\begin{align*}
&\frac d{dr} \Big[ |H_n(r e^{-i \pi /4})|^2 \Big]
=
8 n r |H_{n-1} (r e^{-i \pi / 4})|^2
- 4 n (n-1) 
\Big\{
e^{-i \pi / 4} H_{n-1}(r e^{-i \pi / 4}) H_{n-2}(r e^{i \pi / 4}) \\
&\qquad+
e^{i \pi / 4} H_{n-2}(r e^{-i \pi / 4}) H_{n-1}(r e^{i \pi / 4})
\Big\}.
\end{align*}
Next, we replace $H_{n-1}(r e^{- i \pi / 4})$ and $H_{n-1}(r e^{i \pi / 4})$ 
in the second term (in curly brackets) using \eqref{eq:C} again.
After some algebra we obtain
\begin{align*}
&\frac d{dr} \Big[ |H_n(r e^{-i \pi /4})|^2 \Big]
=
8 n r |H_{n-1} (r e^{-i \pi / 4})|^2 
+ 8 n (n-1) (n-2) \Big\{
e^{-i \pi / 4}  H_{n-3}(r e^{-i \pi / 4}) H_{n-2}(r e^{i \pi / 4})\\
&\qquad 
+ 
e^{i \pi / 4} H_{n-2}(r e^{-i \pi / 4}) H_{n-3}(r e^{i \pi / 4})
\Big\}.
\end{align*}
By means of \eqref{eq:B} we finally find
\begin{equation}
\label{eq:D} 
\frac{d}{dr} \Big[ |H_n(r e^{-i \pi /4})|^2 \Big] 
=
8 n r |H_{n-1}(r e^{-i \pi /4})|^2 + 4n(n-1)
\frac{d}{dr}
\Big[ | H_{n-2}(r e^{-i \pi /4}) |^2 \Big].                                       
\end{equation} 
The first term on the right-hand side of \eqref{eq:D}  is non-negative for $r\geq 0$. 
Furthermore, we have $H_0(z) = 1$, $H_1(z) = 2z$ and, hence,
\begin{equation}
\label{eq:E} 
\frac{d}{dr} \Big[ | H_0(r e^{-i \pi / 4}) |^2 \Big] = 0,
\qquad
\frac{d}{dr} \Big[ | H_1(r e^{-i \pi / 4}) |^2 \Big] = 8 r \geq 0.
\end{equation}
Finally, the assertion in \eqref{eq:A} follows from \eqref{eq:D} and \eqref{eq:E} 
using induction on $n$.

\subsection{Error estimation for compactly supported initial wave functions}

Using \eqref{eq:derivative_G_1d_abs} in combination with Lemma~\ref{lemma:f_n_r1_r2} 
shows that
\begin{equation*}
\max_{\substack{x' \in [-L/2, L/2] \\ x \in [a,b]}}
\Big| \frac{\partial^n}{\partial (x')^n} G^{(1)}(x - x', t) \Big|
\leq
\frac{q^{n+1}}{\sqrt{\pi}}
\big| H_n \big( R e^{-i \pi / 4} \big) \big|,
\end{equation*}
where $H_n$ are the Hermite polynomials,
$R = q \big[ (L/2) + \max(|a|, |b|) \big]$ and
$q=1/\sqrt{2 t}$.
Moreover, applying Leibniz's rule to
\[
f(x, x', t) = G^{(1)}(x-x', t) u(x'), 
\;\; x \in [a,b], \;\; x' \in [-L/2, L/2], \;\; t > 0,
\]
yields
\begin{equation*}
\frac{\partial^n f(x, x', t)}{\partial (x')^n}
=
\sum_{\ell=0}^n \binom{n}{\ell} 
\bigg[
\frac{\partial^\ell}{\partial (x')^\ell} G^{(1)}(x-x', t)
\bigg]
\bigg[
\frac{\partial^{n-\ell}}{\partial (x')^{n-\ell}}
u(x')
\bigg]
\end{equation*}
which in turn gives
\begin{subequations}
\label{eq:euler_maclaurin_utils}
\begin{equation}
\begin{aligned}
\label{eq:euler_maclaurin_utils_1}
&\max_{x \in [a,b]} \bigg| \frac{\partial^n f(x,x',t)}{\partial (x')^n} \Big|_{x'=L/2} 
- \frac{\partial^n f(x,x',t)}{\partial (x')^n} \Big|_{x'=-L/2} \bigg| \\
&\qquad \qquad\leq
2 \sum_{\ell=0}^n \binom{n}{\ell} \frac{q^{\ell+1}}{\sqrt{\pi}} \big|H_\ell(R e^{-i \pi / 4})\big|
\max_{x' \in \{-L/2, L/2\}} \big| u^{(n-\ell)}(x') \big|
\end{aligned}
\end{equation}
as well as
\begin{equation}
\label{eq:euler_maclaurin_utils_2}
\max_{\substack{x' \in [-L/2, L/2]\\x \in [a,b]}}
\bigg| 
\frac{\partial^n f(x,x',t)}{\partial (x')^n}
\bigg|
\leq
\sum_{\ell=0}^n \binom{n}{\ell} \frac{q^{\ell+1}}{\sqrt{\pi}} \big|H_\ell(R e^{-i \pi / 4})\big| \max_{x' \in [-L/2, L/2]} \big| u^{(n-\ell)}(x') \big|.
\end{equation}
\end{subequations}

The error of the discrete Green's function approximation~\eqref{eq:E_compact} depends implicitly
on $x \in [a,b]$.
For every fixed $x \in [a,b]$ it can be estimated via formula
\eqref{eq:euler_maclaurin_error} in Theorem~\ref{theo:euler_maclaurin}.
The required factors $\lambda_\ell$, $\ell=1, \dots, m$ and $\nu_m$
are also dependent on $x$.
Using \eqref{eq:euler_maclaurin_utils_1} and \eqref{eq:euler_maclaurin_utils_2} 
we find $\lambda_\ell^*$ and $\nu_m^*$ such that
$\lambda_\ell^* > \max_{x \in [a,b]} \lambda_\ell(x)$
and $\nu_m^* > \max_{x \in [a,b]} \nu_m(x)$. 
According to Theorem~\ref{theo:euler_maclaurin} the error is bounded by
\begin{equation*}
\max_{x \in [a,b]} |\tilde{\psi}(x,t) - \psi(x,t)| 
\leq
\sum_{\ell=1}^m \lambda_\ell^* h^{2 \ell} + \nu_m^* h^{2m + 2},
\end{equation*}
where $h = L/J$.
Thus, we have found constants $C_m(J)$ such that
\[
\|\tilde{\psi}(\cdot, t) - \psi(\cdot, t) \|_{L^\infty[a,b]} \leq C_m(J).
\]


\bibliographystyle{elsarticle-num} 
\bibliography{mybib}





\end{document}